\documentclass[twocolumn,twocolappendix]{aastex7}

\accepted{accepted by ApJ}

\shorttitle{What Are Extreme Debris Disks? }
\shortauthors{Su et al.}

\begin{document}

\title{Extreme Debris Disks: Insights into Violent Collisions in Planet Formation and Destruction}

\author[orcid=0000-0002-3532-5580]{Kate Y. L. Su}
\affiliation{Space Science Institute, 4750 Walnut Street, Suite 205, Boulder, CO 80301, USA}
\email[show]{ksu@spacescience.org}

\author[orcid=0009-0001-9360-2670]{Attila Mo\'or}
\affiliation{Konkoly Observatory, HUN-REN Research Centre for Astronomy and Earth Sciences, MTA Centre of Excellence, Konkoly-Thege Mikl\'os \'ut 15-17, 1121 Budapest, Hungary}
\affiliation{CSFK, MTA Centre of Excellence, Budapest, Konkoly Thege Mikl\'os \'ut 15-17., H-1121, Hungary}
\email[]{moor.attila@csfk.org}

\author[orcid=0000-0001-7157-6275]{\'Agnes K\'osp\'al}
\affiliation{Konkoly Observatory, HUN-REN Research Centre for Astronomy and Earth Sciences, MTA Centre of Excellence, Konkoly-Thege Mikl\'os \'ut 15-17, 1121 Budapest, Hungary}
\affiliation{CSFK, MTA Centre of Excellence, Budapest, Konkoly Thege Mikl\'os \'ut 15-17., H-1121, Hungary}
\affiliation{Max-Planck-Insitut f\"ur Astronomie, K\"onigstuhl 17, 69117 Heidelberg, Germany}
\email[]{kospal.agnes@csfk.org}

\author[orcid=0000-0003-2303-6519]{George H. Rieke}
\affiliation{Department of Astronomy and Steward Observatory, University of Arizona, Tucson, AZ 85721, USA}
\affiliation{Lunar and Planetary Laboratory, The University of Arizona, Tucson, AZ 85721, USA}
\email[]{grieke@arizona.edu}

\author[0000-0003-4623-1165]{Antranik A. Sefilian}
\affiliation{Department of Astronomy and Steward Observatory, University of Arizona, Tucson, AZ 85721, USA}
\email[]{asefilian@as.arizona.edu}

\author[0000-0002-1226-3305]{Renu Malhotra}
\affiliation{Lunar and Planetary Laboratory, The University of Arizona, Tucson, AZ 85721, USA}
\email[]{malhotra@arizona.edu}

\author[orcid=0000-0001-7962-1683]{Ilaria Pascucci}
\affiliation{Lunar and Planetary Laboratory, The University of Arizona, Tucson, AZ 85721, USA}
\email[]{pascucci@arizona.edu}

\author[orcid=0000-0003-4393-9520]{Alan P. Jackson}
\affiliation{Department of Physics, Astronomy, and Geosciences, Towson University, 8000 York Road, Towson, MD, USA}
\email[]{alanjackson@towson.edu}

\author[orcid=0000-0001-6015-646X]{P\'eter \'Abrah\'am}
\affiliation{Konkoly Observatory, HUN-REN Research Centre for Astronomy and Earth Sciences, MTA Centre of Excellence, Konkoly-Thege Mikl\'os \'ut 15-17, 1121 Budapest, Hungary}
\affiliation{CSFK, MTA Centre of Excellence, Budapest, Konkoly Thege Mikl\'os \'ut 15-17., H-1121, Hungary}
\affiliation{Department of Astrophysics, University of Vienna, T\"urkenschanzstr. 17, A-1180 Vienna, Austria}
\email[]{abraham.peter@csfk.org}

\author[orcid=0000-0002-4276-3730]{Nicholas P. Ballering}
\affiliation{Space Science Institute, 4750 Walnut Street, Suite 205, Boulder, CO 80301, USA}
\affiliation{Department of Astronomy, University of Wisconsin-Madison, Madison, WI 53706, USA}
\email[]{nballering@SpaceScience.org}

\begin{abstract}


Debris disks are dusty structures around mature stars, primarily identified by infrared excesses in the stellar spectra. A subset, known as extreme debris disks (EDDs), exhibits stochastic infrared variability, believed to result from large-scale violent collisions that contribute to the formation or destruction of rocky planetary bodies. We analyze JWST and Spitzer mid-infrared spectra for 21 EDDs to investigate the connection between impact-produced dust mineralogy and the age and dynamical state of these systems during various stages of planet formation and evolution. Our findings indicate that EDDs contain significantly more optically thin, small dust grains compared to those typically seen in protoplanetary and debris disks. Predominantly submicron in size, these grains are thermally altered, as shown by their high levels of silica and crystalline silicate composition. Along with stochastic infrared variability, these features define EDDs as a subclass of debris systems where dust is generated from large collisions between Moon- and Mars-sized bodies. Our results not only provide diagnostic information about the physical conditions of violent events during terrestrial planet formation, aiding in differentiating the complex outcomes of these impacts, but also offer a potential marker for identifying planetary systems experiencing dynamical instability.

\end{abstract}


\section{Introduction}
\label{sec:intro}

Stars are born with circumstellar disks of gas and dust, which serve as the cradles for planet formation and evolution. Giant planets are estimated to form within $\sim$10 Myr before the dispersal of gas in protoplanetary disks (PPDs) \citep{pascucci09}, while ice giants and terrestrial planets can continue to grow in the later gas-poor debris disk (DD) phase \citep{armitage24}. Dust in DDs is commonly referred to as second-generation because it primarily arises from collisions between leftover planetesimals that failed to form planets \citep{wyatt08}.  Consequently, the dust levels in DDs reflect the subsequent planet formation activity, as large bodies perturb and collide with smaller ones. An effective approach to studying these collisional processes in exoplanetary systems is to analyze infrared emissions from circumstellar dust (i.e., infrared excess) in low-optical-depth systems like DDs, as the amount of dust is directly linked to these processes \citep{kenyon05}. Our solar system has experienced enhanced collisional activity, especially during chaotic phases, including the giant impact that formed the Earth-Moon system \citep{asphaug14_MoonFormationGiantImpact} and intense bombardment during giant planet migration \citep{strom05}. By integrating infrared observations with theoretical models, researchers have identified a subclass of debris disks known as extreme debris disks (EDDs), characterized by exceptionally large (re-emitting more than $\sim$1\% of the stellar luminosity) and time-variable amounts of warm ($\gtrsim$a few hundred K) dust found around only about $\sim$1\% of DDs of similar age surrounding relatively young stars \citep{balog09,kennedy_wyatt13,meng17}. EDDs are thought to represent the chaotic phases occurring within exoplanetary systems.

The substantial warm emissions -- coupled with significant variability -- link EDDs to the rapid evolution of dust debris in the terrestrial zone. Two key pieces of evidence suggest that large-scale collisions involving large asteroids and planetary embryos give rise to the EDD phenomenon. First, the disk emissions in EDDs display stochastic variability on timescales of months to years, in contrast to the stable emissions observed in typical DDs, which are thought to be sustained by much more frequent collisions among smaller, $\sim$km-sized planetesimals. Most EDDs are found around stars younger than $\sim$200 Myr, with about two-thirds exhibiting variable infrared emissions indicative of rapid dust production and clearing \citep{moor21}. Theoretical studies indicate that violent collisions of large bodies can generate dense debris clouds orbiting the star, leading to complex semi-regular variability influenced by viewing geometry and the dynamical and collisional evolution of the debris clouds \citep{jackson12,watt24_postimpact_edd_evolution}. This “wiggle” behavior in infrared light curves has been extensively documented in three EDDs \citep{su19,moor22_tyc4209}. In rare inclined orientations, eclipsing dust clumps have been observed to be comparable in size to the star (e.g., \citealt{su22_hd166}), further validating the scale of these impacts.

Secondly, EDDs show very prominent solid-state features in the mid-infrared arising from $\sim$sub-\micron\ amorphous/crystalline grains. The majority of DDs have infrared spectra that display, at most, only subtle features \citep{chen06,ballering14,chen14}, resulting from erosive collisions between smaller, km-sized planetesimals and the rapid loss of small grains due to stellar radiation pressure. Because of the short lifetimes of small grains in DDs, strong spectral features indicate an elevated level of dust-producing collisions. Theoretical simulations suggest that collisions between large, gravity-dominated bodies are violent events that generate powerful shocks, vaporizing a significant portion of the colliding bodies \citep{kraus12,davies20}. This vapor is expected to quickly condense into dust \citep{johnson12a,johnson15}. These violent impacts are further substantiated by three EDDs that exhibit a distinct $\sim$9 \micron\ silica feature linked to freshly condensed silica smoke \citep{rhee08,fujiwara12}, produced by hypervelocity impacts \citep{lisse09}. In addition to direct vapor condensation, annealing at high temperatures ($\gtrsim$1200 K) -- a localized condition resulting from large-scale impacts -- can also transform amorphous silicate material into crystalline forms, such as forsterite (Mg-rich olivine) grains, which have been observed in several EDDs \citep{weinberger11,olofsson12,meng14}. As suggested by \citet{morlok14}, the mineralogical dichotomy (silica-rich vs.\ forsterite-rich) observed in EDDs may be related to the initial conditions following impacts, as traced by these thermally altered grains. This makes the study of impact-produced dust mineralogy a valuable tool for investigating giant impacts occurring around other stars.

The discovery of EDDs occurred near the end of the Spitzer cryogenic mission; therefore, only a handful of systems with high-quality mid-infrared spectra are available. With the launch of JWST, we are now in a position to further explore the mineralogical dichotomy in EDDs by more than doubling the sample size. In this work, we analyze 16 EDDs using JWST MIRI/MRS data, with 12 of these having the complete 4.9--27.9 \micron\ mid-infrared spectra obtained for the first time. 
By including an additional five EDDs observed by Spitzer, we revisit the mineralogical dichotomy (silica-rich vs.\ forsterite-rich) reported in earlier Spitzer studies, and provide a global view of EDD properties linked to impact-produced dust mineralogy, placing them within a broader context compared to earlier PPDs and typical DDs by highlighting three characteristics that define this subclass. 

The rest of the paper is organized as follows. The JWST observations, data reduction, and methodology for extracting dust features are briefly discussed in Section \ref{sec:data}, with additional details and basic sample properties provided in Appendix \ref{sec:jwstsample}. In Section \ref{sec:analysis}, we explore the unique characteristics observed in EDDs, highlighting elevated amounts of small and thermally altered grains (Sections \ref{sec:edds_I} and \ref{sec:edds_II}), as well as infrared variability (Section \ref{sec:edds_III}). Further technical details are elaborated in Appendices \ref{sec:dustindices} and \ref{sec:irvariability}. We discuss the broader implications of our findings concerning different phases of planet formation and evolution, comparing them to theoretical expectations in Section \ref{sec:discussion}, and conclude in Section \ref{sec:conclusion}.

\section{Observations and Results}
\label{sec:data}

\subsection{Mid-infrared Spectra from JWST and Spitzer } 

This work reports mid-infrared spectral analysis for 16 EDDs using JWST MIRI/MRS data from the GTO and GO programs (PID 1206 -- PI Rieke, PID 1282 -- PI Henning, and PID 3189 -- PI Su), collectively designed to provide a global view of EDD properties related to impact-produced dust mineralogy. Four of the targets have high-quality mid-infrared spectra obtained by Spitzer/IRS $\sim$20 years ago, while the remaining twelve have mid-infrared spectra acquired by JWST/MIRI for the first time. Table \ref{tab:jwst_obs} lists 
the JWST observations, including the date, on-source exposure time, target identifications, and short names used in the work hereafter. All data were obtained using the MIRI/MRS mode with all four channels and subbands, employing 4-point point-source dithers, resulting in a continuous spectrum ranging from 4.9 to 27.9 \micron. No dedicated background observations were obtained.

We used the JWST Calibration Pipeline version 1.15 \citep{bushouse23_jwstpipeline} and the Calibration Reference Data System context \texttt{pmap\_1298} to reduce the data using default parameter settings, following the same steps as those in the work of \citet{su25_hd23514} to flag additional bad pixels. We inspected the source extension (in terms of full-width at half-maximum, FWHM) and compared it to that of calibration stars with similar brightness across all spectral cubes and wavelengths, confirming that no significant extended structures were observed around the targets. A 1D spectrum was extracted from each of the cubes using the default aperture of 2 PSF FWHM to minimize resampling artifacts, along with a large sky annulus, both of which increase linearly with wavelengths \citep{law25_mrscalibration}. 

\begin{figure*}
    \centering
    \includegraphics[width=0.49\linewidth]{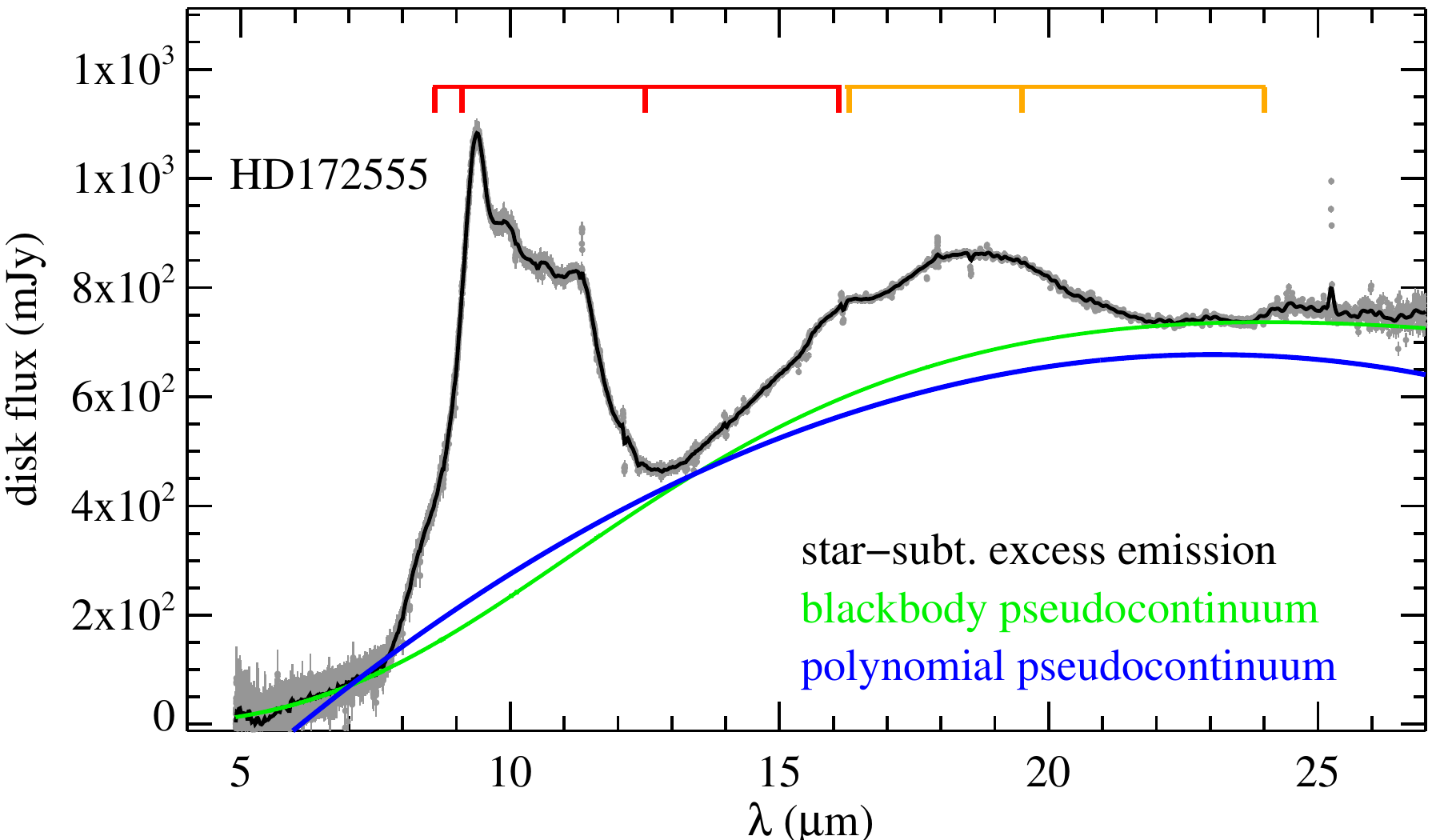}
    \includegraphics[width=0.49\linewidth]{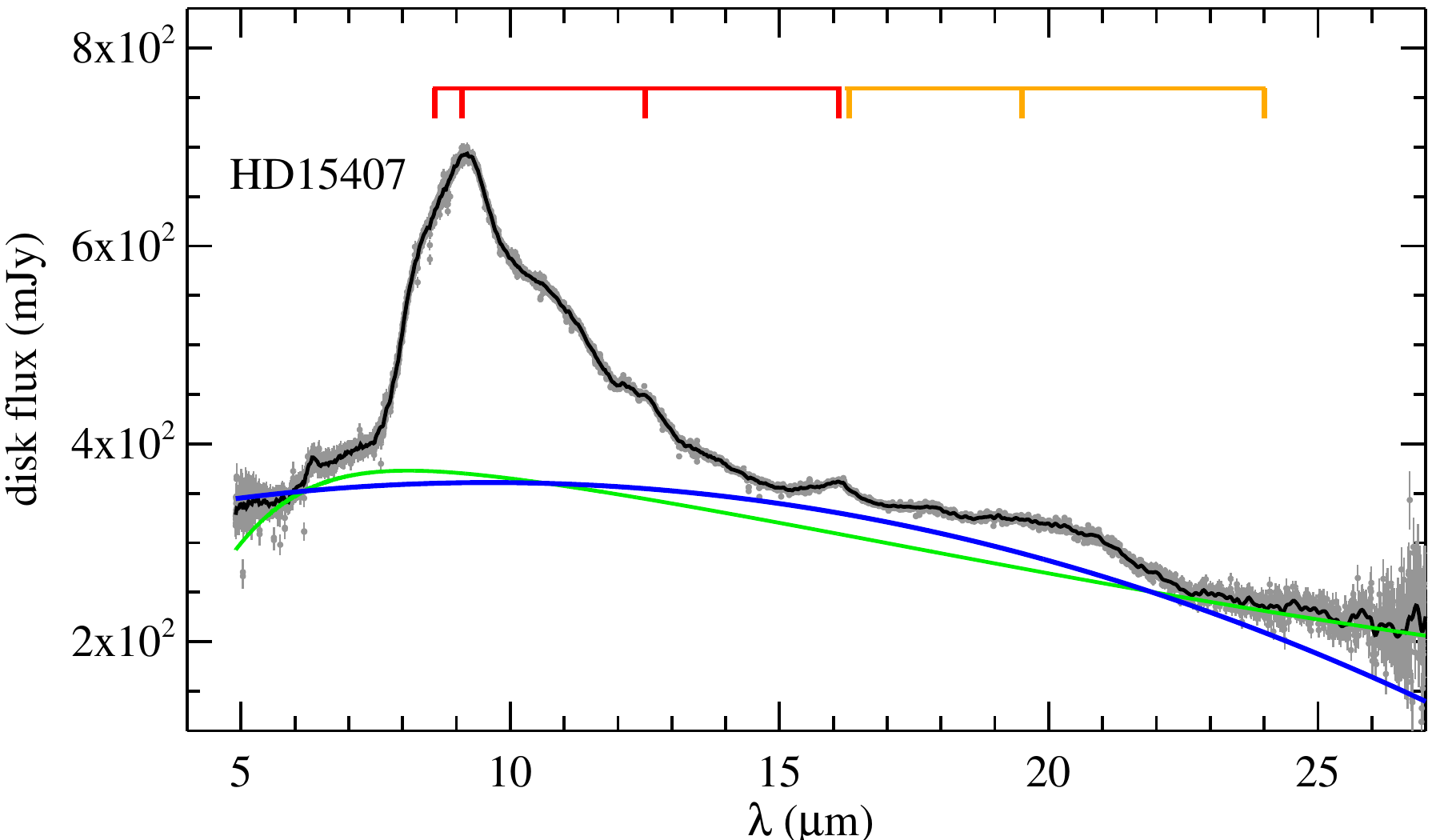}
    \includegraphics[width=0.49\linewidth]{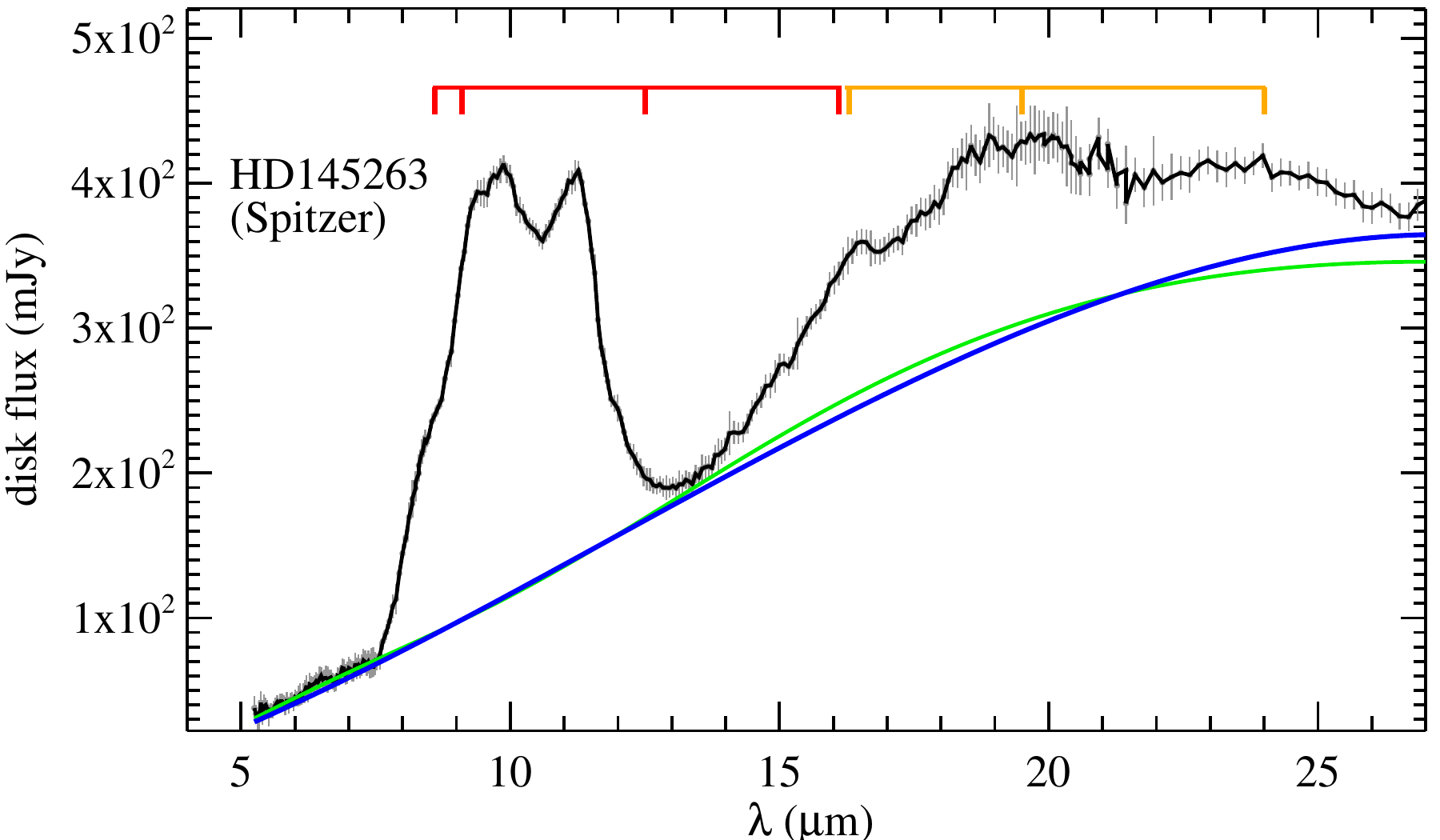}
    \includegraphics[width=0.49\linewidth]{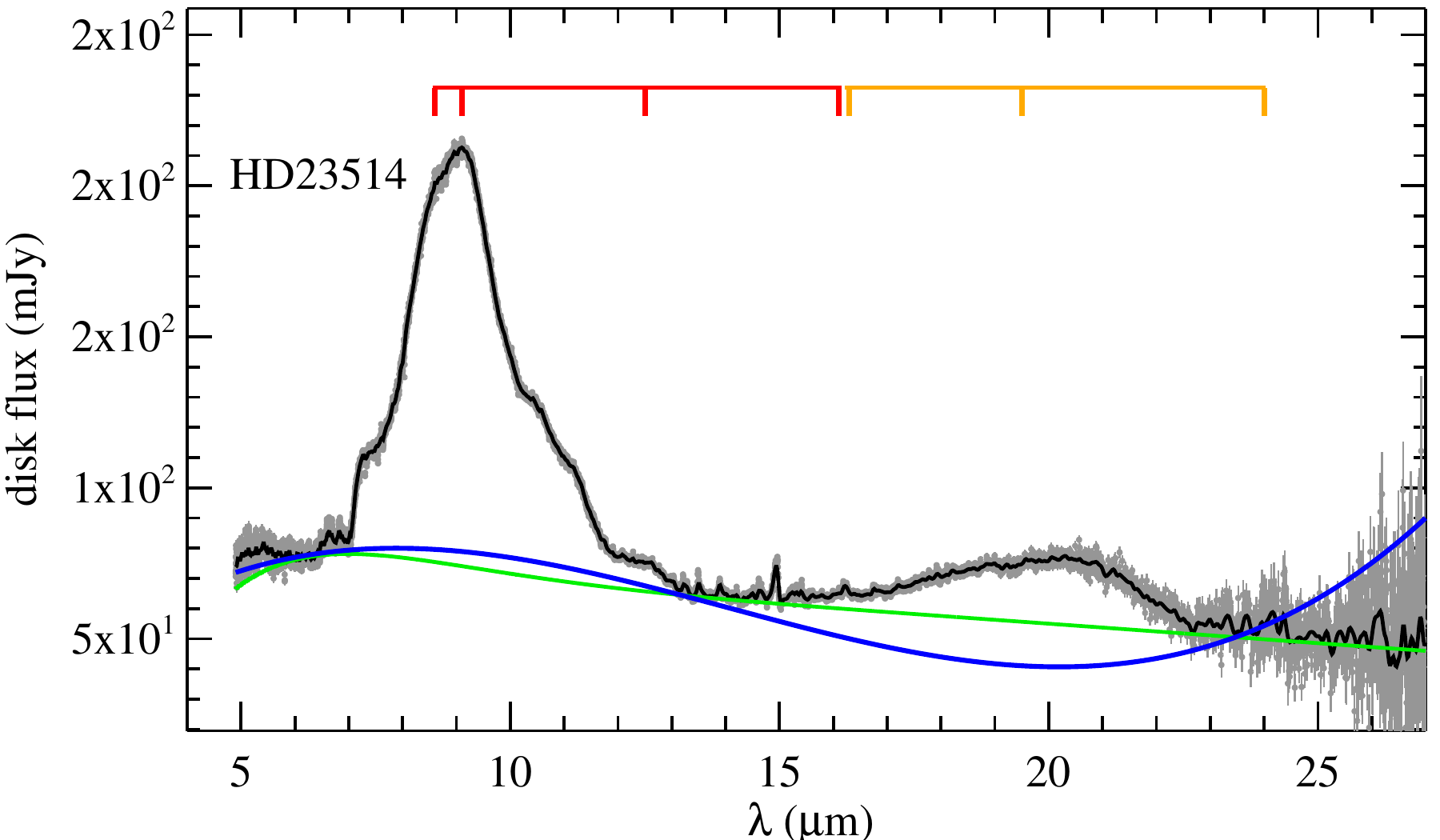}
    \includegraphics[width=0.49\linewidth]{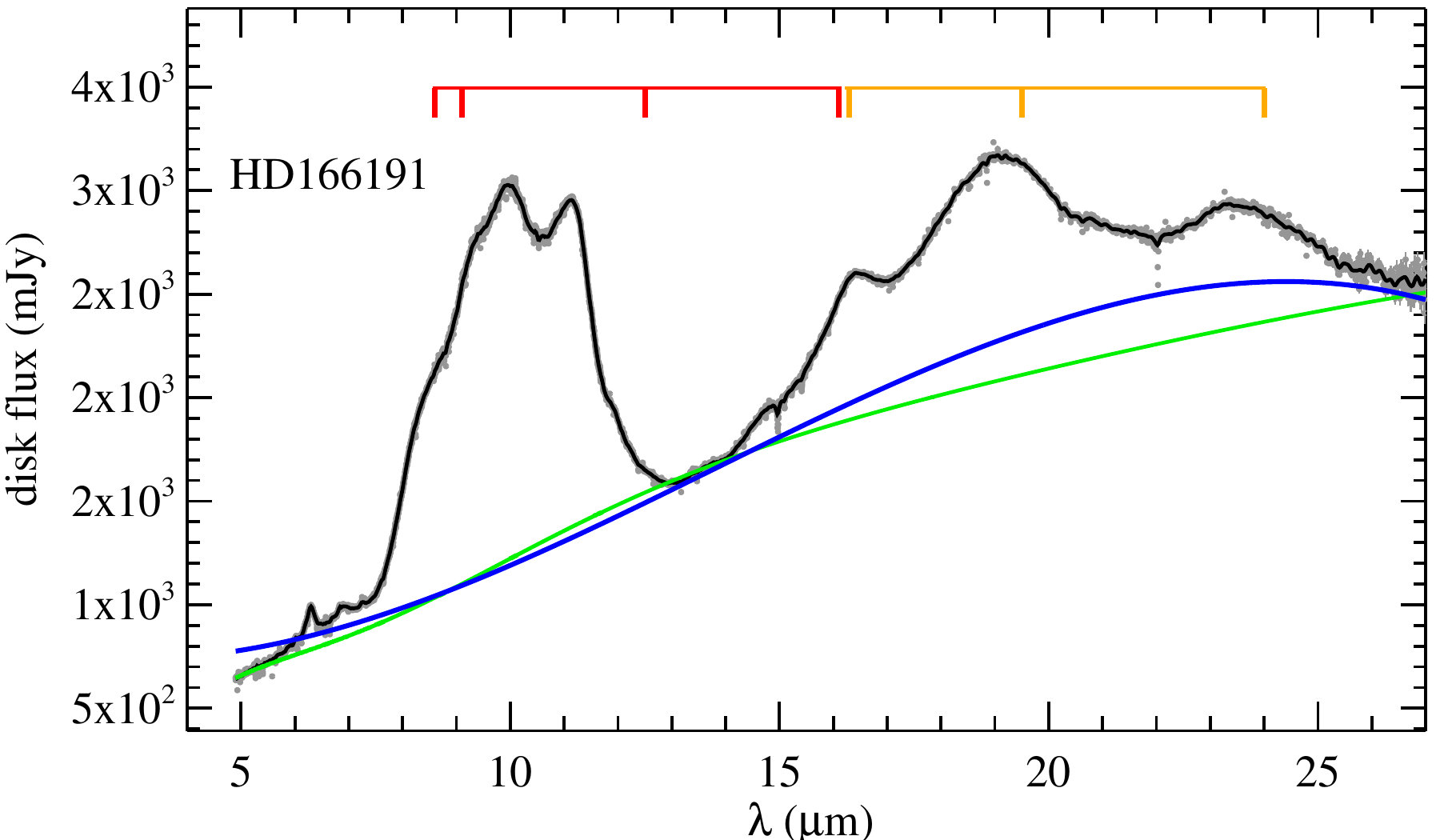}
    \includegraphics[width=0.49\linewidth]{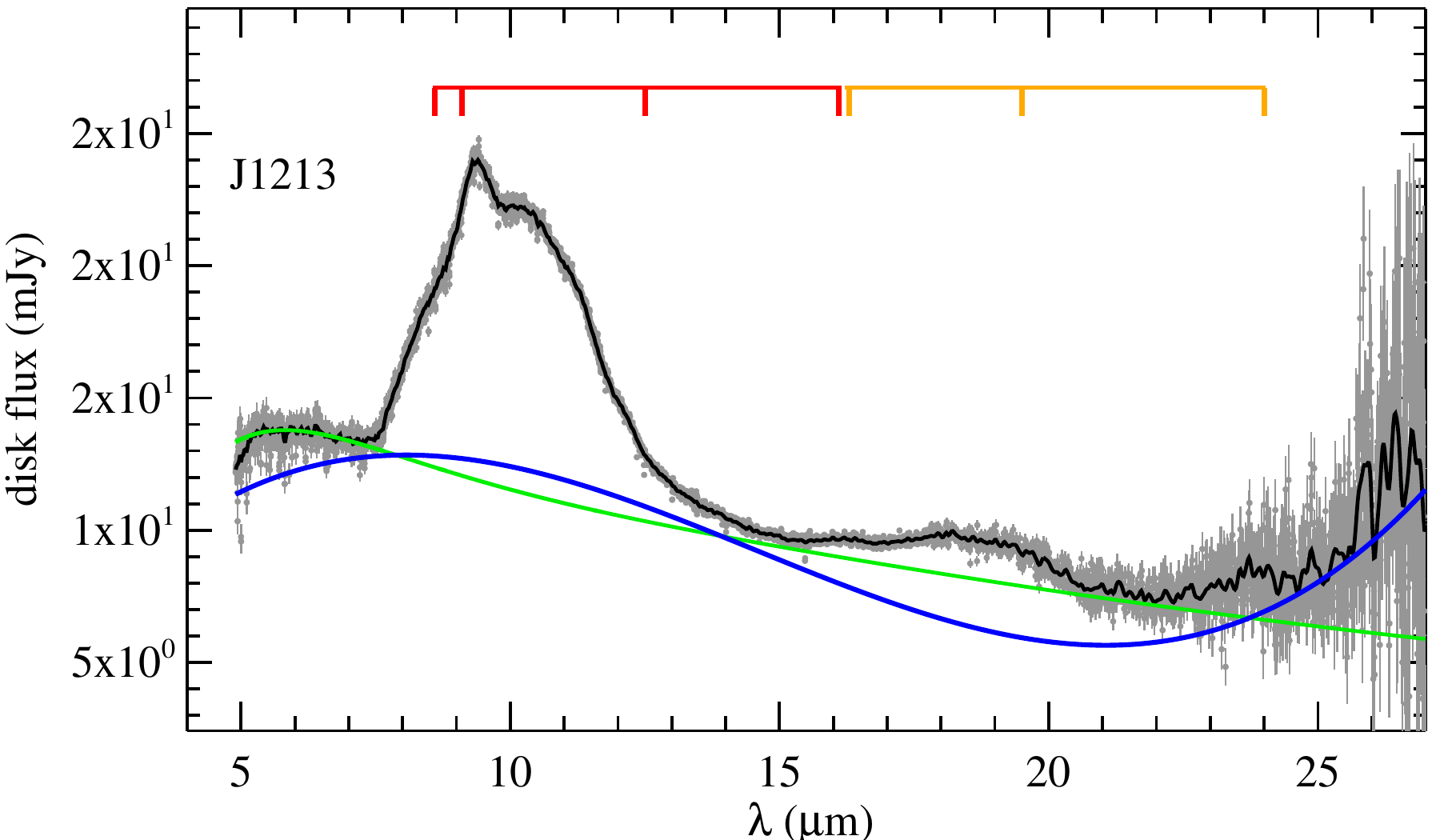}
    \includegraphics[width=0.49\linewidth]{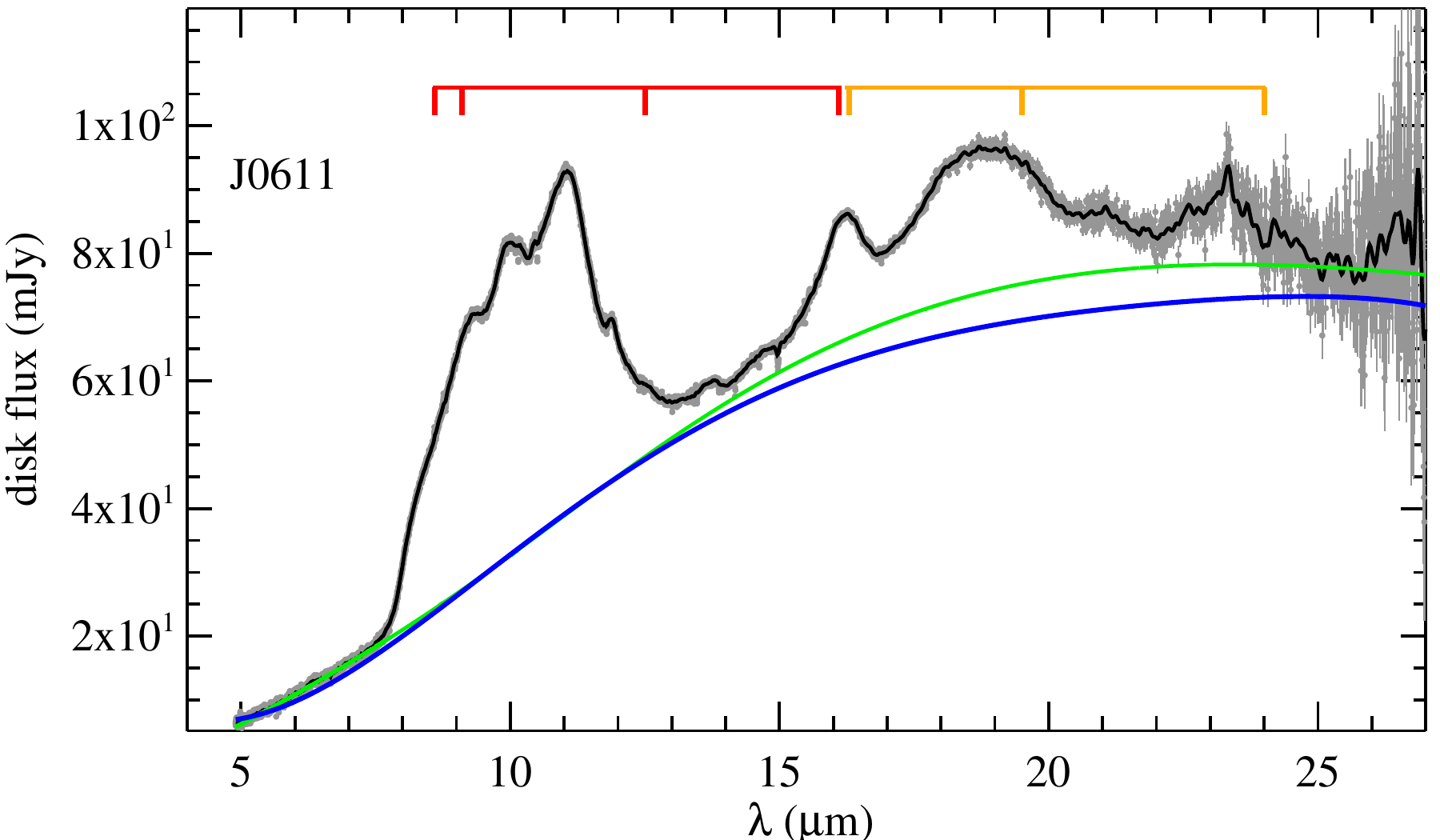}
    \includegraphics[width=0.49\linewidth]{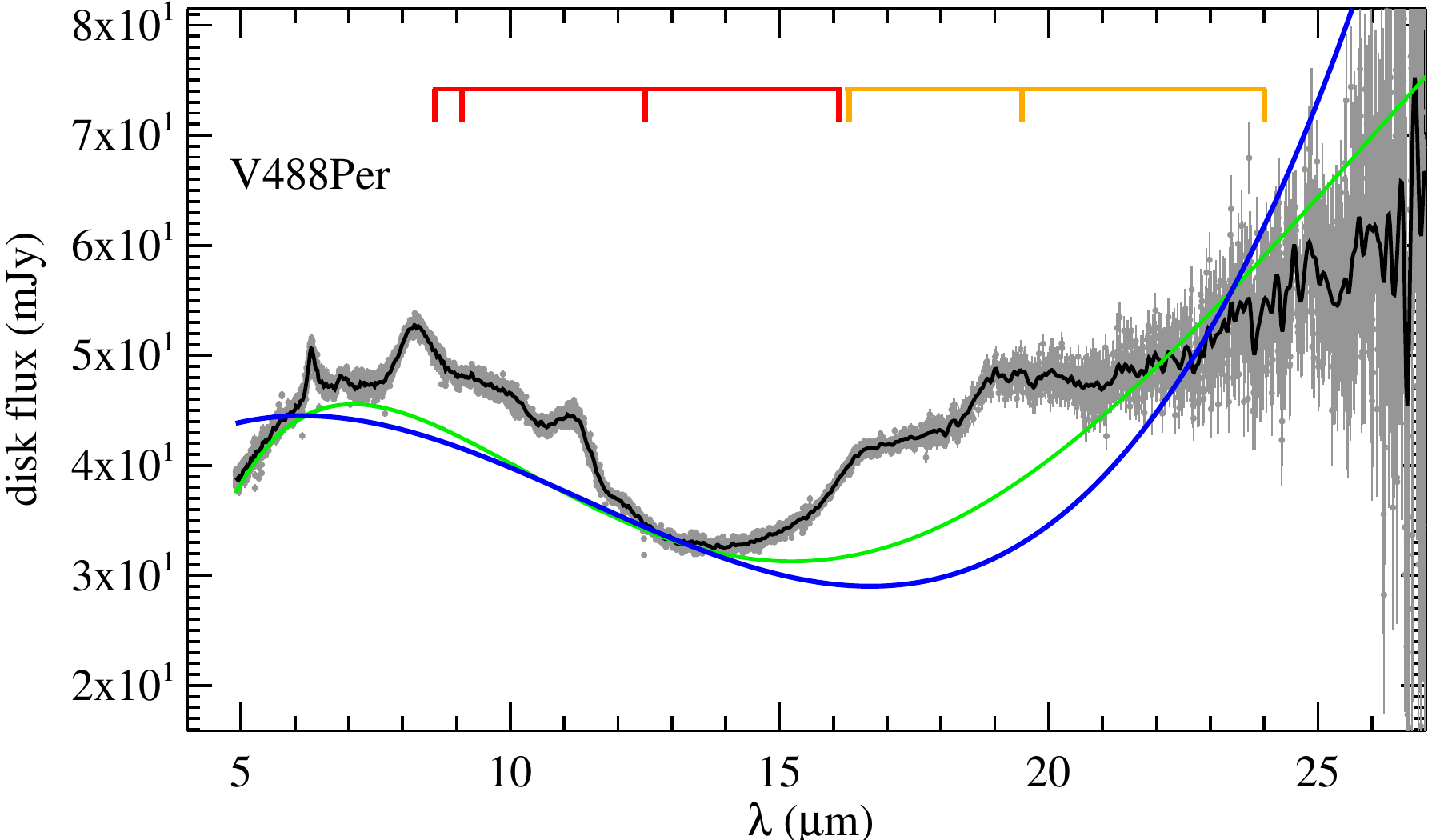}
    \caption{Star-subtracted mid-infrared spectra of the EDDs are shown in black with error bars in gray. JWST MIRI/MRS spectra were binned to $R\sim$300 for clarity while Spitzer IRS spectra were shown in their native resolution. The two pseudocontinua are shown in green (a combination of blackbody functions) and blue (a polynomial function). The wavelengths for prominent peaks of silica dust are marked on the top in red while signatures of crystalline olivine (Fe-bearing forsterite) in the 20 \micron\ region are marked in orange. The eight systems shown here are classified as silica-rich EDDs where the top four are known to be rich in silica dust using Spitzer data, while the bottom four are identified by this work. }
    \label{fig:disksed_linearly}
\end{figure*}

To provide a better perspective on the dust mineralogy in EDDs, we also analyzed five other EDDs with high-quality mid-infrared spectra from Spitzer/IRS extracted from the CASSIS website \citep{cassis_ref}. Although the Spitzer/IRS spectra have much lower spectral resolution than the JWST data, they are sufficient for our analysis method (described below). The basic properties for the whole EDD sample are detailed in Appendix \ref{sec:jwstsample}, including stellar parameters, ages, multiplicity, and the presence of cold dust, with cited references. We use these properties to search for correlations with the observed dust mineralogy -- particularly stellar age -- to probe evolutionary trends. Figure \ref{fig:disksed_linearly} shows the disk emission after subtracting the stellar contribution for the entire sample.  

\setcounter{figure}{0}
\begin{figure*}
    \includegraphics[width=0.49\linewidth]{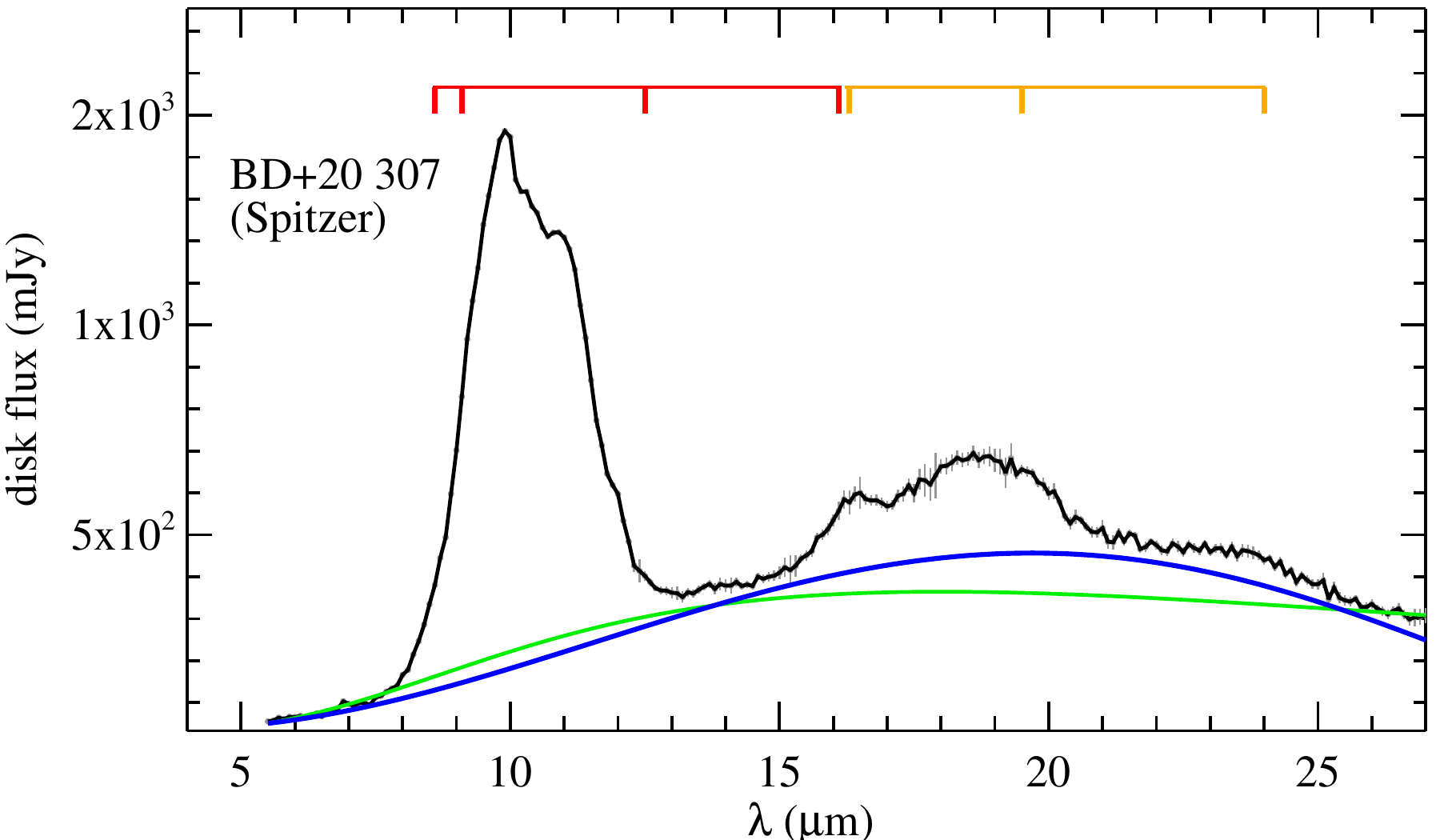}
    \includegraphics[width=0.49\linewidth]{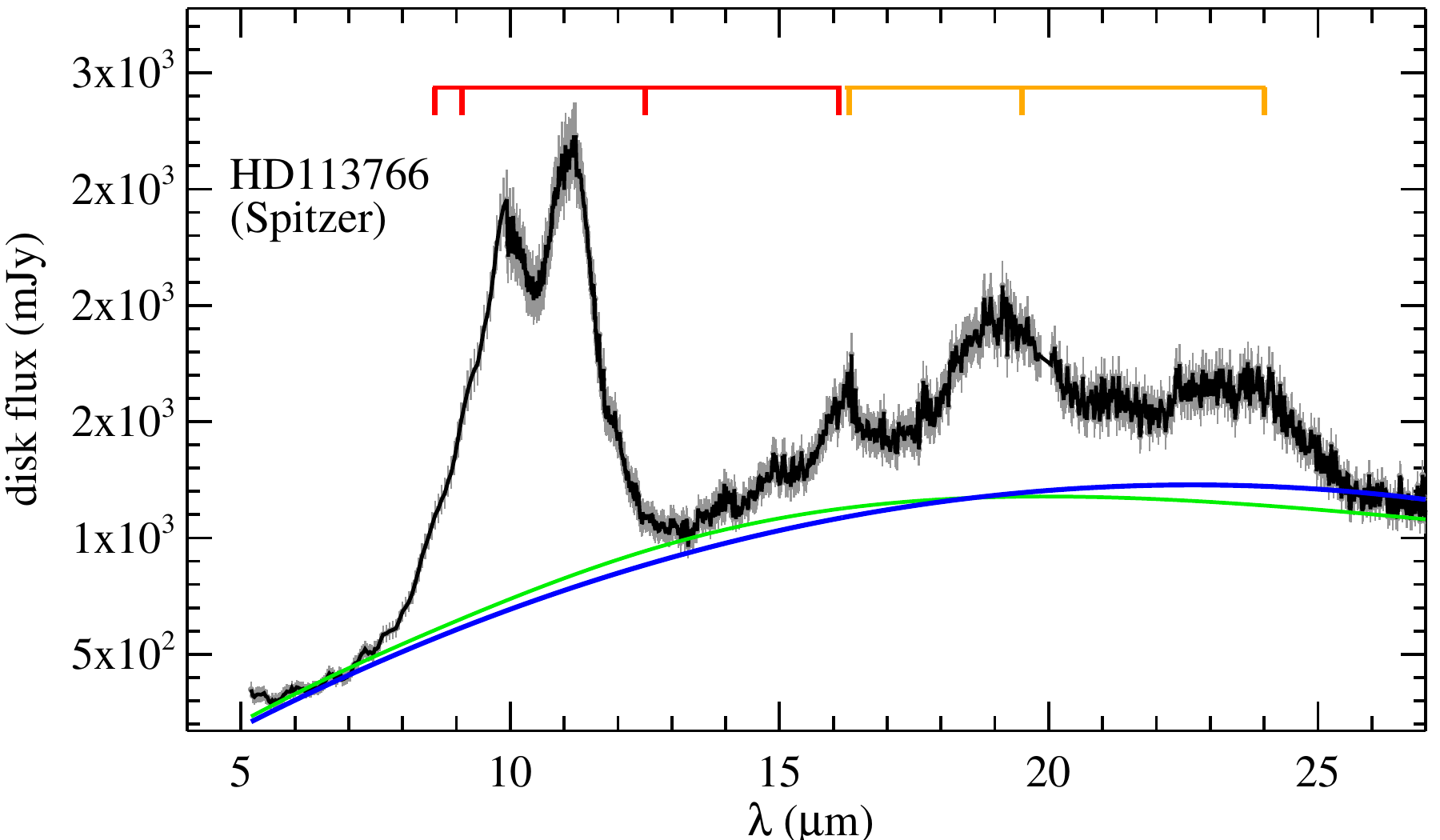}
    \includegraphics[width=0.49\linewidth]{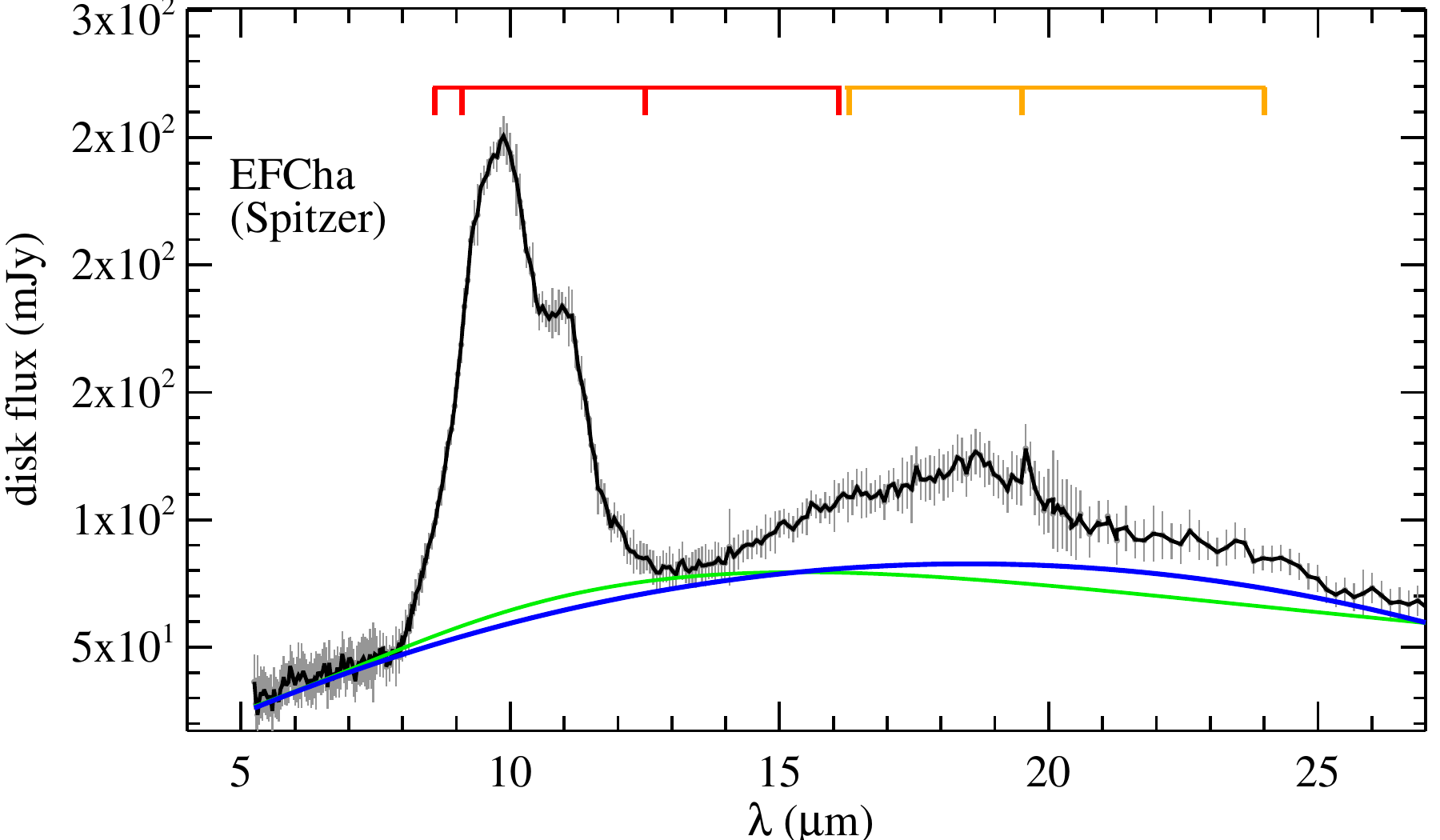}
    \includegraphics[width=0.49\linewidth]{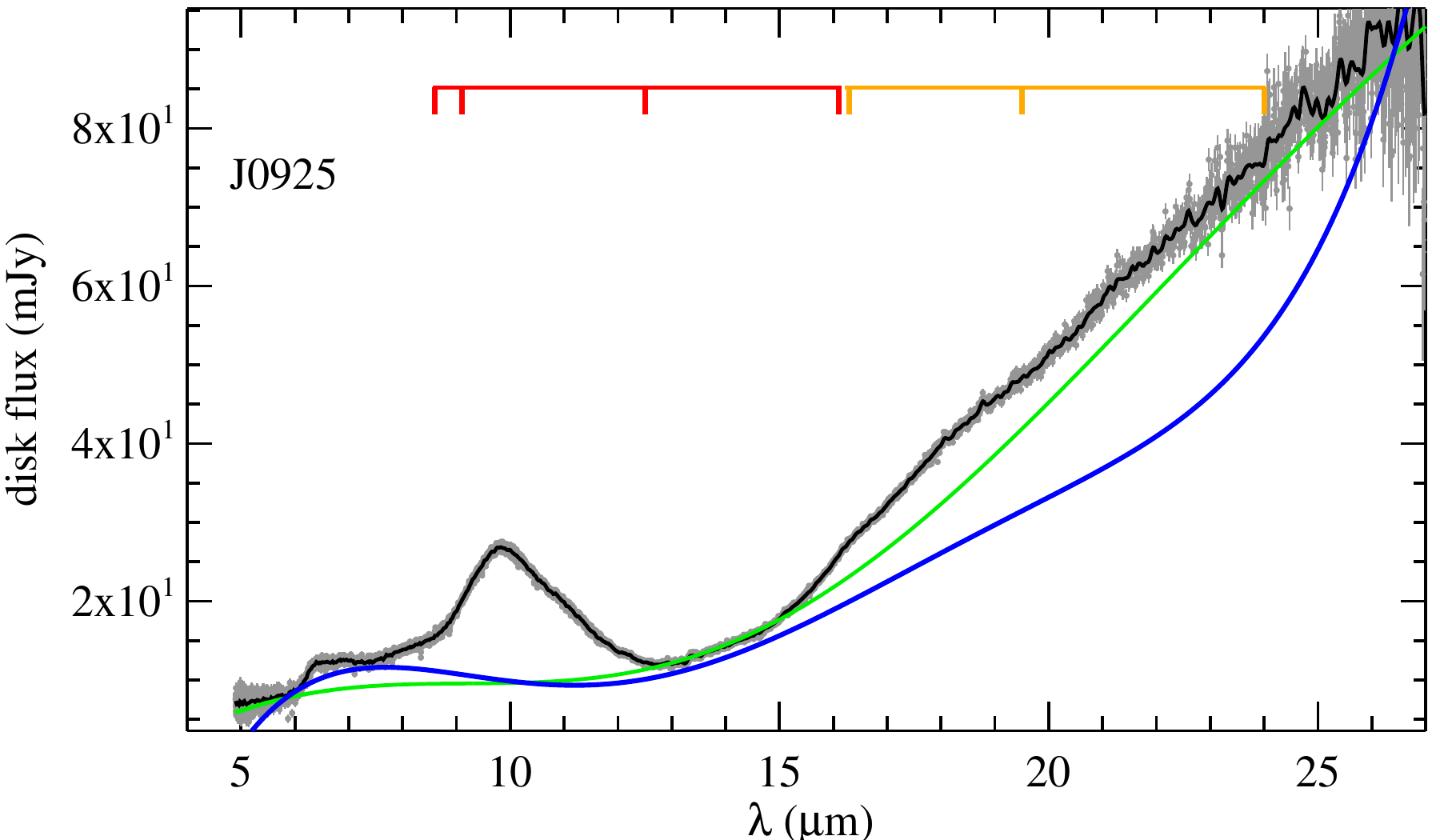}
    \includegraphics[width=0.49\linewidth]{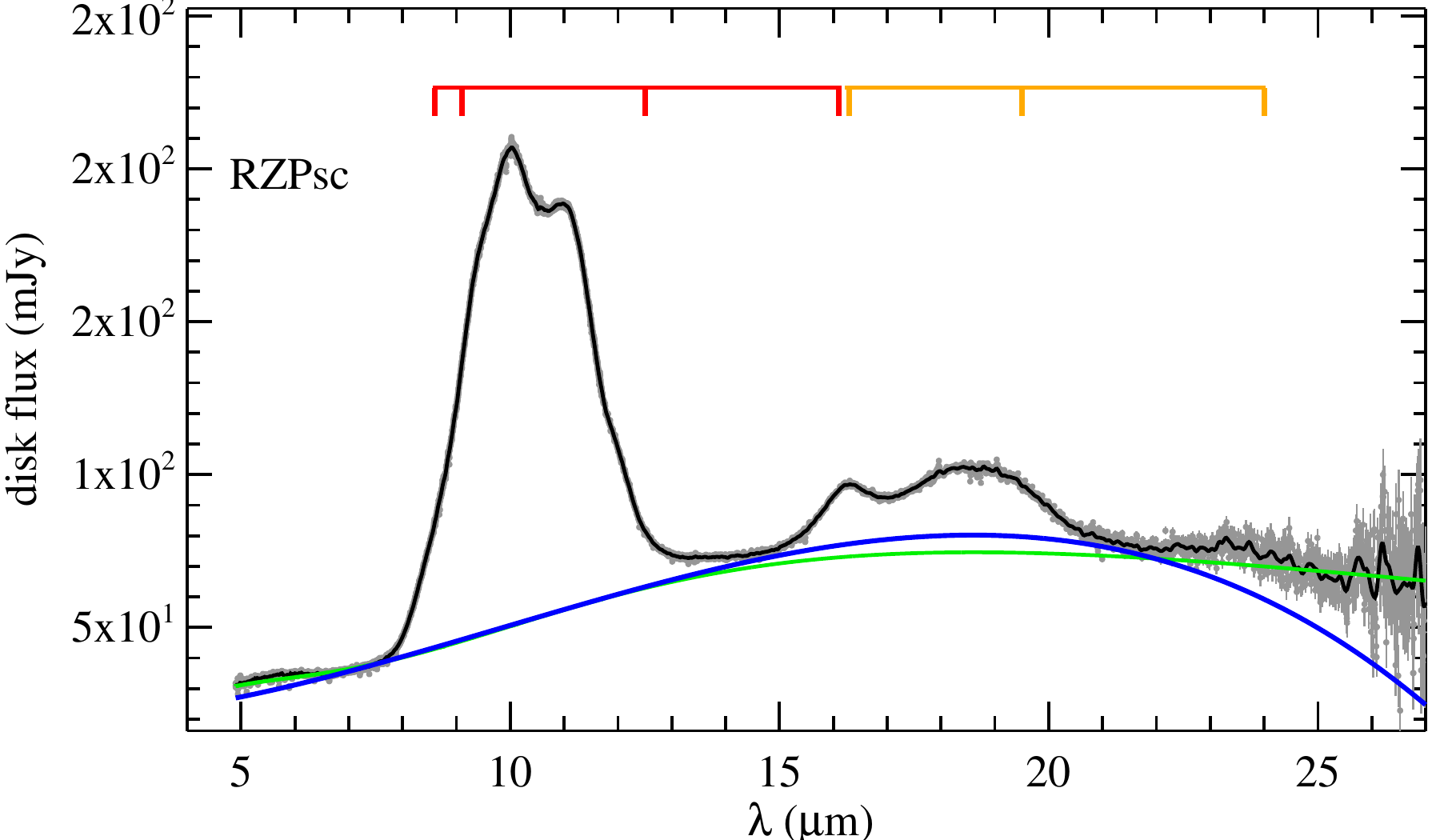}
    \includegraphics[width=0.49\linewidth]{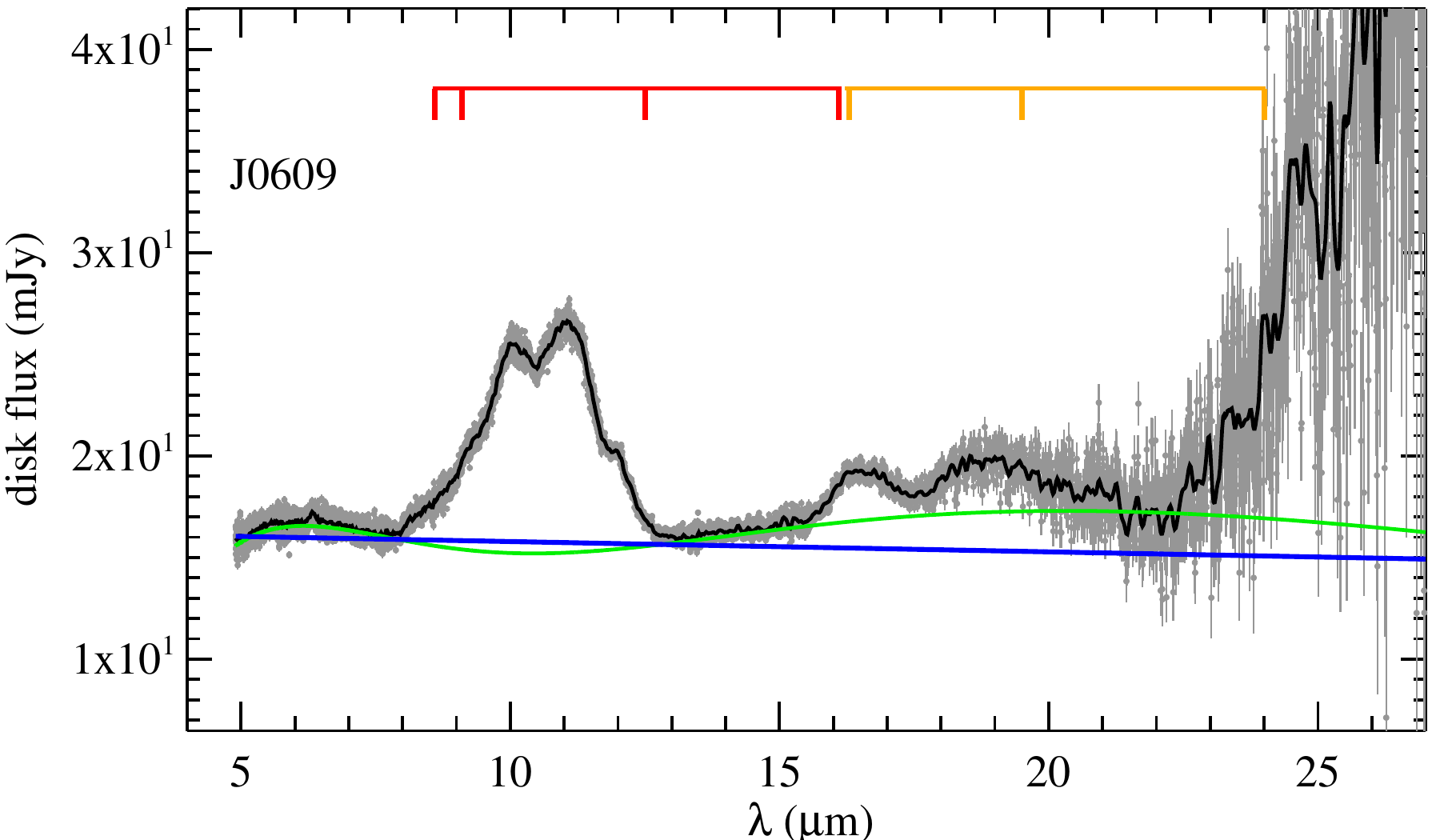}
    \includegraphics[width=0.49\linewidth]{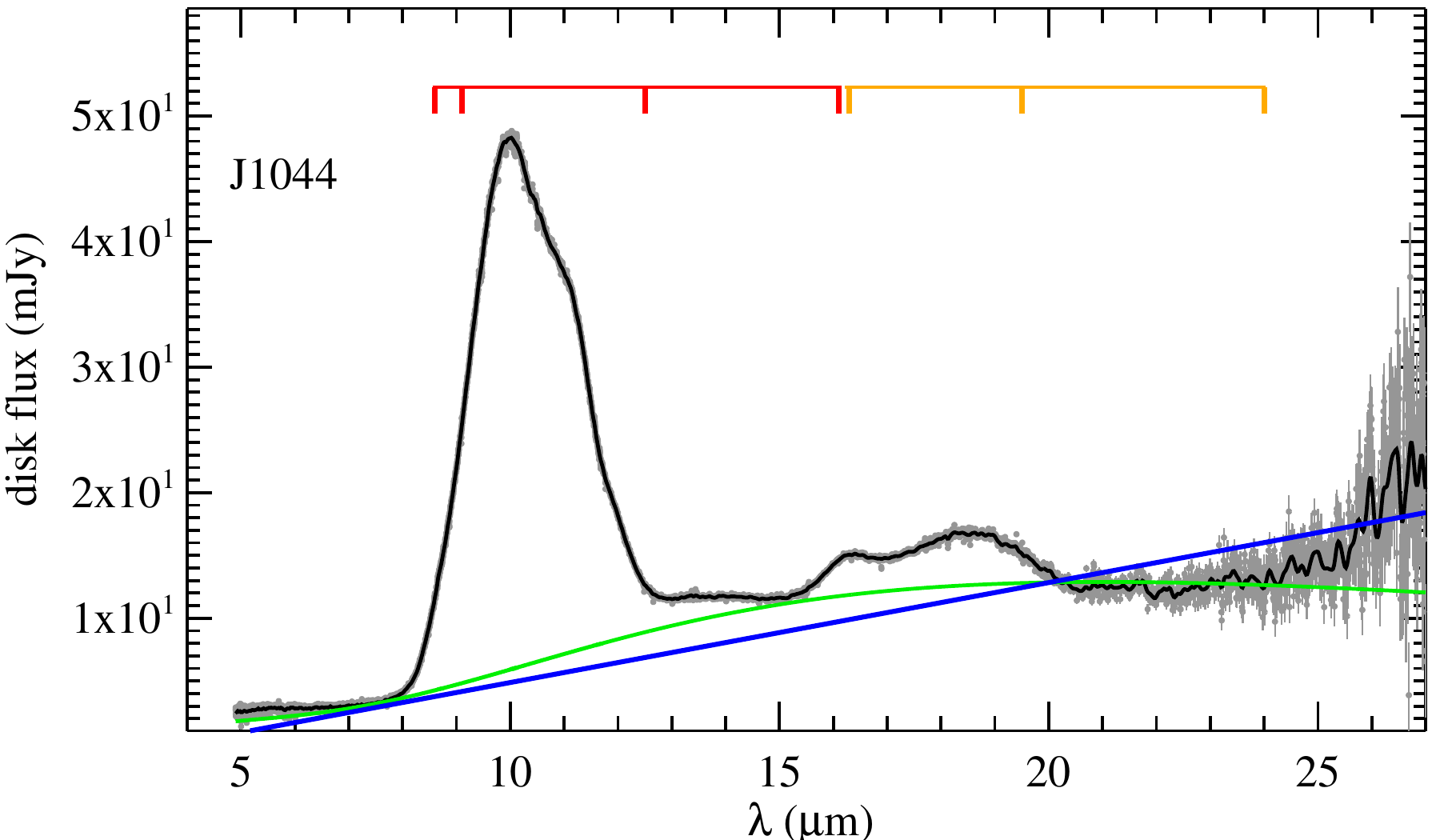}
    \includegraphics[width=0.49\linewidth]{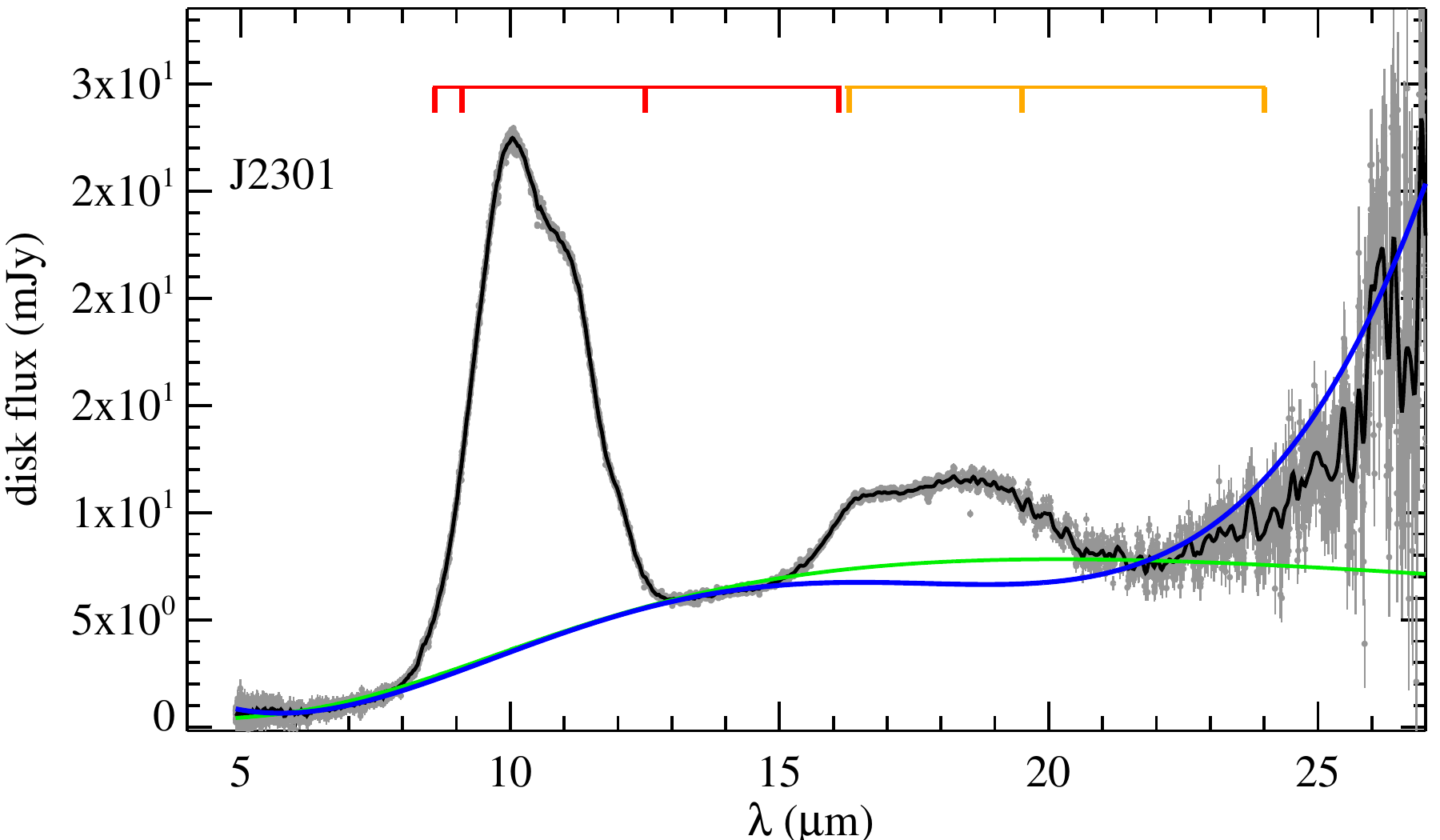}
    \caption{Continuous Figure \ref{fig:disksed_linearly} for silica-poor systems classified by this work. We note that some systems (J0609, J2301 and V488\,Per) suffer from poor background subtraction at wavelengths longer than $\sim$23 \micron\ due to the faintness of the sources and increasing telescope emission.}
    \label{}
\end{figure*}
\setcounter{figure}{0}
\begin{figure*}
    \includegraphics[width=0.49\linewidth]{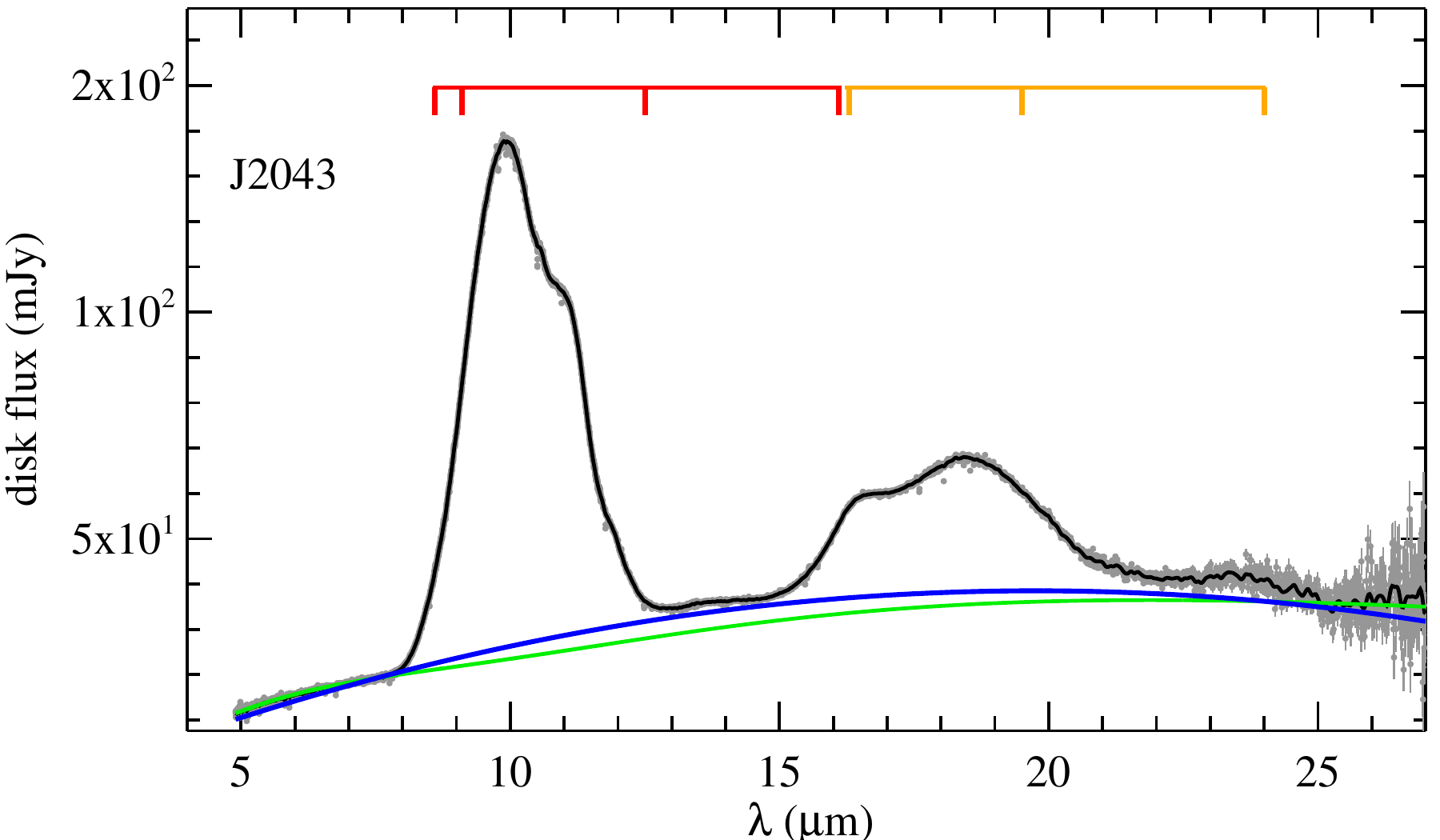}
    \includegraphics[width=0.49\linewidth]{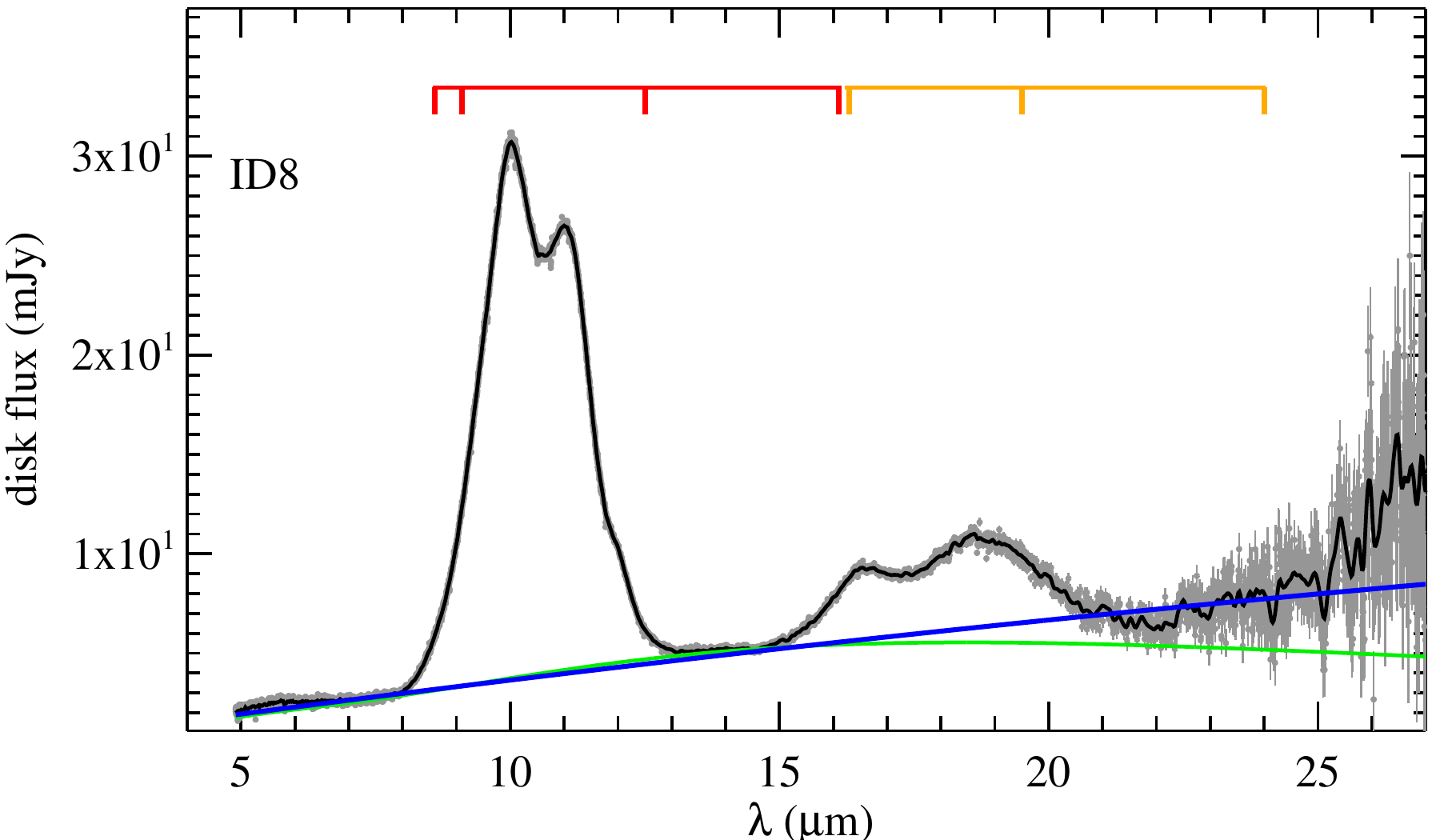}
    \includegraphics[width=0.49\linewidth]{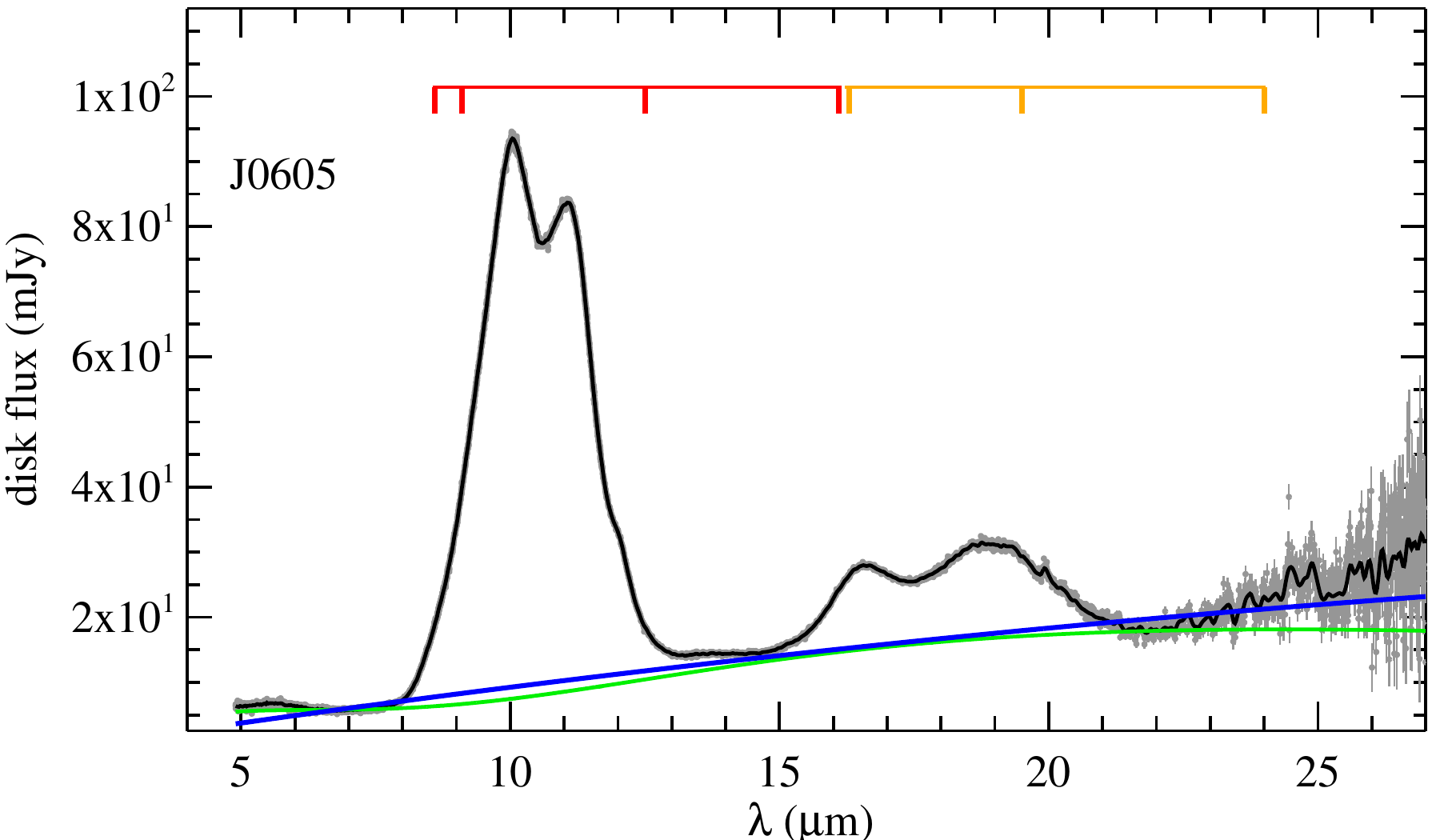}
    \includegraphics[width=0.49\linewidth]{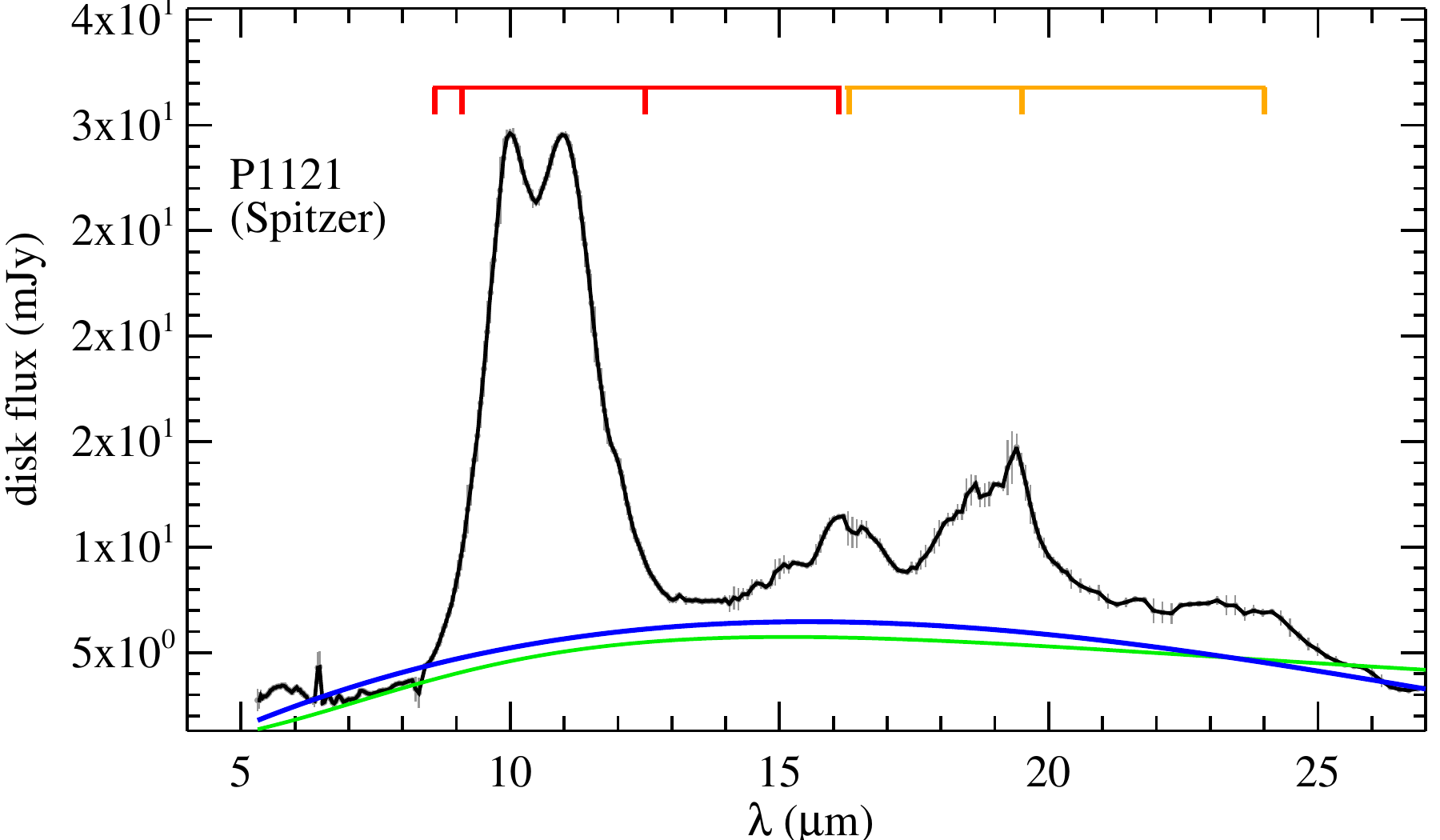}
    \includegraphics[width=0.49\linewidth]{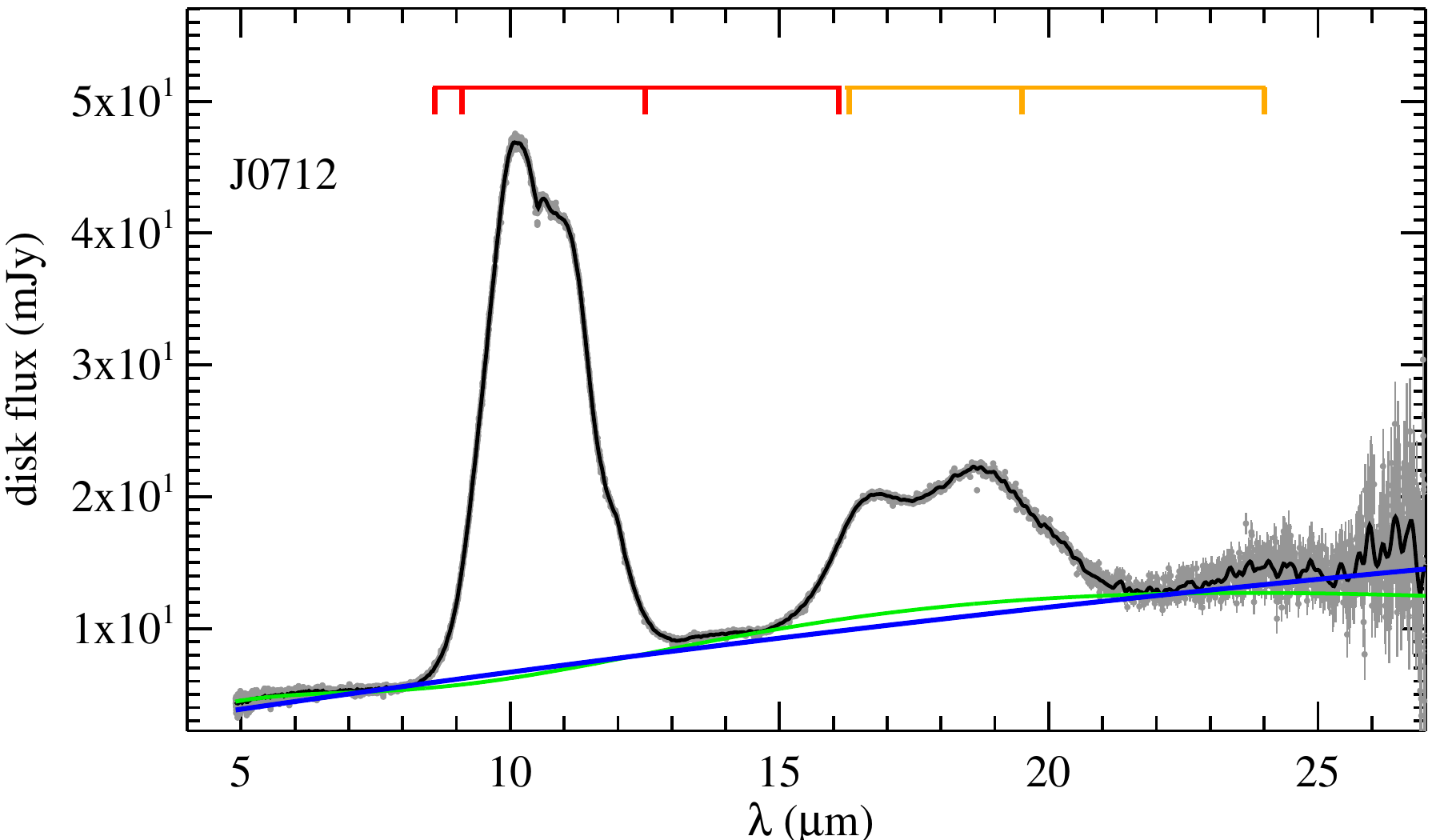}
    \caption{Continuous Figure \ref{fig:disksed_linearly}.}
    \label{}
\end{figure*}

\subsection{Methodology -- 10 \micron\ Feature Properties}

Our objective is to gain a general understanding of the dust mineralogy produced by impacts within a few au by analyzing the observed 10 \micron\ feature, broadly defined between 7.2 and 13.5 \micron. To achieve this, we adopted a straightforward methodology that can be quickly and uniformly applied to a large sample of systems, avoiding debates over the precise features emitted by a given dust composition or mineral. 
We revisit the silica-rich vs.\ forsterite-rich mineralogical dichotomy by adopting a ``dust emissivity" approach to identify systems that show silica-rich signatures in their 10‑$\mu$m features. The term emissivity is not the conventional definition in thermodynamics (a value between 0 and 1), see Appendix \ref{sec:mathforms} for the definition used in this work. Dust emissivity is defined as the ratio of the feature-producing spectrum to the underlying continuum and can be used to reveal the composition of small grains via their distinct solid-state features (e.g., \citealt{watson09}). Because we focus on dust within a few au (the terrestrial zone), we restrict our analysis to the 10 \micron\ region, which probes warm dust; its features are less sensitive to grain impurities, structure (e.g., porosity), and continuum placement. The methodology and the rationale for our choices are detailed in Appendix \ref{sec:dustindices}. The derived dust emissivity in the 10 \micron\ region for the entire sample is shown in Figure \ref{fig:10um_emissivity}.

We quantified the properties of the 10 \micron\ dust emissivity using its strength and morphology. The feature strength, referred to as $W_{10}$, is the equivalent width of the 10 \micron\ feature and serves as a tracer of the amount of optically thin grains (i.e., the total emitting area, which is proportional to the mass) in a system. We also utilized the FWHM and the dust indices of crystalline silicates and silica to quantify the  morphology of the feature. The computed FWHM acts as a proxy for the dominant grain sizes and the levels of crystallinity that produce the feature. Generally, the narrower the feature, the smaller the grain size \citep{przygodda03,kessler-silacci06_c2d}. Furthermore, it also reflects the dust's crystallinity; as the width increases, so does the proportion of the crystalline contribution \citep{watson09, olofsson09_c2d}. We stress that FWHM and $W_{10}$ serve as proxies, but not in an absolute sense (for details, see Appendix \ref{sec:dustindices}). 

Additionally, we employed dust indices to characterize the amount of processed dust in the observed 10 \micron\ dust emissivity. A key aspect of this analysis is defining a pristine 10 \micron\ feature that lacks processed dust and using it as a standard to determine how much an observed spectrum differs by contrasting the specific wavelength range in which processed dust features appear. The greater the contrast (i.e., departure from a nominal value of 1), the more prevalent a specific mineral is in the spectrum (see Appendix \ref{sec:mathforms} for the mathematical formulation of the defined indices and mineral features). We used the dust emissivity derived from the observed JWST spectrum of LkCa\,15 \citep{su25_hd23514} as the pristine standard and computed the dust indices for three minerals: crystalline olivine ($O_{10}$) and pyroxene ($P_{10}$), and silica ($S_{10}$), as defined by \citet{watson09}. To better characterize silica dust, we introduced a short-wavelength index centered at 8.50 \micron\ ($S_{10,s}$) for quantifying the silica feature, alongside a long-wavelength index ($S_{10}$) centered at 12.46 \micron. Finally, to minimize overlapping features among the three minerals, we reduced the width of the range over which the dust indices were computed compared to the one used in \citet{watson09}. The suitability of using LkCa\,15 as the pristine reference and associated uncertainty are further justified in Appendix \ref{sec:labindices}.

The derived 10 \micron\ properties for the combined sample of 21 EDDs, including systematic uncertainties, are summarized in Table \ref{tab:measuredindices}. To calibrate the dust-index values, we also computed indices for a wide variety of materials with dust absorption opacity measured in the mid-infrared for comparison, including 11 silica-like species, five materials that experienced different degrees of processing, various synthesized materials, and observed and modeled ISM-like compositions (Table \ref{tab:lab_indices}). For a broader context regarding processed dust in circumstellar environments, we also computed the dust indices for a subset of PPDs and DDs to better define the unique properties presented in EDDs. 
Specifically, we used the silica index on the blue‑side shoulder of the 10 \micron\ feature ($S_{10,s}$), derived from silica‑like compositions and from silica‑rich EDDs and PPDs identified in previous spectral‑decomposition studies, to set the threshold for identifying new silica‑rich systems (see Appendix \ref{sec:labindices} for details). Using this metric, we identify four additional silica-rich systems (V488\,Per, J0611, J1213, and HD\,166191); together with the four previously identified silica-rich EDDs from Spitzer, eight of 21 EDDs (38$^{+11}_{-9}$\%)  are  silica-rich. Because real dust is a mixture of minerals (different silica polymorphs, varying impurities and porosities), the labels "silica-rich" and "silica-poor" are relative to the sample and chosen threshold; including or excluding edge cases near the boundary does not affect our overall conclusions (discussed below). 

\begin{figure*}
    \centering
    \includegraphics[width=0.495\linewidth]{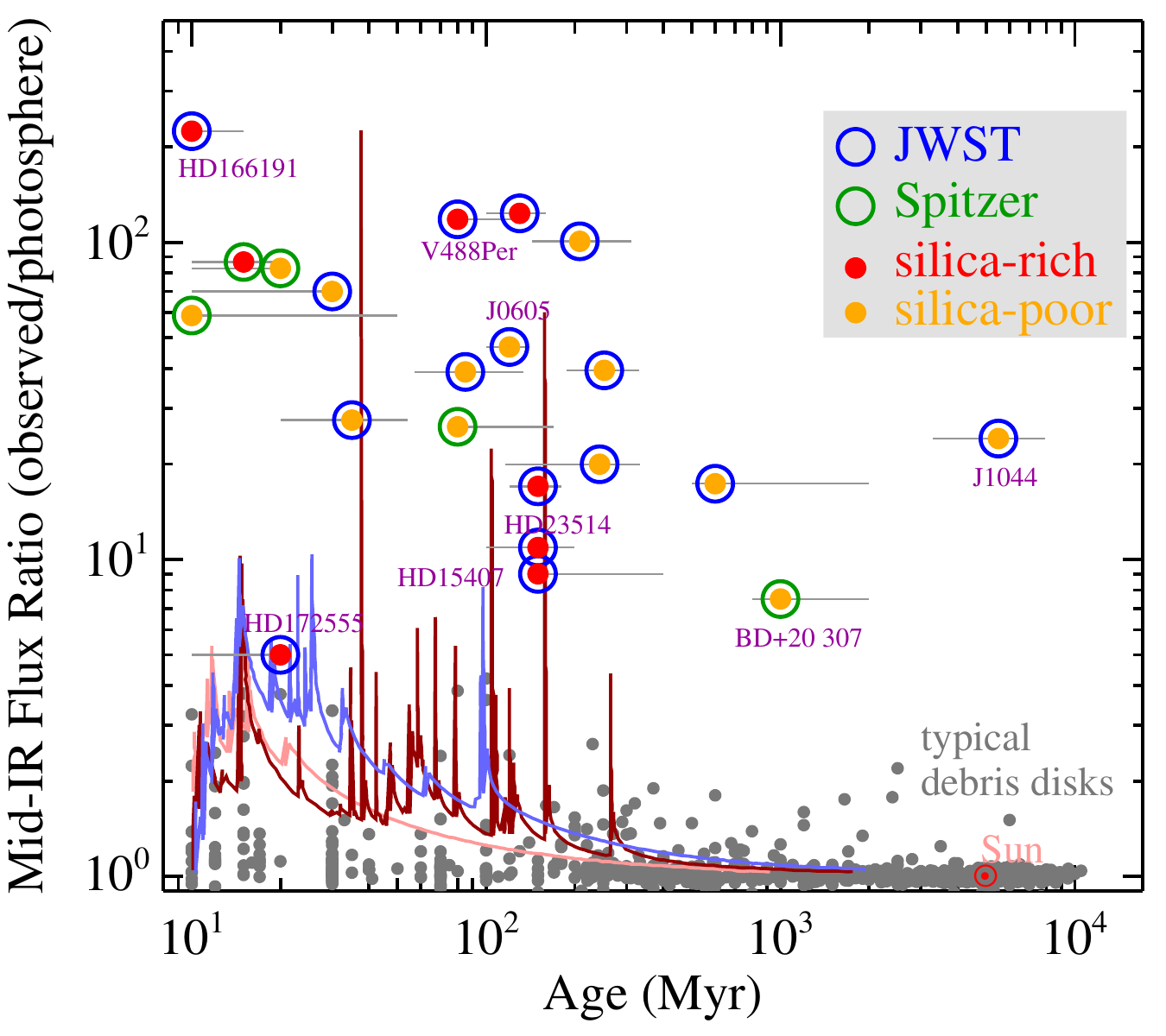}
    \includegraphics[width=0.495\linewidth]{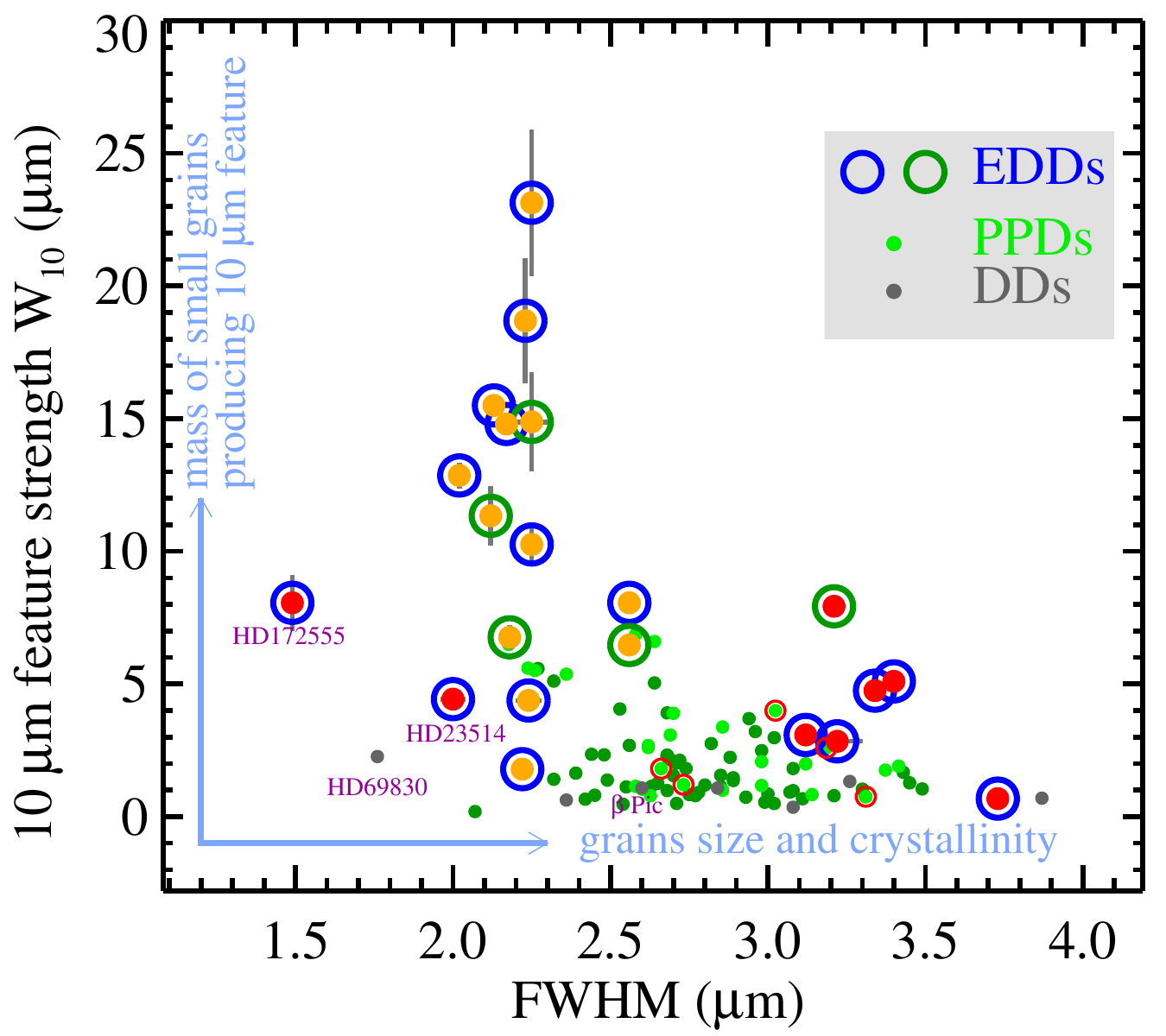}
    \caption{EDD properties in the context of other disks -- debris disks (DDs) and protoplanetary disks (PPDs). The left panel shows the evolution of warm dust in DDs, depicting the amount of dust as measured by 22/24 \micron flux relative to the star adopted from \citealt{chen_su_xu20}.  This warm dust is largely depleted as stars age, except in EDDs (large symbols with horizontal bars indicating age uncertainties). The colored lines represent simulated dust evolution models for terrestrial planet formation \citep{genda15}, showing that giant impact activities peak at $\sim$10--100 Myr and are significantly reduced after $\sim$300 Myr. The right panel displays the measured width (FWHM) and strength ($W_{10}$) of the 10 \micron\ feature. In both panels, EDDs rich in silica dust are marked with red dots, while those less so (silica-poor but rich in crystallinity silicates) are marked with orange dots. Measurements for PPDs and DDs are shown as small dots in different colors (gray for DDs; green for PPDs, with dark green indicating the values measured by \citet{watson09}).}
    \label{fig:edds_context}
\end{figure*}

These measured properties, together with other system parameters, are shown in Figures \ref{fig:edds_context} and \ref{fig:indices} and are discussed in the next section, which highlights the unique characteristics of EDDs.

\section{Analysis}
\label{sec:analysis}

\subsection{Elevated Amount of Optically-thin Small Grains }
\label{sec:edds_I}

EDDs were originally identified by the substantial presence of warm dust surrounding mainly Sun-like stars, which exhibit high mid-infrared fluxes at 22/24 \micron, exceeding their stellar emission by a factor of ten to several hundred (Figure \ref{fig:edds_context}, left panel). The large abundance of small grains in EDDs is reflected in the equivalent width of the 10 \micron\ feature, $W_{10}$ (Figure \ref{fig:edds_context}, right panel). EDDs have much higher $W_{10}$ values than those of PPDs and DDs. The prominence of the 10 \micron\ feature largely depends on grain sizes and emitting temperatures. Smaller grains display sharper features than the muted characteristics of larger grains. Furthermore, when exposed to the same radiation field, smaller grains heat to higher temperatures than their larger counterparts \citep{aannestad_purcell73_ismgrains,kruegel03_physics_ism_dust}, resulting in decreased relative feature strengths between 10 and 20 \micron\ as temperatures drop (see Figure D1 of \citealt{su25_hd23514}). In PPDs, $W_{10}$ values are constrained by the optically thin disk atmosphere, despite having higher dust content. Consequently, achieving large $W_{10}$ values in PPDs is challenging, except in cases of extremely flared disks or those with large inner holes (e.g., PDS 70; \citealt{jang24_pds70_dustmineralogy}). When a measurable 10 \micron\ emission feature appears, it must originate from relatively optically thin conditions.

It is widely recognized that the dominant grain size responsible for the 10 \micron\ feature in PPDs (i.e., grains in the low-optical depth disk surface) is of order a few microns \citep{oliverira11_ppd_dust_mineralogy,varga26}, likely due to a balance between grain growth, fragmentation, and vertical mixing in a gas-rich environment \citep{birnstiel10}. As noted earlier, the measured FWHM not only reflects the grain size but also indicates the fraction of mixed crystalline silicates in composite porous particles \citep{voshchinnikov_henning08}. Interestingly, roughly one-third of the EDDs have large FWHMs that overlap with the extensive FWHM range observed in PPDs (Figure \ref{fig:edds_context}). As shown in the next subsection, these EDDs are rich in both silica and crystalline silicates, resulting in a broader FWHM. In other words, the measured FWHMs indicate that the solid-state features of the majority of EDDs are produced by grains smaller than those in PPDs (i.e., sub-\micron-sized grains). 

Low $W_{10}$ values in DDs are understandable, as grains experience collisional destruction followed by rapid loss due to radiation pressure in a low optical-depth environment. Typical debris disks rarely exhibit prominent 10 \micron\ solid-state features, consistent with a paucity of warm, small grains. The reason is twofold: either they lack small grains, or the grains are too cold, situated tens of au away from the star. When the dust temperature falls below a few hundred degrees (i.e., beyond a few au), the bulk emission shifts toward longer wavelengths. Consequently, the 10 \micron\ feature becomes weaker, even in the presence of small grains. The prominence of the solid-state feature in EDDs suggests that the emission arises from optically thin small grains within a few au, which are subject to rapid loss if they are smaller than the sizes affected by radiation pressure blowout. Almost all EDDs in our sample show elevated small‑grain populations (measured by $W_{10}$), consistent with recent large dust‑producing collisions (between bodies $\gtrsim$ a few hundred km) occurring within the terrestrial‑planet formation zone. That phase is expected to wane after a few hundred Myr, consistent with only three known EDDs older than 300 Myr among our 21 systems (left panel of Fig. \ref{fig:edds_context}); implications of the old EDDs are discussed in Section \ref{sec:discussion}.

\subsection{Elevated Amount of Thermally Altered Grains }
\label{sec:edds_II}

\begin{figure*}
    \centering
    \includegraphics[width=0.495\linewidth]{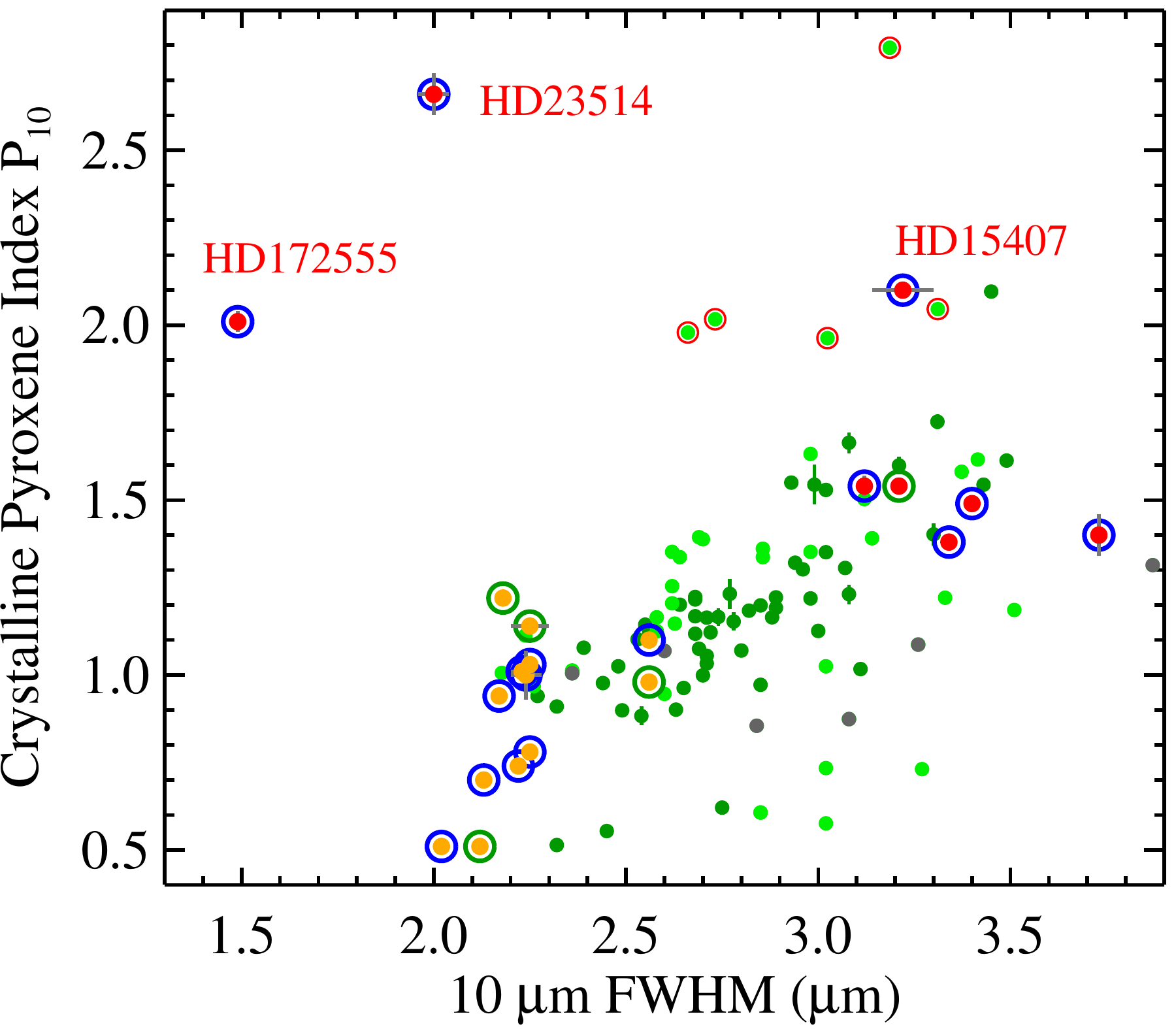}
    \includegraphics[width=0.495\linewidth]{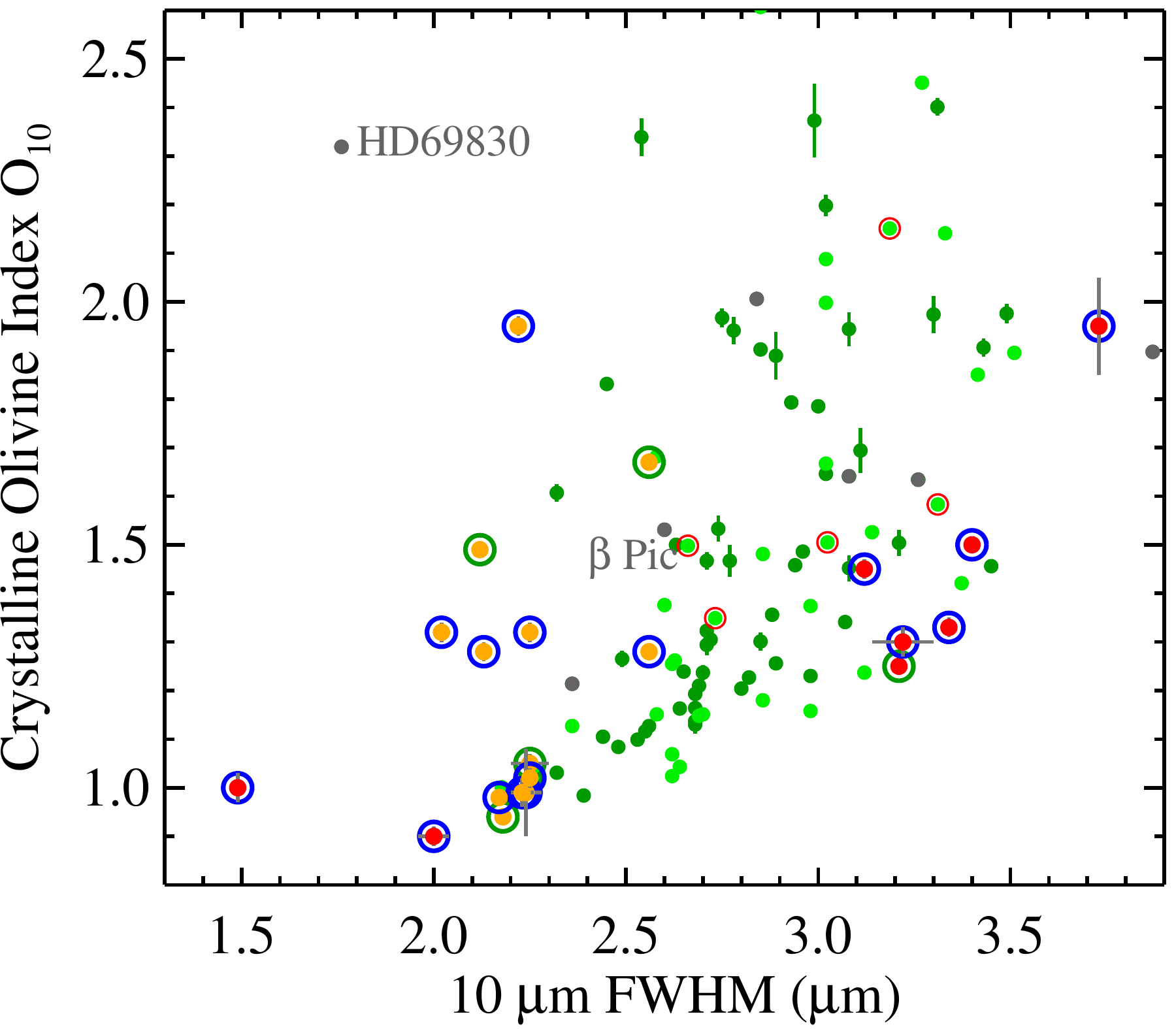}
    \includegraphics[width=0.495\linewidth]{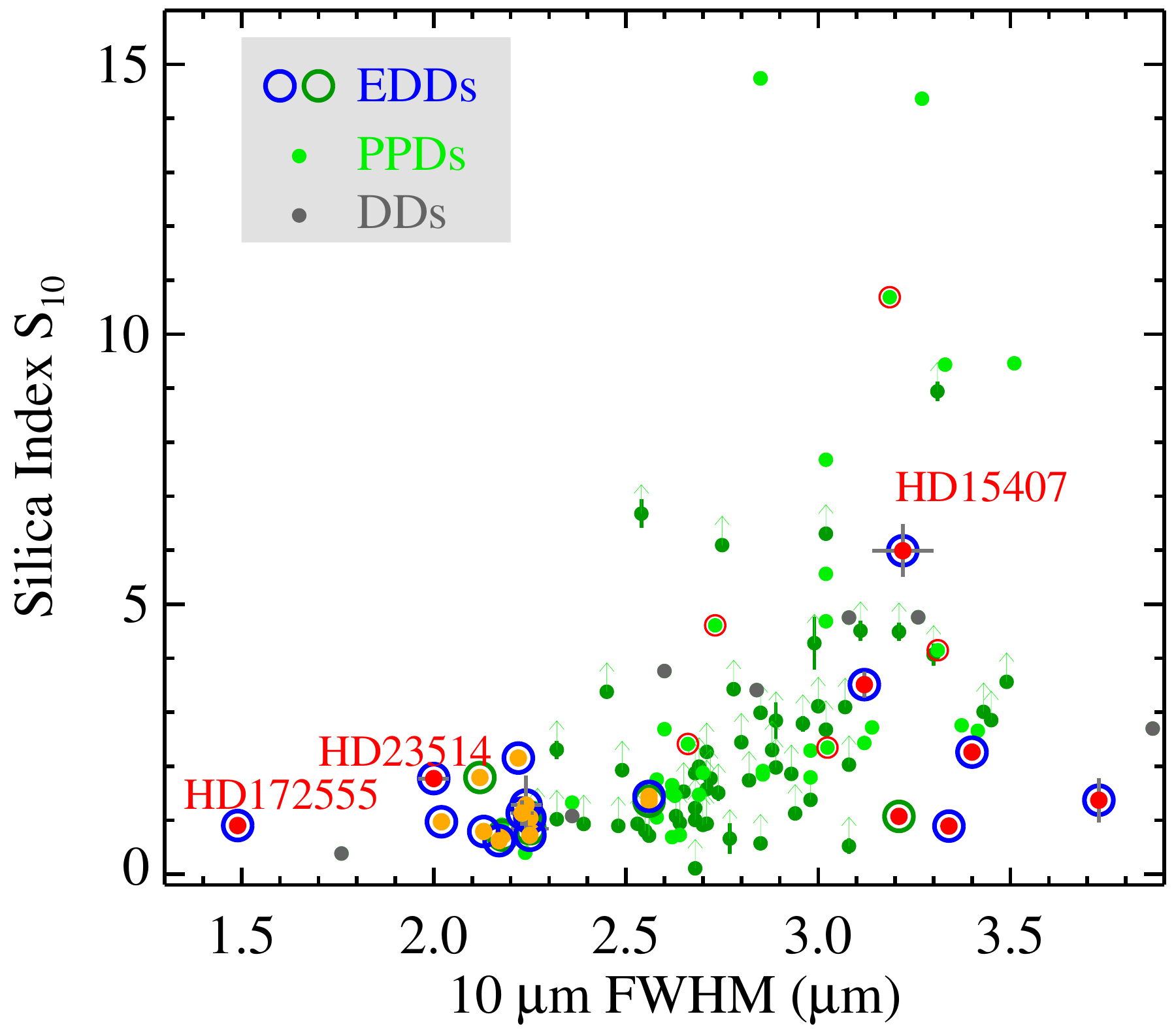}
    \includegraphics[width=0.495\linewidth]{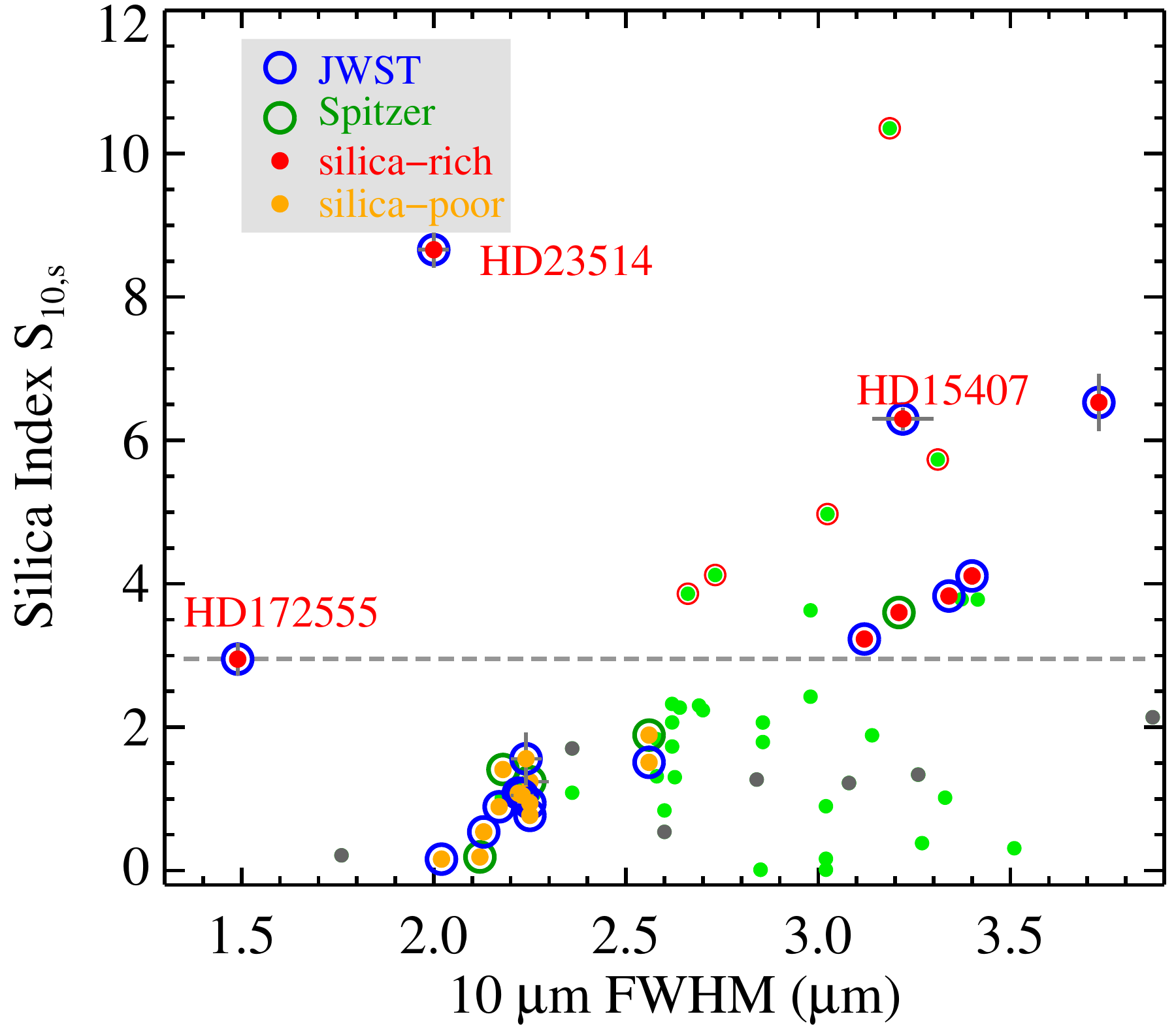}
    \caption{Dust indices in the 10 \micron\ region: $P_{10}$ and $O_{10}$ (upper two panels) for crystalline silicates and $S_{10}$ and $S_{10,s}$ (lower two panels) for silica, for EDDs and a selected sample of PDDs and DDs (smaller dots). 
    The PPD indices from \citet{watson09} are shown in dark green and marked with upward arrows in the $S_{10}$ panel since they were computed using a wider wavelength width (see Appendix \ref{sec:mathforms}). Thin, red circles mark the five silica-rich PPDs from \citet{sargen09_silia}. The horizontal dashed line in the $S_{10,s}$ panel marks the thresholds for silica-rich systems. }
    \label{fig:indices}
\end{figure*}

Among a handful of EDDs discovered by Spitzer, there appears to be a mineralogical dichotomy in the impact-produced dust: silica-rich vs.\ forsterite-rich \citep{meng15}. Classic examples of silica-rich EDDs include HD\,172555 \citep{lisse09}, HD\,15407 \citep{fujiwara12}, and HD\,23514 \citep{rhee08}, all characterized by a prominent $\sim$9 \micron\ feature likely produced by vapor condensates resulting from hypervelocity collisions between planetary embryos \citep{johnson12b}. Forsterite-rich systems such as ID\,8 and P\,1121 \citep{olofsson12,su19} are interpreted as thermally annealed products of amorphous silicates transforming into crystalline silicates, primarily forsterite (Mg-rich olivine), at high temperatures \citep{fabian00,thompson19_annealingsilicates}. Comparative studies of dust mineralogy in meteorites \citep{morlok14_processedmaterial} suggest that while both types are highly processed and have experienced intense heating, silica-rich materials are more highly shocked, whereas forsterite-rich materials are less so \citep{morlok14}. It is interesting to note that the gas phase condensation temperature for silica-like materials is typically around 1200 to 1800 K and can reach up to a couple of thousand K at high pressures. In contrast, for forsterite, the temperature typically ranges from 1000 to 1200 K, suggesting that the events responsible for the production of silica are typically more energetic (resulting in higher temperatures) than those for forsterite.

With twelve additional high‑quality JWST mid‑infrared spectra of EDDs, we can now reassess the apparent dichotomy in impact‑produced mineralogy by comparing the relative amounts of highly processed dust. Figure \ref{fig:indices} presents four dust indices for the 21 EDDs alongside values for selected PPDs and DDs; we use the PPD indices to provide global context for thermally processed dust in EDDs. Spectral decompositions of Spitzer PPD spectra show that dust in the disk surface layer has $\sim$10--40\% crystallinity by mass fraction \citep{sargent09,oliverira11_ppd_dust_mineralogy}. Although most PPD studies focus on Mg‑rich crystalline silicates (enstatite and forsterite), silica (amorphous quartz) contributes a comparable mass fraction in some systems \citep{sargent06,olofsson10_c2d_spectraldecomp,oliverira11_ppd_dust_mineralogy}; a small subset (five of $\sim$80 PPDs) shows distinct features at $\sim$9, 12.6, and 20 \micron\ attributed to crystalline silica \citep{sargen09_silia}. EDDs exhibit a similar spread in dust indices to PPDs, suggesting that the prominent 10 \micron\ features are dominated by thermally processed dust. As detailed in Appendix \ref{sec:labindices}, we use the $S_{10,s}$ index -- calibrated from previously identified silica‑rich systems -- to select silica‑rich objects. In both PPDs and EDDs, silica‑rich systems preferentially occupy the high‑$P_{10}$ region, while the $O_{10}$ and $S_{10}$ distributions are more mixed. Another interesting observation from these dust indices is the positive correlation between $S_{10,s}$ and the combined crystalline silicate indices ($\sqrt{P_{10}^2+O_{10}^2}$, Figure \ref{fig:silica_vs_crystallinity}). All silica-rich systems, including PPDs, exhibit high crystallinity, reflecting the fact that they are all highly processed materials.

The varying proportions of silica polymorphs mixed with other crystalline silicates complicate accurate identification. This highlights the limitation of using only the 10 \micron\ region for dust‑index analysis; extending the analysis to the $\sim$20 \micron\ region also faces challenges because feature positions and strengths are highly sensitive to composition (Al/Mg/Fe ratios) and grain porosity, introducing additional degeneracies that future spectral‑decomposition frameworks must address. 
Instead of using the original dichotomous terminology, we classified the EDD dust mineralogy as silica-rich vs.\ silica-poor, with the latter representing the initial forsterite-rich classification. Although we classify systems into two broad themes based on the observed mineralogy, it is evident that there exists a wide spectrum of dust compositions and varying levels of mixing in all circumstellar environments. In our broad definition, a silica-rich system is one where the observed features resemble those of silica dust with sufficient abundance (greater than the typical values observed in PPDs). This does not necessarily imply that silica-rich systems lack any forsterite or enstatite dust, or vice versa (see detailed comparison in Appendix \ref{sec:labindices}). Our goal is to identify EDDs enriched in silica dust and to uncover their implications by correlating them with other characteristics of EDDs. Based on the threshold of the $S_{10,s}$ silica index,  38$^{+11}_{-9}$\% of EDDs (eight out of 21) are classified as silica-rich, while fewer than 10\% of PPDs (6$^{+4}_{-2}$\%; five out of $\sim$80, see Appendix \ref{sec:labindices}) fall into the same category using Spitzer data\footnote{ \citet{varga26} analyzed 26 PPDs with high-quality JWST data, finding annealed (i.e., crystalline) silica in nine systems -- a fraction of 35$^{+10}_{-8}$\%. They emphasize that such robust detections require high signal-to-noise JWST data. The incidence of silica-rich PPDs is not the focus of this study and has no impact on our general conclusion.}. This likely suggests an enrichment of silica dust in EDDs compared to the precursor material found in PPDs, implying that advanced thermal processes in EDDs play a crucial role in the formation of silica dust. As shown in the left panel of Figure \ref{fig:edds_context}, both types of dust mineralogy are found around EDDs associated with stars younger than 300 Myr, while only silica-poor EDDs are observed around older stars (though the sample size is small). Another interesting feature revealed in the right panel of Figure \ref{fig:edds_context} is that silica-poor systems tend to have relatively large $W_{10}$ values, indicating a greater mass of the optically thin small grains. We will explore the implications of these findings in Section \ref{sec:discussion}. 

\begin{deluxetable*}{rcrccccccrl}
\tablewidth{0pt}
\caption{Derived properties of EDDs: 10 \micron\ dust indices and mid-infrared variability}
\label{tab:measuredindices}
\tablehead{
\colhead{Object$^1$}  &  \colhead{FWHM} &  \colhead{$W_{10}$} & \colhead{$P_{10}$} & \colhead{$O_{10}$} & \colhead{$S_{10}$} & \colhead{$S_{10,s}$} & \colhead{silica} &  \colhead{$f_d^2$}    &  \colhead{4.6 \micron\ var.$^3$}   & \colhead{10/20 \micron\ var.$^4$} \\ 
\colhead{} & \colhead{($\mu$m)}   & \colhead{($\mu$m)} & \colhead{} & \colhead{} & \colhead{} & \colhead{} &  \colhead{rich}   & \colhead{$\times10^{-2}$} &  \colhead{max/typ.}  &  \colhead{} 
}
\startdata
          HD\,172555  &  1.49$\pm$0.01  &  8.06$\pm$1.05  &  2.01$\pm$0.03  &  1.00$\pm$0.03  &  0.90$\pm$0.01  &  2.95$\pm$0.23  &  Y   &  0.07  & sat.$^\ddagger$ &  brightening$^\nparallel$   \\ 
               ID\,8  &  2.13$\pm$0.01  & 15.50$\pm$0.27  &  0.70$\pm$0.01  &  1.28$\pm$0.02  &  0.79$\pm$0.06  &  0.54$\pm$0.02  &  N   &  2.67  &    142/41  &  variable$^{\ast}$\\ 
           HD\,23514  &  2.00$\pm$0.04  &  4.43$\pm$0.28  &  2.66$\pm$0.06  &  0.90$\pm$0.02  &  1.77$\pm$0.10  &  8.66$\pm$0.25  &  Y   &  1.65  &      31/8  &  fading$^\nparallel$\\ 
           HD\,15407  &  3.22$\pm$0.08  &  2.84$\pm$0.04  &  2.10$\pm$0.02  &  1.30$\pm$0.03  &  5.99$\pm$0.49  &  6.30$\pm$0.16  &  Y   &  0.66  &      10/5  &  fading at $\lambda >$8 $\mu$m \\ 
           V488\,Per  &  3.73$\pm$0.01  &  0.68$\pm$0.04  &  1.40$\pm$0.06  &  1.95$\pm$0.10  &  1.37$\pm$0.41  &  6.53$\pm$0.40  &  Y   &  16.4  &   693/198  &  fading at $\lambda <$8 $\mu$m and W4 \\ 
               J0609  &  2.22$\pm$0.02  &  1.79$\pm$0.08  &  0.74$\pm$0.01  &  1.95$\pm$0.02  &  2.15$\pm$0.01  &  1.08$\pm$0.03  &  N   &  15.0  &      29/8  &  \\ 
               J0605  &  2.25$\pm$0.01  & 23.14$\pm$2.77  &  0.78$\pm$0.01  &  1.32$\pm$0.02  &  1.04$\pm$0.02  &  0.77$\pm$0.04  &  N   &  3.54  &     58/15  &  fading at N-band \\ 
               J0712  &  2.02$\pm$0.01  & 12.86$\pm$0.48  &  0.51$\pm$0.01  &  1.32$\pm$0.02  &  0.97$\pm$0.05  &  0.16$\pm$0.03  &  N   &  3.00  &      25/7  &  \\ 
               J0611  &  3.40$\pm$0.02  &  5.10$\pm$0.06  &  1.49$\pm$0.01  &  1.50$\pm$0.01  &  2.26$\pm$0.05  &  4.11$\pm$0.12  &  Y   &  5.09  &     46/11  &  fading at WISE W3/W4\\ 
               J0925  &  2.24$\pm$0.04  &  4.38$\pm$0.16  &  1.00$\pm$0.07  &  0.99$\pm$0.09  &  1.28$\pm$0.55  &  1.56$\pm$0.37  &  N   &  3.00  &    173/37  &  brightening at VISIR \\ 
               J1044  &  2.23$\pm$0.03  & 18.69$\pm$2.36  &  1.01$\pm$0.02  &  0.99$\pm$0.03  &  1.13$\pm$0.24  &  1.05$\pm$0.01  &  N   &  2.43  &     87/23  &  \\ 
               J1213  &  3.12$\pm$0.01  &  3.08$\pm$0.23  &  1.54$\pm$0.03  &  1.45$\pm$0.02  &  3.51$\pm$0.24  &  3.23$\pm$0.10  &  Y   &  1.02  &      21/5  &  \\ 
               J2043  &  2.25$\pm$0.01  & 10.26$\pm$0.83  &  1.03$\pm$0.01  &  1.02$\pm$0.02  &  0.72$\pm$0.18  &  0.94$\pm$0.01  &  N   &  5.21  &      20/5  &  brightening at VISIR \\ 
          HD\,166191  &  3.34$\pm$0.02  &  4.75$\pm$0.08  &  1.38$\pm$0.02  &  1.33$\pm$0.02  &  0.89$\pm$0.14  &  3.83$\pm$0.12  &  Y   &  9.15  &   sat.$^\ddagger$  & variable$^{\ast}$  \\ 
             RZ\,Psc  &  2.56$\pm$0.01  &  8.06$\pm$0.04  &  1.10$\pm$0.01  &  1.28$\pm$0.01  &  1.44$\pm$0.01  &  1.51$\pm$0.01  &  N   &  7.95  &     72/16  &  variable$^{\ast}$ \\ 
               J2301  &  2.17$\pm$0.01  & 14.80$\pm$0.35  &  0.94$\pm$0.01  &  0.98$\pm$0.01  &  0.62$\pm$0.01  &  0.89$\pm$0.05  &  N   &  0.65  &    321/87  &  fading at $\lambda >$8 $\mu$m \\ 
BD+20\,307$^\dagger$  &  2.25$\pm$0.05  & 14.88$\pm$1.87  &  1.14$\pm$0.02  &  1.05$\pm$0.02  &  0.84$\pm$0.02  &  1.24$\pm$0.10  &  N   &  3.20  &    293/57  &  brightening$^\nparallel$ \\ 
   P\,1121$^\dagger$  &  2.12$\pm$0.01  & 11.34$\pm$1.11  &  0.51$\pm$0.02  &  1.49$\pm$0.01  &  1.79$\pm$0.10  &  0.19$\pm$0.10  &  N   &  2.20  &    216/49  &  \\ 
HD\,113766$^\dagger$  &  2.56$\pm$0.01  &  6.47$\pm$0.39  &  0.98$\pm$0.01  &  1.67$\pm$0.01  &  1.35$\pm$0.18  &  1.89$\pm$0.05  &  N   &  3.49  & sat.$^\ddagger$   &  \\ 
HD\,145263$^\dagger$  &  3.21$\pm$0.01  &  7.93$\pm$0.01  &  1.54$\pm$0.01  &  1.25$\pm$0.01  &  1.07$\pm$0.04  &  3.60$\pm$0.06  &  Y   &  1.94  &      25/5  &  \\ 
   EF\,Cha$^\dagger$  &  2.18$\pm$0.01  &  6.76$\pm$0.46  &  1.22$\pm$0.01  &  0.94$\pm$0.01  &  0.72$\pm$0.13  &  1.41$\pm$0.02  &  N   &  0.18  &  sat.  &  \\ 
\enddata
\tablecomments{~~$^1$: Object short name is used here and the full ALLWISE ID is given in Table \ref{tab:jwst_obs}.; $^2$: The infrared fractional luminosity ($f_d$) values include the dust feature from the mid-infrared spectra, which are more accurate than the previous estimates based on broadband photometry; $^3$: WISE W2 variability in terms of maximum (peak-to-peak) and typical percentages relative to the minimum disk flux; $^4$: comparison between previous 10/20 \micron\ photometry/spectroscopy and the JWST/MIRI/MRS data. $^\dagger$: Using Spitzer/IRS data.  $^\ddagger$: WISE W2 Variability analysis was adopted from the Spitzer/IRAC I2 data (Appendix \ref{sec:irvariability}). $^{\ast}$: multiple mid-infrared spectroscopic data suggest both brightening and fading behaviors (Appendix \ref{sec:irvariability}). $^\nparallel$: based on published studies: \citet{su25_hd23514} for HD\,23514, \citet{samland25_hd172555} for HD\,172555 and \citet{thompson19} for BD+20\,307. }
\end{deluxetable*}

\subsection{Infrared Variability }
\label{sec:edds_III}
 
Infrared variability resulting from changes in the emitting area (i.e., total dust cross-section) greater than a few times the stellar size is another hallmark commonly observed in EDDs. Put in perspective, the required total cross-sectional area for $f_d =$ 0.01 at 1 au is roughly 3$\times10^{25}$ cm$^2$; therefore, 10\% variations require changes to the area of $\pm$3$\times10^{24}$ cm$^2$. Both short-term (weekly to monthly) and long-term (yearly) variations are recognized \citep{su19,moor21}. Dedicated monitoring by the warm Spitzer mission, using 3--5 \micron\ photometry for a sample of two dozen EDDs, shows that most variations are stochastic \citep{rieke21_v488per,su22_hd166,su23_rzpsc,su25_hd23514}, while only three systems exhibit significant semi-regular monthly modulations in their infrared light curves (ID\,8, P\,1121 and TYC\,4209-1322-1; \citealt{su19,moor22_tyc4209}). The "wiggly" nature in these EDDs is related to the complex, dynamical, and collisional interplay between the brighter vapor condensates initially and the fainter boulder-grinding disks that emerge later following a violent impact \citep{su19,watt24_postimpact_edd_evolution}. In rare highly-inclined systems, irregular optical transit profiles are observed as large dust clumps, formed in the aftermath of a major impact, pass in front of the star \citep{gaidos19_hd240779,melis21,su22_hd166,su23_rzpsc,kenworthy23_asassn21qj,zakamska26_asassn24fw}.

We focused on long-term behaviors by conducting an analysis of the general infrared variability around these EDDs, utilizing WISE and other mid-infrared photometric data. WISE 3--5 \micron\ photometry tracks the evolution of warm dust every six months between 2014 and 2024, while 10/20 \micron\ photometry from IRAS, Spitzer, AKARI, MSX, and/or ground-based VLT, Gemini and Subaru monitors the evolution of silicate features over several decades. The variability behavior for each of the objects is detailed in Appendix \ref{sec:irvariability}. Since infrared observations directly track variations in the debris cross-sections (constraining the minimum change required), we aim to connect the observed dust mineralogy with disk variability to investigate potential correlations.

For each system, we derived two indicators to represent the relative infrared variability using WISE 4.6 \micron\ data\footnote{Several systems are saturated in the WISE W1/W2 data, so we adopted archival Spitzer I1/I2 data instead.} (see Table \ref{tab:measuredindices}): the maximum and typical variations relative to the minimum disk flux at 4.6 \micron, which is assumed to be the nominal background flux level sustained by collisions among existing small planetesimals. The typical variation reflects the stochastic nature of collisions, while the maximum variation represents the greatest degree of change observed in the data over the monitored period. As detailed in Appendix \ref{sec:wise4.6um}, we find that the two variability indicators are strongly correlated, suggesting that they are likely driven by a common physical process related to the aftermath of giant impacts. The observed flux change can also be expressed as the minimum change in dust cross-section ($\Delta\Sigma$) required to match the maximum flux change under the assumptions that (1) the emission is optically thin and (2) all dust is at the same temperature. Using the warmer temperature used in the pseudocontinuum, we found that $\Delta \Sigma$ ranges from to 2.7 to 233 times that of the stellar surface (assuming a size comparable to the Sun) in our sample (details see Appendix \ref{sec:wise4.6um}), suggesting that the variation originates from the circumstellar environment rather than from the star or any emerging planetary bodies. 

\begin{figure}
    \centering
    \includegraphics[width=\linewidth]{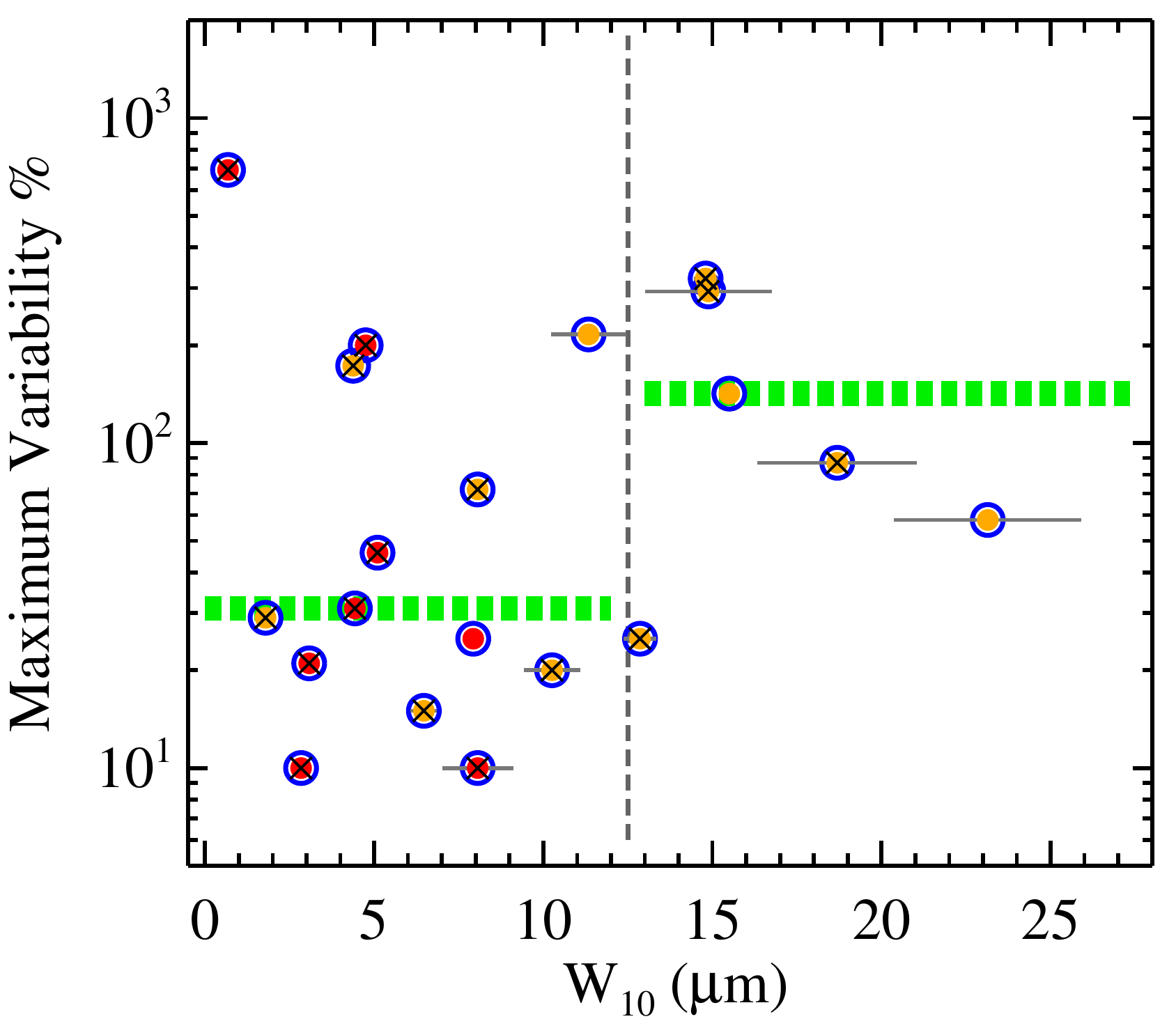}
    \caption{Correlation between the observed 4.6 \micron\ disk variability (y-axis) and EDD properties: $W_{10}$ (x-axis), impact-dust mineralogy (red: silica-rich; orange: silica-poor), and multiplicity (crosses). Dividing EDDs by $W_{10}$ shows a potential trend: high‑$W_{10}$ ($\gtrsim$12.5 \micron) systems -- exclusively silica‑poor -- tend to exhibit greater variability (a factor of $\sim$5), as indicated by the median values (green dashed lines). }
    \label{fig:var_w10}
\end{figure}

We observe no strong correlation between disk variability, as measured by both relative flux change and dust cross-section, and factors such as derived dust indices, stellar age, and multiplicity (Figure \ref{fig:WISEvariability}). In systems younger than 300 Myr ($\sim$85\% of the sample), we observe a diverse range of variability levels and dust mineralogy. Older systems (age $>$300 Myr) seem to occupy a region of greater variability compared to younger EDDs; however, the sample size of older systems is small (only three). The only correlation we identified is that high $W_{10}$ systems (a proxy for the mass of optically-thin small grains), which are exclusively silica-poor, tend to exhibit greater variability by a factor of 5 as measured in the relative flux space (Figure \ref{fig:var_w10}), but not in the dust cross-section (Figure \ref{fig:WISEvariability}). This trend is unlikely to be related to age, as no correlation is observed between stellar age and silica-poor systems. We will explore this further in Section \ref{sec:discussion}. 

Multiplicity is a notable characteristic of EDDs, indicating that most EDDs have one or more companions at varying separations -- close (within a few au), intermediate (tens to 500 au), or wide (greater than 500 au; details in Table \ref{tab:EDD_properties}). Nonetheless, our search for correlations between multiplicity and EDD properties revealed no significant trends. Notably, three older EDDs (over 300 million years) with high $W_{10}$ values are found in wide binary systems with high eccentricity, but not all high $W_{10}$ EDDs have stellar companions. Due to the small sample size, the influence of multiplicity on EDD phenomena remains unclear. Future research would benefit from a larger sample, especially with older EDDs and more precise binary property assessments.

Given that the time sampling for the 10/20 \micron\ observations varies across different sources (depending on brightness), we focus on comparing the synthetic photometry derived from the JWST/MIRI/MRS spectrum with existing photometry from IRAS, AKARI, Spitzer, MSX, and ground-based mid-infrared cameras. This comparison allows us to characterize EDDs as exhibiting fading, brightening, or both behaviors at 10/20 \micron\ (Table \ref{tab:measuredindices}). As detailed in Appendix \ref{sec:10_20umVar}, all three behaviors are found, although the fading behavior is detected more frequently than the other two (Figures \ref{fig:midir_var} and \ref{fig:midir_var2}). 
This behavior aligns with the identification criteria for EDDs, which are exceptionally brighter than the general population, suggesting that this phenomenon is associated with the aftermath of an abrupt past event. Except for a few cases where significant brightening events were recorded in the last decade with sufficient infrared temporal data, most major impacts in EDDs occurred prior to their discovery (e.g., \citealt{su22_hd166,kenworthy23_asassn21qj}). We also find no dramatic changes in the morphology of the 10 \micron\ feature over several decades, with variations primarily attributed to the underlying continuum rather than the dust composition, although subtle changes may still occur (Figure \ref{fig:midir_var}). This overall stability is consistent with earlier studies \citep{thompson19,su20,su25_hd23514,samland25_hd172555}. Similar to the 4.6 \micron\ WISE data, there is no significant correlation between 10/20 \micron\ variability behavior and dust mineralogy, regardless of whether the dust is silica-rich or silica-poor.

\section{Implications}
\label{sec:discussion}

\begin{figure*}
    \centering
    \includegraphics[width=0.495\linewidth]{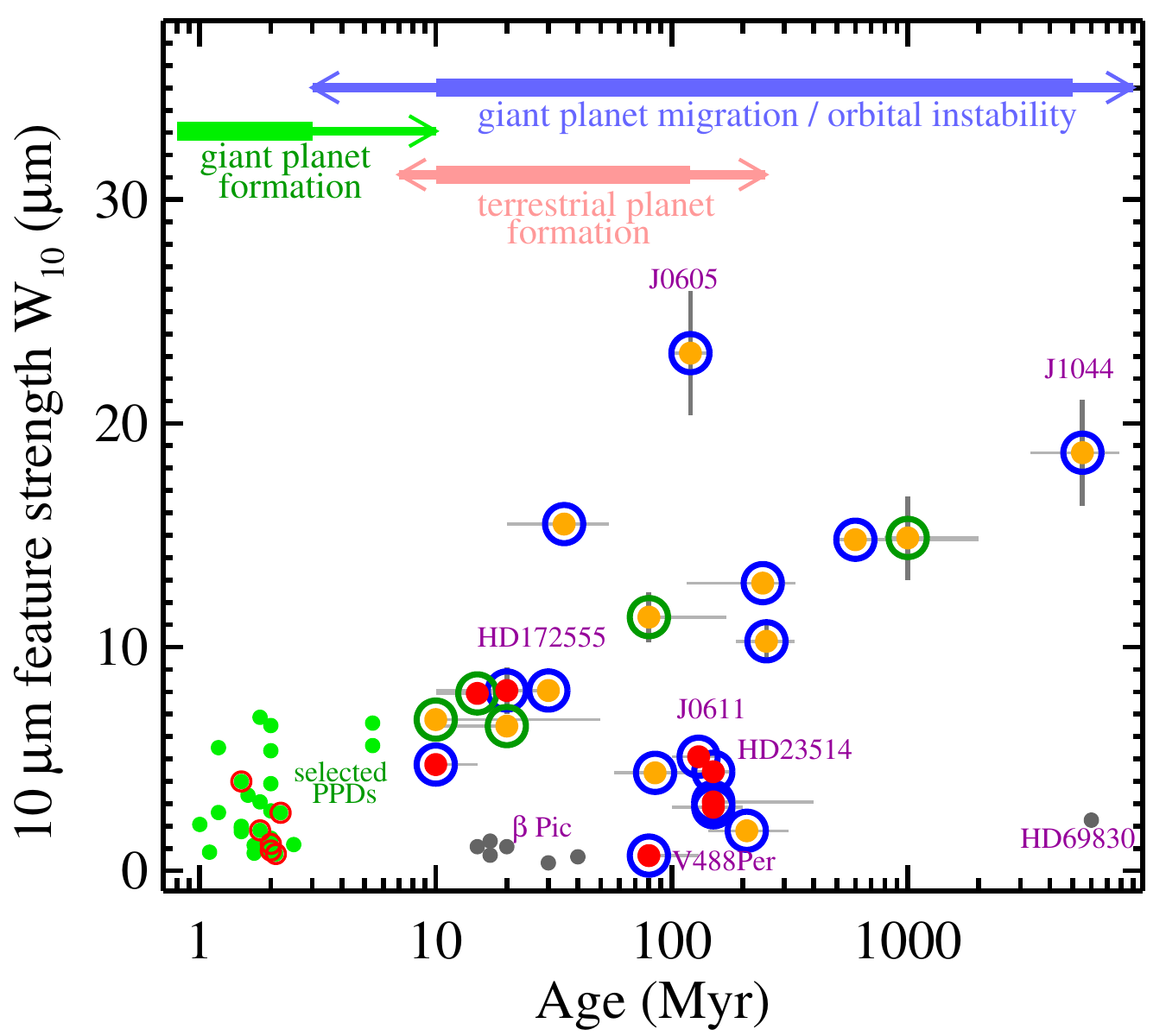}
    \includegraphics[width=0.495\linewidth]{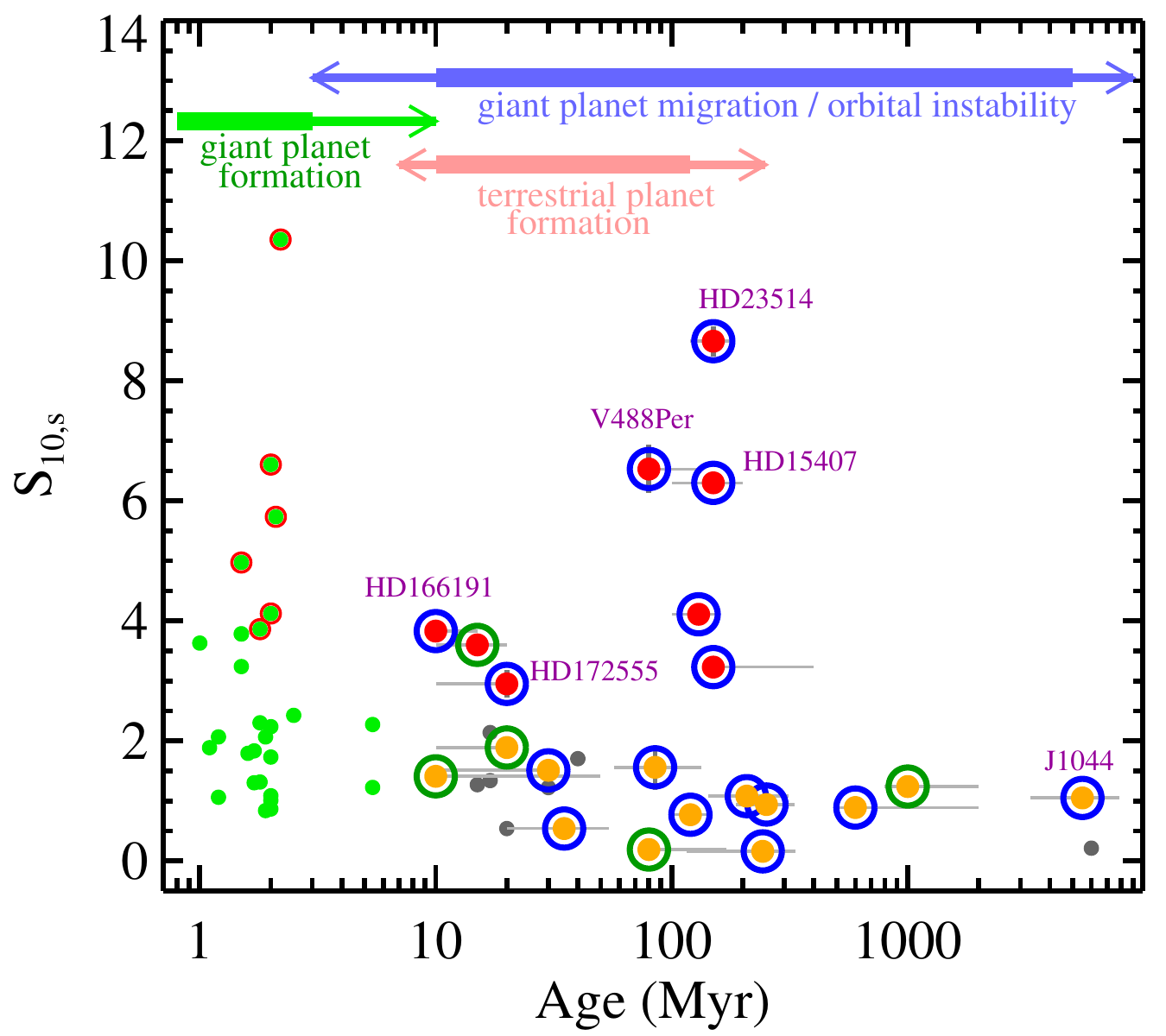}
    \caption{The 10 \micron\ feature strength (left) and abundance of silica dust (characterized by $S_{10,s}$, right) are shown at three different stages of planet formation and evolution: giant planet formation, terrestrial planet formation, and the giant planet migration/orbital instability period, which are indicated at the top of each panel. Symbols and colors are the same as in Figure \ref{fig:edds_context}. Both types of silica-rich and silica-poor EDDs are found during the terrestrial planet formation phase, but no silica-rich systems are found in stars $\gtrsim$300 Myr despite their relatively large $W_{10}$, implying that collisions induced by orbital instability are less violent but more frequent at ages $<$300 Myr. High abundance of silica dust is only associated with the EDDs corresponding to the terrestrial planet formation phase. }
    \label{fig:age_trends}
\end{figure*}

The implications of our results are summarized and illustrated in Figure \ref{fig:age_trends}, where the three different phases of planet formation and evolution -- giant planet formation prior to the dispersal of nebular gas, terrestrial planet formation through planetesimal/embryo collisions, and orbital instability due to the rearrangement of giant planets or other late instabilities -- are roughly categorized by the system's age. While PPDs establish the initial conditions for giant planet formation, EDDs offer the best insight into the dynamically active stages during terrestrial planet formation and subsequent planetary evolution, particularly through the properties of highly processed materials. Although both silica and crystalline silicate dust have undergone significant processing and intense heating, silica-rich materials exhibit higher shock levels than the others (referred to as silica-poor), particularly the forsterite-rich materials, as noted by \citet{morlok14_processedmaterial,morlok14}. Because impact energy strongly depends on mass, we suggest that silica-rich EDDs tend to arise from collisions between large planetary embryos (similar in size to Mars), while silica-poor EDDs are linked to impacts involving somewhat smaller bodies (approximately the size of the Moon) involving less energetic events. Notably, silica-rich EDD systems are exclusively observed around stars younger than $\sim$300 Myr, consistent with the findings of numerical simulations on terrestrial planet formation, as the giant impact phase is expected to significantly diminish after a few hundred million years \citep{chambers13,quintana16}.

Terrestrial planet formation is a messy, complicated process. Impact simulations demonstrate a broad spectrum of outcomes that depend on factors such as collision type (head-on vs.\ grazing), impact velocity, and impactor sizes \citep{stewart12}, and the presence or absence of giant planets \citep{carter15,carter_stewart20}. Therefore, it is not surprising to observe a mixture of silica-rich and silica-poor EDDs during the terrestrial planet formation era. Although the smallest size of debris clumps that can be tracked in impact simulations is much larger than the dust observed through astronomical observations, these simulations can provide a zero-order estimate of the spatial distribution of the resulting debris.  Recent studies \citep{cambioni19,emsenhuber24_giantimpacts} indicate that events involving low-mass (less than Mars-sized) impacting bodies are more likely to generate numerous debris clumps, while high-velocity impacts (greater than $\sim$40 km\,s$^{-1}$) produce fewer clumps, likely due to the stronger shocks or vaporization that occur under such conditions, which inhibit debris formation. Our findings are consistent with these theoretical expectations, as silica-rich EDDs typically exhibit relatively low $W_{10}$, which is attributed to a smaller amount of optically thin small dust. The abundance of silica dust as traced by the $S_{10,s}$ index in EDDs peaks at a few hundred million years (as shown in the right panel of Figure \ref{fig:age_trends}), providing strong observational support for the occurrence of planetary embryo collisions (i.e., the final phase of giant impacts) in theoretical studies.

We further suggest that the relative abundance of silica dust compared to non-silica dust (such as crystalline silicates) might indicate how long it has been since the initial giant impact that created the silica dust. After a giant collision, a new belt of resulting remnants (objects smaller than the original bodies) is expected to form. Subsequent collisions in the belt likely produce silica-poor material, as these impacts are generally less energetic than the original, as typically found in impact simulations. This speculation would only hold if both types of dust have similar lifetimes against radiation pressure blowout, as subsequent dust dynamics could erase the initial imprint.

We propose that silica dust in the silica-rich EDDs forms from direct gas phase condensation resulting from high-energy impact events. Such energetic impacts are expected to vaporize a significant fraction of the impacting bodies \citep{davies20}, resulting in the formation of droplets that condense from rapidly cooling gas.  These condensed materials -- known as spherules -- have been identified in impact products preserved in terrestrial and lunar craters \citep{smit99,johnson14}. In this case, the expansion of the vapor quickly cools the gas and should preferentially form silica in low ambient pressure as amorphous polymorphs such as silica glass, obsidian, or fused quartz. Grinding down the freshly condensed spherules into sub-micron dust occurs rapidly in the dense, impact-produced cloud \citep{johnson12b,su20}. This is consistent with the observation that obsidian and fused quartz are the matching minerals that can reproduce the observed spectra of HD\,172555 and HD\,15407 \citep{lisse09,fujiwara12}. For less energetic events, in addition to direct gas phase condensation, annealing may play an important role in forming crystalline silicates, specifically Mg-rich forms like forsterite \citep{davoisne06_GEMS}. Interestingly, there exists 
an antagonistic relationship between quartz (crystalline silica) and forsterite -- pairs of minerals that rarely or never coexist in igneous rocks in the early Hadean Eon \citep{hazen23_evolutionarymineralogy_VII}. Given the prominence of the features, it is unlikely that the apparent dichotomy (silica-rich vs.\ silica-poor/forsterite-rich) is geologically related to newly formed magma oceans, as these highly processed grains likely formed in interplanetary space -- due to rapid cooling of expanding vapor -- rather than on planetary surfaces. A key test for silica formed in expanding vapor condensates will be the scarcity of high-pressure polymorphs such as coesite and stishovite. Future spectral decomposition analysis of these newly identified silica-rich EDDs will further test this hypothesis.

Our analysis also reveals that silica-poor systems are preferentially linked to higher $W_{10}$ values (as shown in the left panel of Figure \ref{fig:age_trends}), where large $W_{10}$ values imply a more widespread distribution of optically thin debris. Notably, all three older EDDs, which exist beyond the era of terrestrial planet formation, exhibit very high $W_{10}$ values and are silica-poor. These old EDDs likely resulted from planet migration and the rearrangement of an earlier established planetary system, triggered by a late dynamical instability due to planet-planet and planet-planetesimal interactions \citep{raymond10,volk15,quarles_kaib19} or by tidal evolution of planet-satellite interactions \citep{hansen23}. The former cases might be analogous to proposals for a Late Heavy Bombardment in the solar system, when the gas giants migrated outward, creating a more chaotic environment that disturbed smaller bodies and sent them toward the inner terrestrial zone \citep{strom05,gomes05,bottke17_LHB_review}. The latter case might be similar to how some of the moons formed around Saturn \citep{asphaug_reufer13_lateformationSaturnMoons,cuk16_lateformationSaturnMoons}.

Interestingly, recent studies of the solar system suggest that the orbital instabilities shaping its final planetary architecture may have occurred earlier than previously expected \citep{nesvorny18,liu22}. 
In parallel, studies of mature multi-planet systems observed in transit reveal a similar picture. Resonant chains of planets, which are expected to form during gas-disk-driven migration (e.g., \citealt{izidoro17_breakingthechains,wong_lee24} and references therein), are nearly universal among the youngest systems but are progressively dissolved with age. This ``breaking-the-chain" process is reflected in the decline of near-resonant configurations from $\gtrsim$80\% ($<$100 Myr) to $\sim$40\% (100 Myr--1 Gyr) and $\lesssim$ 25\% ($>$1 Gyr), implying a characteristic disruption timescale of $\sim$100 Myr
\citep{hamer24_MMRs_kinematicallyYoung,dai24_PrevalenceYoungResonancePlanets}. A range of mechanisms has been proposed to drive this evolution \citep{izidoro17_breakingthechains,poon20_compactSuperEarthInstabilityGiantImpacts,goldberg22_CompactSystemOrbitalInstabilities,li25_BrokenChainsMajorMinorMergers,charalambous25_breakingchainStellarFlybys}, including interactions with leftover planetary embryos as well as residual planetesimal populations \citep{goldberg_petit25,li25_breakingChainsPlanetesimalFlybys,ogihara_kunitomo25}. The coincidence of this timescale with the observed peak in the incidence of EDDs raises the possibility that EDDs might also be an observable byproduct of the same dynamical instabilities that reshape planetary system architectures.

Additional evidence linking the EDD phenomenon to dynamical instability -- rather than terrestrial planet formation -- arises from the observed disk variability. As shown in Figure \ref{fig:var_w10}, EDDs with high $W_{10}$ ($\gtrsim$12.5 \micron) tend to exhibit greater variability at 3--5 \micron\ -- by a factor of five -- as indicated by the median values for the $W_{10}\gtrsim$12.5 \micron\ and $W_{10}\lesssim$12.5 \micron\ groups. The combination of high $W_{10}$ and the degree of variability suggests that these large $W_{10}$ systems are undergoing significant collisional activities. Furthermore, the fact that they are all silica-poor systems implies that these collisional activities are predominantly driven by minor bodies and/or less energetic grazing events. This is consistent with the fact that hit-and-run type of collisions are most commonly found during dynamical instability \citep{ghosh24_subNeptuneCollisions,cambioni25}. Our findings align with the results from \citet{wu24_pingpongscatering} and \citet{li25_BrokenChainsMajorMinorMergers}, which demonstrate that minor mergers between debris and resonant planets play an important role in enhancing the asymmetry of the departure from exact commensurability between neighboring planets. In this context, silica-poor EDDs with very high $W_{10}$ values may serve as indicators of dynamical instabilities that ultimately shape the final architecture of planetary systems. This intriguing hypothesis can be further tested with a larger sample of high-quality JWST data, particularly by expanding the focus to include older EDDs. 
  
In closing, we briefly discuss two circumstellar disks outside our sample that may or may not be related to the EDD phenomenon. We do not classify the HD\,69830 system as an EDD due to its relatively low dust levels and lack of variability. This 4--10 Gyr-old star, which hosts three super-Earth-sized planets (likely in a chain of mean motion resonances within 0.63 au \citep{lovis06}) and that are packed inside a highly stirred debris disk at $\sim$1 au \citep{wyatt07a}, would have been classified as an EDD in its past. Currently, its debris is approximately 1,400 times more luminous than the zodiacal emission in our solar system, exhibiting high crystallinity from small grains \citep{beichman11}. This contrasts sharply with the properties of our Solar system's zodiacal cloud, which is dominated by grains of a few 100 \micron\ in size. Several theoretical studies have explored its formation pathways \citep{alibert06,payne09}, making it one of the successful examples in the context of ``breaking-the-chain" instability where the three super-Earths locked into orbital resonances as they are today. We stress that the EDD phenomenon can arise from various mechanisms, including the formation of circumplanetary rings resulting from collisions or tidal disruptions of satellites orbiting giant planets.  Notably, a recent observation has identified a giant, gas-rich circumplanetary ring occulting the Gyr-old main-sequence star ASASSN-24fw, where its significant infrared excess of warm dust may qualify it as an EDD \citep{zakamska26_asassn24fw,shah_asassn-24fw}. The size of the ring is similar to that of Saturn's largest dusty ring, discovered by Spitzer \citep{verbiscer_2009_SaturnLargestRing}. Its exact connection with the EDD phenomenon requires further investigation.

\section{Conclusion }
\label{sec:conclusion}

We examined a sample of 21 EDDs with high-quality mid-infrared spectroscopic data (Section \ref{sec:data}) to enhance our understanding of the relationship between dust mineralogy and EDD properties (Section \ref{sec:analysis}). This includes disk variability, which reflects the dynamical state following major impact events (Appendix \ref{sec:irvariability}), and stellar age, which signifies the different stages of planet formation and evolution (Section \ref{sec:discussion}). Our findings suggest that EDDs consistently contain a significantly higher amount of low optical-depth dust grains compared to the typical levels observed in PPDs and DDs (Appendix \ref{sec:labindices}). These grains are predominantly sub-\micron\ in size, whereas PPDs are primarily dominated by micron-sized grains. Additionally, our analysis reveals that silicate dust in EDDs exhibits high crystallinity in both pyroxene and olivine similar to that found in PPDs, indicating that it has undergone advanced processing compared to primordial ISM-like material. The high crystallinity in silicates observed in PPDs has been recognized for decades and is mainly attributed to thermal alterations resulting from rapid heating and cooling in gas-rich environments (see references in \citealt{jang24_spatialdist_crystaline}). We argue that the high crystallinity observed in EDDs is not inherited from early-stage PPDs; if it were, we would have observed similar crystallinity in DDs. Instead, this high crystallinity is generated anew during recent transient violent events, as evidenced by the combination of short loss timescales and large quantities.

In addition to the large quantity of small grains (traced by the equivalent width of the 10 \micron\ feature $W_{10}$), another important factor that distinguishes EDDs from PPDs is the abundance of silica dust.  One-third of the EDDs ($\sim$38\%) exhibit significantly elevated levels of silica dust, which tend to broaden the measured FWHM of the 10 \micron\ feature (Appendix \ref{sec:labindices}). Four dust indices, $P_{10}$, $O_{10}$, $S_{10}$, and $S_{10,s}$, are used to quantify the 10 \micron\ dust morphology, with the former two describing the silicate dust crystallinity and the latter two addressing the silica fraction. By benchmarking dust indices measured from natural silica-rich materials and previously identified silica-rich systems, we categorize the EDD sample into silica-rich and silica-poor systems by establishing thresholds in the $S_{10,s}$ index. In our dichotomous terminology, a silica-rich system is characterized by observed features that closely resemble those of silica dust in sufficient abundance, exceeding the typical values found in circumstellar disks. The exact percentage of silica-rich PPDs is uncertain (largely because they were not the focus of early studies) but is likely under 10\%, suggesting that advanced thermal processing in EDDs primarily drives the formation of silica dust.

A final characteristic commonly found in EDDs is that they often exhibit stochastic infrared variability on both short- (weekly to monthly) and long-term (yearly) timescales due to significant changes in the distribution of circumstellar dust. We quantified the disk variability using both decade-long WISE data at 3--5 \micron\ (Appendix \ref{sec:wise4.6um}) and 10/20 \micron\ photometric data (Appendix \ref{sec:10_20umVar}), where the WISE data probe the hotter part of the debris, while the 10/20 \micron\ photometry traces both the 10-\micron\ features resulting from small grains and warm dust particles. We found no strong correlation between infrared variability and dust content -- measured as the infrared fractional luminosity -- nor with stellar age and multiplicity (Figure \ref{fig:WISEvariability}), noting that only three EDDs in the sample are older than 300 Myr, and nearly all EDDs have companions at close, intermediate and wide separations (Table \ref{tab:EDD_properties}).  Surprisingly, the silica richness or poorness of the system does not impact the degree of variation; a mixture of behaviors, including fading, brightening, or both, is observed. Comparisons between the synthetic photometry derived from JWST spectra (obtained in 2023/2024) and other mid-infrared photometry from 1983 to 2022 indicate that fading behaviors are more frequently detected within the sample. Additionally, no significant morphological change is observed in the 10-\micron\ feature, suggesting that the changes are primarily due to the amounts of dust rather than changes in dust composition. Overall, these findings indicate that the EDD dust mineralogy was inherited from the initial events that created the EDD phenomenon and has remained stable since its formation. The observed disk variability is linked to the aftermath of a past event, which may be the initial violent collision or subsequent related events triggered by that collision.

EDDs provide crucial insights into terrestrial planet formation and subsequent evolution, particularly following the initial phase of giant planet formation in a gas-rich environment. The highly processed material, in the form of crystalline silicates and silica dust, offers diagnostic information about the physical conditions of giant impact events during terrestrial planet formation, aiding in differentiating the complex outcomes of these impacts. We propose that silica dust in EDDs forms through gas-phase condensation during violent energetic events that experience greater shock levels than those primarily responsible for forming crystalline silicates, where both direct condensation and the annealing of existing dust play significant roles. As a result, silica-rich EDDs arise from collisions between planetary embryos, similar to the Mars-sized objects modeled in impact simulations, while silica-poor EDDs originate from collisions between smaller objects (i.e., Moon-sized objects) and/or less energetic impacts, such as low-velocity grazing events. Our hypothesis is well-supported by impact simulations, which indicate that giant impacts involving planetary-embryo-sized objects generate less dispersible debris, as a significant portion is eventually accreted back to the emerging body. This finding aligns with observations that silica-rich systems exhibit relatively lower $W_{10}$ values compared to their silica-poor counterparts. Additionally, we find that the prevalence of silica-rich EDDs peaks at a few hundred million years and diminishes after 300 Myr, providing robust observational evidence for the frequency of giant impacts amongst planet-sized bodies, and signaling the final phase of terrestrial planet formation, as envisioned by theoretical simulations.

While no clear correlation exists between the EDD dust mineralogy (silica-rich vs.\ silica-poor) and infrared variability at 3--5 \micron, our analysis identifies a potential trend when differentiating EDDs by the amount of feature-producing small grains (i.e., $W_{10}$). The dynamical state of a planetary system, as measured by infrared variability, is five times higher for EDDs with large $W_{10}$ (roughly one-third of the sample) than for those with small $W_{10}$, with all large $W_{10}$ systems being exclusively silica-poor. The correlation between high $W_{10}$ values and the high degree of variability indicates that these silica-poor, large $W_{10}$ systems are undergoing significant collisional activities, likely due to orbital instability caused by the dynamical interactions of unseen planets. This is further supported by the fact that three out of seven of these highly active, silica-poor EDDs are older than 300 Myr, beyond the expected epoch of terrestrial planet formation.  It is noteworthy that the emergence of these silica-poor, large $W_{10}$ systems occurs around $\sim$100 Myr, although our sample is small. This leads us to propose that silica-poor EDDs with large $W_{10}$ values might be signposts for planetary systems undergoing significant planetary rearrangement -- for example, a phenomenon known as``breaking the chain", which has been identified to explain the architectural arrangement of mature multiple-planet systems. Our reasoning is threefold. First, no silica-rich EDDs have been found beyond the era of terrestrial formation ($>$300 Myr). Second, silica-poor EDDs likely result from collisions between Moon-sized bodies or less energetic events that produce minimal vapor condensates. The presence of large $W_{10}$ values necessitates numerous such collisions, leading to increased dynamical activity, as indicated by infrared variability. Finally, both observational evidence and theoretical studies suggest that the phase of orbital instability that reshapes planetary architecture likely occurs within a few hundred million years. Our results therefore serve to motivate future theoretical investigations aimed at linking ``breaking-the-chains" models with observable dust production, a topic that has received little attention to date.
 
While our study provides observational evidence that enhances the characterization and definition of a subclass of debris disks, the EDDs, and their connections to terrestrial planet formation as well as systems experiencing orbital instabilities, many unexplored areas remain related to their direct relationships with these active research fields. Although impact simulations align with our general conclusions, many details need further investigation. For example, most simulations do not directly compute the vapor population because they lack or simplify shocked thermodynamic physics. How vapor condenses to dust (composition and size distribution), how composition is altered by melting, and how the resulting material subsequently evolves are important, because these processes directly determine the solid-state features observed in EDDs over time. Most EDDs are known to exist within binary or multiple systems; however, the absence of detailed information about companion stars -- such as their mass and orbital properties -- limits our ability to draw conclusive insights into their roles in the EDD phenomenon. This challenge is further compounded by the scarce information on other EDD properties, including the presence of existing planets and the cold components of distant planetesimal populations. Additionally, the small sample size (fewer than two dozen) significantly restricts our capacity to draw comprehensive conclusions. Expanding the sample to target specific parameter spaces and providing detailed characterization of known individual systems will further test the proposed connections.

\appendix
 
\restartappendixnumbering

\section{Extreme Debris Disk Sample}
\label{sec:jwstsample}

\begin{deluxetable*}{lcccrccl}
    \footnotesize
    \tablecaption{JWST Observation Log }
     \label{tab:jwst_obs}
\tablehead{
\colhead{ALLWISE ID} & \colhead{Short Name}  &\colhead{PID} & \colhead{Obs. Date} & \colhead{BMJD}   & \colhead{Exp$^\dagger$} &  \colhead{Reference} 
}
\startdata
 J104416.70-451613.9 &     J1044  & 3189 &      2024-06-05 &    60466.495 &      888 &        [1]  \\ 
 J181030.32-233400.5 &HD\,166191  & 3189 &      2024-05-04 &    60434.184 &       99 &        [2,3]  \\ 
 J061103.54-471129.2 &     J0611  & 3189 &  2024-04-01 &    60401.107 &      222 &        [4]  \\ 
 J010942.07+275701.7 &   RZ\,Psc  &  3189 &  2024-07-28 &    60520.001 &      222 &        [5]  \\ 
 J060513.59-191308.4 &     J0605  &  3189 &  2024-03-06 &    60375.608 &      444 &        [4]  \\ 
 J060917.00-150808.5 &     J0609  &  3189 &  2024-03-06 &    60375.567 &      222 &        [1]  \\ 
 J204315.23+104335.3 &     J2043  &  3189 &  2024-06-22 &    60483.087 &      444 &        [1]  \\ 
 J092521.90-673224.8 &     J0925  & 3189 &  2024-05-07 &    60437.758 &      222 &        [1]  \\ 
 J230112.67-585821.9 &     J2301  &  3189 &  2024-08-16 &    60538.662 &      888 &        [6]  \\ 
 J121334.13-053543.4 &     J1213  &  3189 &  2024-06-22 &    60483.332 &      888 &        [4]  \\ 
 J071206.54-475242.3 &     J0712  &  3189 &  2024-03-16 &    60385.209 &      888 &        [1]  \\ 
 J080902.49-485817.2 &     ID\,8  &  1206 &  2024-01-07 &    60316.715 &      888 &        [7]  \\ 
 J032818.69+483947.9 & V488\,Per  & 1206 & 2024-02-25 &    60365.240 &      177 &        [8]   \\
 J034638.40+225510.7 & HD\,23514  &  1206 & 2023-09-23 &    60210.761 &      133 &        [9,10] \\
 J023050.76+553253.3 & HD\,15407  &  1206 & 2023-09-23 &    60210.808 &      111 &        [11]   \\
 J184526.93-645218.1 &HD\,172555  &  1282 &  2023-09-18 &    60205.475 &      654 &        [12,13]   \\
\enddata
    \tablecomments{~~~$^\dagger$ on-source exposure time per subband per channel. Reference: [1] \citealt{moor24_edds_visir}, [2] \citealt{kennedy14}, [3] \citealt{su22_hd166}, [4] \citealt{moor21}, [5] \citealt{su23_rzpsc}, [6] \citealt{melis21}, [7] \citealt{su19}, [8] \citealt{rieke21_v488per}, [9] \citealt{rhee08}, [10] \citealt{su25_hd23514}, [11] \citealt{fujiwara12}, [12] \citealt{lisse09}, [13] \citealt{samland25_hd172555}. }
\end{deluxetable*}

Although dozens of EDD candidates have been identified using the all-sky WISE and Gaia catalogs (e.g., \citealt{contardo_hogg24}), the EDD nature of the 21 systems presented here has been confirmed by additional follow-up observations that determined their stellar properties and ruled out background contamination as false identifications of infrared excess. In particular, these systems all show indications of prominent 10 \micron\ features, either from previous mid-infrared spectroscopy  (mainly Spitzer/IRS) or narrow-band photometry (VLT/VISIR, \citealt{moor24_edds_visir}). Table \ref{tab:jwst_obs} lists the JWST observations, including the program ID, observation date, and on-source exposure time. Table \ref{tab:EDD_properties} gives other system properties, including those observed by Spitzer, and references used in this study. Most of the stellar properties, such as spectral type, temperature, mass, age and multiplicity, were adopted from the literature when available, while we derive the values for those that are not. The uncertainties in the system age depend on the diagnostic methods (i.e., stellar lithium content, gyrochronology, magnetic activity, and kinematic properties); the listed values represent our evaluation of the literature values combined with our own assessments. 

Table \ref{tab:EDD_properties} also identifies the five EDDs for which cold ($\lesssim$150 K) dust is required either to fit their far-infrared SEDs or because their disk emission is resolved in the millimeter (analogous to the Kuiper belt). Identifying a cold component typically requires far-infrared data, which many WISE-discovered EDDs lack. Consequently, whether a system hosts distinct (hot/warm/cold) planetesimal belts cannot be reliably determined from the SED alone \citep{kennedy_wyatt14}. We list this cold-component information only for completeness and for use in future individual studies. HD\,166191, HD\,113766, and RZ\,Psc have ALMA millimeter observations that detect dust emission confined within tens of au \citep{worthen26_hd166,matra25_reasons,su23_rzpsc}. HD\,172555 and HD\,145263 also have ALMA detections — the former is resolved as a compact disk ($<$10 au; \citealt{roumeliotis26_hd172555}) and the latter is unresolved \citep{lieman-sifry16}. 
Importantly, their millimeter fluxes are consistent with the bulk SED emission at temperatures of $\sim$300–400 K, not with the cold dust typically inferred for Kuiper-belt analogs.

Multiplicity has also been found to be a common feature in EDDs, as suggested by previous studies \citep{zuckerman15,moor21,moor24_edds_visir}. Excluding EDDs that are farther than 250 pc and less-studied systems, most EDDs have one or more companions at either very close (within a few au), intermediate (a few tens of au to 500 au) or wide ($>$500 au) separations (Table \ref{tab:EDD_properties}). Table \ref{tab:comp_properties} details the companions' properties collected from the literature for completeness and, like the cold‑component information, serves as motivation for future detailed study. Similarly, no planetary objects have been reported as associated with these systems. These stellar properties (age, luminosity, and multiplicity) are further used to search for correlations with the derived properties of the EDDs (see Appendix \ref{sec:irvariability} for details). 

The mid-infrared disk spectra shown in Figure \ref{fig:disksed_linearly} were derived by subtracting the stellar contribution using Kurucz atmospheric models estimated by the stellar type, temperature, and distance. We note that the MIRI/MRS spectra of HD\,172555 and HD\,23514 have been published previously by \citet{samland25_hd172555} and \citet{su25_hd23514}. Additionally, both ID\,8 and V488\,Per have multiple epochs of MIRI/MRS spectra, but we only use the most recent ones for this work. We derived the infrared fractional luminosity ($f_d$) as listed in Table \ref{tab:measuredindices} by including the contributions of the prominent dust features, which provide a more accurate estimate than the previous values commonly derived from broadband photometry.

Although our mineralogy analysis solely focuses on the 10 \micron\ region, all EDDs exhibit solid-state features in the associated 20 \micron\ range, which is important for identifying the exact mineral composition and the Mg/Fe ratio in crystalline silicates \citep{deVries18}. Detailed spectral decomposition is needed to extract this information (e.g., \citealt{juhasz09,olofsson10_c2d_spectraldecomp}), which is beyond the scope of the current work. We also note that a handful of the systems display noticeable features that are not characterized in the current study. Both HD\,166191 and V488\,Per show a 6.3 \micron\ feature arising from complex hydrocarbons  (also called polycyclic aromatic hydrocarbons, PAHs)  \citep{jensen22_complexhydrocarbons} that are first detected in debris disk systems (Figure \ref{fig:disksed_linearly}). Infrared gas emission lines have been reported in HD\,172555 and HD\,23514 \citep{samland25_hd172555,su25_hd23514}, and here we report hot CO gas emission is also detected in HD\,15407 for the first time. Additionally, CO$_2$ gas absorption features near 15 \micron\ were also detected in HD\,166191, RZ\,Psc, and J0611, with the former two systems known to host highly inclined disks as shown through resolved imaging \citep{su23_rzpsc,worthen26_hd166}, and cold CO emission has been detected in HD\,166191 by ALMA \citep{worthen26_hd166}. A detailed analysis of these gas properties in EDDs will be reported in a future paper.

\begin{deluxetable*}{rccccccccccccc}
     \tablewidth{0pt}
    \footnotesize      
    \tablecaption{EDD properties}
    \label{tab:EDD_properties}
\tablehead{
\colhead{Name} & \colhead{D} & \colhead{SpT} & \colhead{$T_\ast$} & \colhead{$M_\ast$} & \colhead{$L_\ast$} & \colhead{Age} & \colhead{Multip.} & \colhead{Close}  &  \colhead{Inter-}  & \colhead{Wide}  &  \colhead{Silica}& \colhead{Cold} & \colhead{Large} \\  
     \colhead{}          & \colhead{(pc)} & \colhead{} & \colhead{(K)} & \colhead{($M_\sun$)} & \colhead{($L_\sun$)}  &\colhead{(Myr)}&\colhead{} &  \colhead{}       &  \colhead{mediate}       &  \colhead{}       &  \colhead{rich}       &  \colhead{Comp.}  & \colhead{ $W_{10}$} 
}
\startdata   
    HD\,166191 & 100.8 & G0V & 6000 & 1.6 & 4.10 & 10 & $\checkmark$ &  & $\checkmark^a$ & $\checkmark$ & $\checkmark$ & $\checkmark$ &   \\
EF\,Cha & 104.5 & A9IV & 7500 & 1.5 & 2.77 & 10 & $\checkmark$ &  & $\checkmark$ &  &  &  &   \\
HD\,113766 & 122.5 & F5V & 5520 & 1.5 & 7.00 & 15 & $\checkmark$ &  & $\checkmark$ &  &  & $\checkmark$ &   \\
HD\,145263 & 141.9 & F2V & 6500 & 1.6 & 4.12 & 15 & \nodata &  &  &  & $\checkmark$ &  &   \\
 HD\,172555  & 28.3 & A7V & 8000 & 1.8 & 8.70 & 20 & $\checkmark$ &  &  & $\checkmark$ & $\checkmark$ &  &   \\
RZ\,Psc & 184.7 & G8V & 5500 & 1.1 & 0.80 & 30 & $\checkmark$ &  & $\checkmark$ &  &  & $\checkmark$ &   \\
      ID\,8  & 358.7 & G6V & 5500 & 0.9 & 0.72 & 35 & \nodata &  &  &  &  &  & $\checkmark$  \\
  V488\,Per  & 173.5 & K0V & 4760 & 0.8 & 0.31 & 80 & $\checkmark$? &  &  & $\checkmark$ & $\checkmark$ & $\checkmark$ &   \\
P\,1121 & 459.2 & F9V & 5750 & 1.1 & 1.50 & 80 & \nodata &  &  &  &  &  & $\checkmark$  \\
    J0925  & 98.9 & K4.5V & 4510 & 0.6 & 0.22 & 85 & $\checkmark$? &  &  &  &  & $\checkmark$ &   \\
    J0605  & 262.1 & G5V & 5660 & 1.0 & 0.94 & 120 & $\times$ &  &  &  &  &  & $\checkmark$  \\
    J0611  & 184.1 & G9V & 5350 & 0.9 & 0.65 & 130 & $\checkmark$ &  &  & $\checkmark$ & $\checkmark$ &  &   \\
    J1213  & 242.1 & F6V & 6350 & 1.2 & 2.55 & $>$150 & $\checkmark$ &  &  & $\checkmark$ & $\checkmark$ &  &   \\
  HD\,23514  & 139.1 & F5V & 6500 & 1.3 & 3.00 & 150 & $\checkmark$ &  & $\checkmark$ &  & $\checkmark$ &  &   \\
  HD\,15407  & 49.3 & F5V & 6750 & 1.3 & 3.60 & 150 & $\checkmark$ &  &  & $\checkmark$ & $\checkmark$ &  &   \\
    J0609  & 348.3 & G9V & 5410 & 0.9 & 0.49 & 208 & $\checkmark$ &  &  & $\checkmark$ &  &  &   \\
    J0712  & 174.3 & K3.5V & 4680 & 0.7 & 0.28 & 243 & $\checkmark$? &  &  &  &  &  & $\checkmark$  \\
    J2043  & 117.5 & K4V & 4640 & 0.7 & 0.21 & 252 & $\checkmark$ &  &  & $\checkmark$ &  &  &   \\
    J2301  & 158.1 & G9V & 5300 & 0.9 & 0.65 & 600 & $\checkmark$ &  &  & $\checkmark$ &  &  & $\checkmark$  \\
BD+20\,307 & 117.3 & G0V & 5540 & 1.0 & 3.00 & 1000 & $\checkmark$ & $\checkmark$ &  & $\checkmark$ &  &  & $\checkmark$  \\
    J1044  & 199.1 & K0.5V & 5240 & 0.8 & 0.41 & 5500 & $\checkmark$ &  &  & $\checkmark$ &  &  & $\checkmark$  \\
\enddata
 \tablecomments{Symbols $\checkmark$ for yes and $\times$ for no in stellar multiplicity with companions in three different separations: close ($<$a few au), intermediate (a few tens of au to 500 au, and wide ($>$500 au). ``?" symbol means the presence of companions is inferred from Gaia astrometry. ``..." means systems are too far to have useful constraints on stellar multiplicity. Large $W_{10}$ systems are the one with values larger than 12.5 \micron. $^a$ There are two close ($\sim$1\arcsec, faint sources around HD\,166191 reported by \citet{schneider13} and one has been found co-moving with the star (private communication).}
\end{deluxetable*}

\begin{deluxetable}{rccrrcccl} 
    \tablecaption{Companion Properties }
    \label{tab:comp_properties}
    \tablehead{
    \colhead{Companion} & \colhead{SpT} & \colhead{M$_\ast$} & \multicolumn{2}{c}{Separations}  & \colhead{$e$} & \colhead{$e^{dn}$} & \colhead{$e^{up}$} & \colhead{Ref.} \\
    \colhead{for Target}   & \colhead{}    & \colhead{(M$_\sun$)} & \colhead{(\arcsec)} & \colhead{(au)} & \colhead{} & \colhead{} & \colhead{} & \colhead{} 
    }
\startdata
HD\,113766  &  F6V   &  \nodata       &  1.39   &   151.4  &  0.50   &  0.37    & 0.85    &  [1,5] \\
HD\,172555  &  K5Ve  &  \nodata       &  71.34  &   2054  &  \nodata   &  \nodata    &  \nodata   &  [13]\\ 
RZ\,Psc     & mid-M  & 0.12        &   0.13   &    23.0    & \nodata   &  \nodata  &  \nodata  &  [2]  \\
V488\,Per   & \nodata  &  \nodata    &  71.00       &   12277    & \nodata  & \nodata    & \nodata   &  [3] \\
J0611      &  M5.5  & 0.096$\pm$0.01 &  24.64  &  4535   &  \nodata  & \nodata   & \nodata  &  [4] \\
J1213      &  M4.5  &  0.19       &   4.17  &  1010  &  0.99   &  0.72    & 1.00    &  [4,5] \\
HD\,23514   &   M7   &  0.06$\pm$0.01 &   2.65   &  368       & \nodata  &  \nodata  & \nodata  &  [6] \\ 
HD\,15407   &  K2.5  & \nodata   & 21.27   &  1048   &  0.63   &  0.33    & 0.86    &  [5,7] \\
J0609      &  M4.5  &  \nodata    &  1.96   &  682   &  \nodata &  \nodata  & \nodata  &  [8] \\
J2043      &  M3.5  &  \nodata    &  18.97  &  2228   &  0.64   &  0.43    & 0.87    &  [5,8]  \\
J2301      &  M4Ve  &  \nodata    &  49.55  &  7836   &  \nodata &  \nodata  & \nodata  &  [9]  \\
BD+20\,307  &   WD   & 0.48--0.58   &   8.40   &  982   &  0.99   &  0.74    & 1.0     &  [5,10,11]   \\
J1044      &   G8V  & \nodata    &  51.56  &   10266  &  0.99   &  0.76    & 1.0     &  [5,8] \\
EF\,Cha    &   \nodata   &  \nodata     &  1.44    &   147.6   &  0.96   &  0.63    & 1.0     &  [5,12] \\
\enddata
\tablecomments{Companion properties -- spectral type (SpT), mass ($M_\ast$), projected separation, and eccentricity ($e$) with lower and upper bounds ($e^{dn}$ and $e^{up}$) were collected from references: [1] \citealt{chen11}, [2] \citealt{kennedy20_mnras_496_75_RZpsc_companion}, [3] \citealt{zuckerman15}, [4] \citealt{moor21}, [5] \citealt{hwang2022}, [6] \citealt{rodriguez12_hd23514}, [7] \citealt{melis10}, [8] \citealt{moor24_edds_visir}, [9] \citealt{melis21}, [10] \citep{hartman2020}, [11] \citealt{fusillo2019}, [12] \citealt{briceno2017}, [13] \citealt{alonso-floriano15}. }
\end{deluxetable}

\section{Dust Emissivity and 10 \micron\ Dust Indices } 
\label{sec:dustindices}

\subsection{Pseudocontinuum Estimate and Dust Emissivity}
\label{sec:pseudocontinuum}

Under optically thin conditions, dust emission equals the dust opacity (which depends on composition and grain size) multiplied by the Plank function integrated over the emitting temperatures. Opacities are measured in the laboratory for a limited range of grain sizes and compositions or computed with theoretical models that rely on imperfect assumptions. Detailed spectral‑decomposition studies using Spitzer data have sometimes reached discrepant conclusions owing to differences in adopted opacities or in dust temperature distribution arising from different disk structures (e.g., \citealt{juhasz09}). With the new JWST spectra of EDDs, we revisit the reported mineralogical dichotomy  (silica‑rich vs. forsterite‑rich) reported in earlier Spitzer studies (e.g., \citealt{meng15}). Rather than attempting definitive mineral identifications, we adopt a ``dust emissivity" approach to robustly identify systems showing silica-rich signatures in the 10-\micron\ region.

The dust emissivity ($W_{\lambda}$) is defined as the ratio between the feature producing spectrum and underlying continuum, i.e.,  $W_{\lambda} = (F_\lambda -  F_{\lambda,c}$) / $F_{\lambda,c}$ where $F_{\lambda}$ is the observed spectrum and $F_{\lambda,c}$ is the continuum. Solid-state features arise from vibrational motions (stretching, bending, lattice modes) of atoms, ions, or molecules in a solid's fixed -- ordered or partially ordered -- structure; they are stronger for small particles, therefore are more useful to infer chemical composition. The purpose of defining a continuum is to separate emission from small grains, which produce prominent solid-state features, from emission by large grains so those features may be used to constrain composition. If all grains share the same temperature distribution, dividing the observed spectrum by the Planck function integrated over that temperature profile (single, two‑component, or more complex) largely isolates the small‑grain emission. In practice this separation is imperfect: small grains contribute a low-level smooth continuum in addition to discrete features, and some compositions (e.g., amorphous carbon) produce no distinct -- or only very broad -- features. Furthermore, it is unlikely all grains share the same temperature profile. For this reason the continuum is termed a ``pseudocontinuum" \citep{harker23_spitzer_comets}: it is not a true continuum but is useful for largely removing the temperature dependence.

Although silicate‑like materials produce features in both the 10 and 20 \micron\ regimes, these wavelengths often probe different disk regions in PPDs: the 10 \micron\ feature is sensitive to dust in the inner au, while the 20 \micron\ feature traces material at tens to hundreds of au from the star (e.g., \citealt{jang24_spatialdist_crystaline}, and references therein). This is consistent with early Spitzer studies showing that a two‑temperature combination satisfactorily reproduces the mid‑infrared silicate features arising from the optically thin regions of PPDs \citep{bouwman08,sargen09_silia,sargent09,olofsson10_c2d_spectraldecomp}. A power‑law temperature distribution (e.g., \citealt{juhasz10_herbig}), which is effectively similar to the polynomial approximation used by \citet{watson09} in PPDs, is also a valid alternative. We adopt both functional forms to determine the dust emissivity and assess the uncertainty introduced by the choice of functional form. 

The anchor points used to define the pseudocontinuum also affect the derived dust emissivity. Ideally they should lie in regions free of solid‑state features; in practice this is difficult -- especially near 20 \micron, where the amorphous‑silicate feature is broad and essentially impossible to avoid within the limited wavelength range covered by JWST/MIRI.  Consequently, we use second- or third-order polynomials rather than the fifth‑order function used for fitting the Spitzer data by \citet{watson09}. The pseudocontinuum is anchored to the disk spectrum at 5.6--7.0, 13.02--13.50, 14.32--14.83, and 24.5--26.0 \micron\ with equal weighting; however, we reduce the weights of the last two anchors for systems with low levels of flat continuum from silica dust or with noisy data beyond 23 \micron. The latter is particularly problematic for J0609, J2301, and V488\,Per, where the cooler temperatures used in the pseudocontinuum -- derived from the limited wavelength coverage of JWST -- are subject to large uncertainty. Among the 21 EDDs, the warm temperatures have a median of $\sim$750 K (range 350--900 K) and the cooler temperatures $\sim$225 K (range 100--300 K); the lowest warm temperature is found in HD\,172555, while the lowest cooler temperature is found in V488\,Per and J0925. For the five EDDs observed with Spitzer IRS, a long wavelength anchor between 35.07 and 35.92 \micron, when available, was also used for pseudocontinuum determination. We stress that the two temperatures used for the pseudocontinuum span plausible extremes, with each having an independent normalization, and do not imply the disk consists of two narrow rings. 
Nonetheless, these temperatures indicate that (1) even if cold ($<$100 K) dust that emits prominently at 20 \micron\ exists (we have no evidence for it given the derived temperatures), its contribution to the 10 \micron\ region is insignificant, and (2) the 10 \micron\ features predominantly originate from warm dust within a few au (i.e., the terrestrial zone).

As shown in Figure \ref{fig:disksed_linearly}, the two pseudocontinua are very similar in the 10‑\micron\ region for most targets but can differ substantially in the 20‑\micron\ region. This is especially true for systems with large warm–cool temperature contrasts ($\sim$500 K), such as V488\,Per and J0925; without a reliable anchor for the cooler dust, the pseudocontinuum at 20 \micron\ is highly uncertain. Solid‑state features of crystalline silicates in the 15--25 \micron\ region can aid mineral identification and help constrain the Fe/Mg ratio. However, feature strengths and profiles depend not only on Fe/Mg but are also highly sensitive to grain porosity and morphology, introducing significant degeneracies in compositional inferences \citep{min2007}. For these reasons, we restrict our analysis to the 10‑\micron\ dust emissivity and defer in‑depth spectral decomposition (including the 15--25 \micron\ region) to future work.

\subsection{Mathematical Forms}
\label{sec:mathforms}

\begin{figure*}
    \centering
    \includegraphics[width=0.23\linewidth]{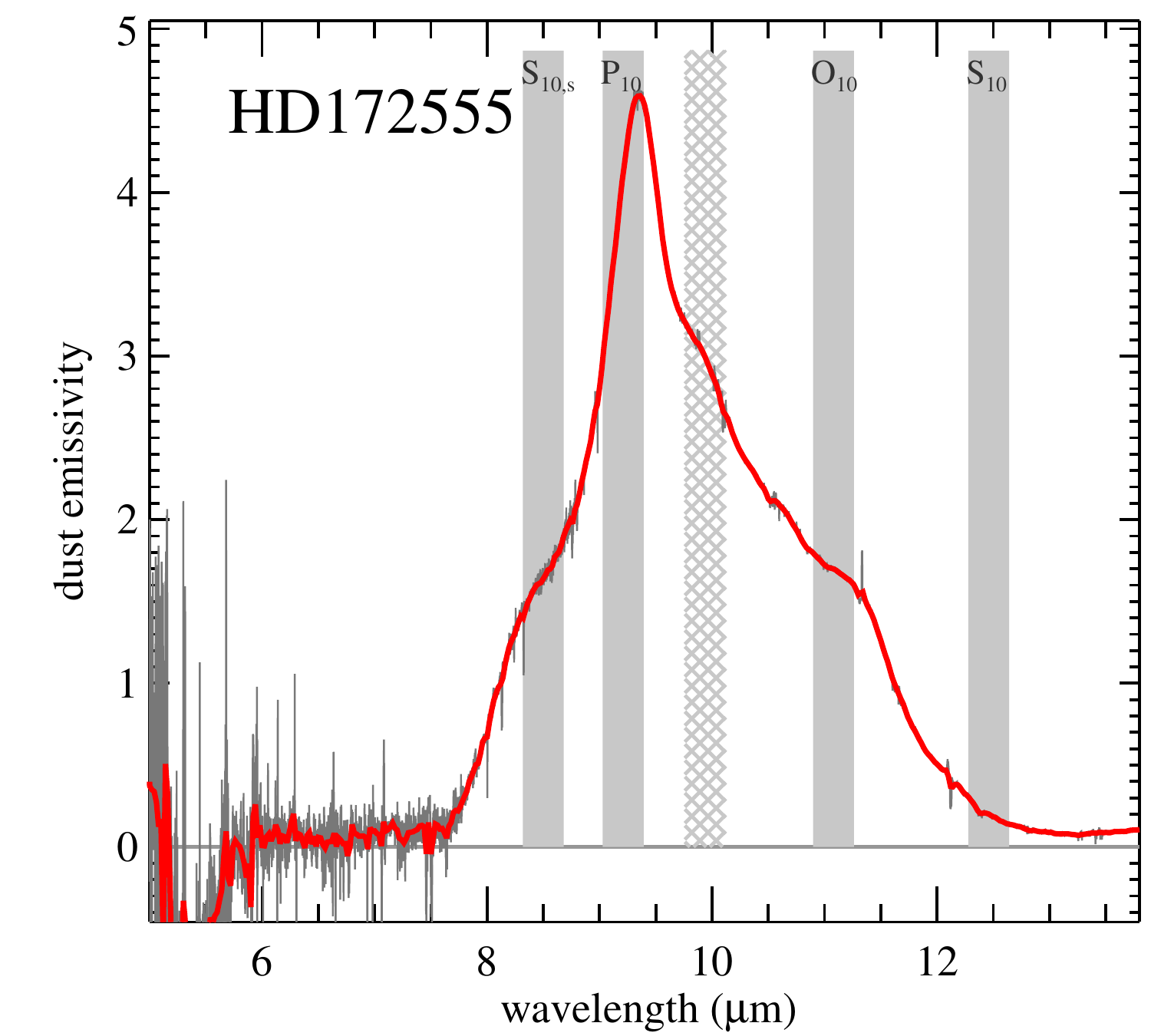}
    \includegraphics[width=0.23\linewidth]{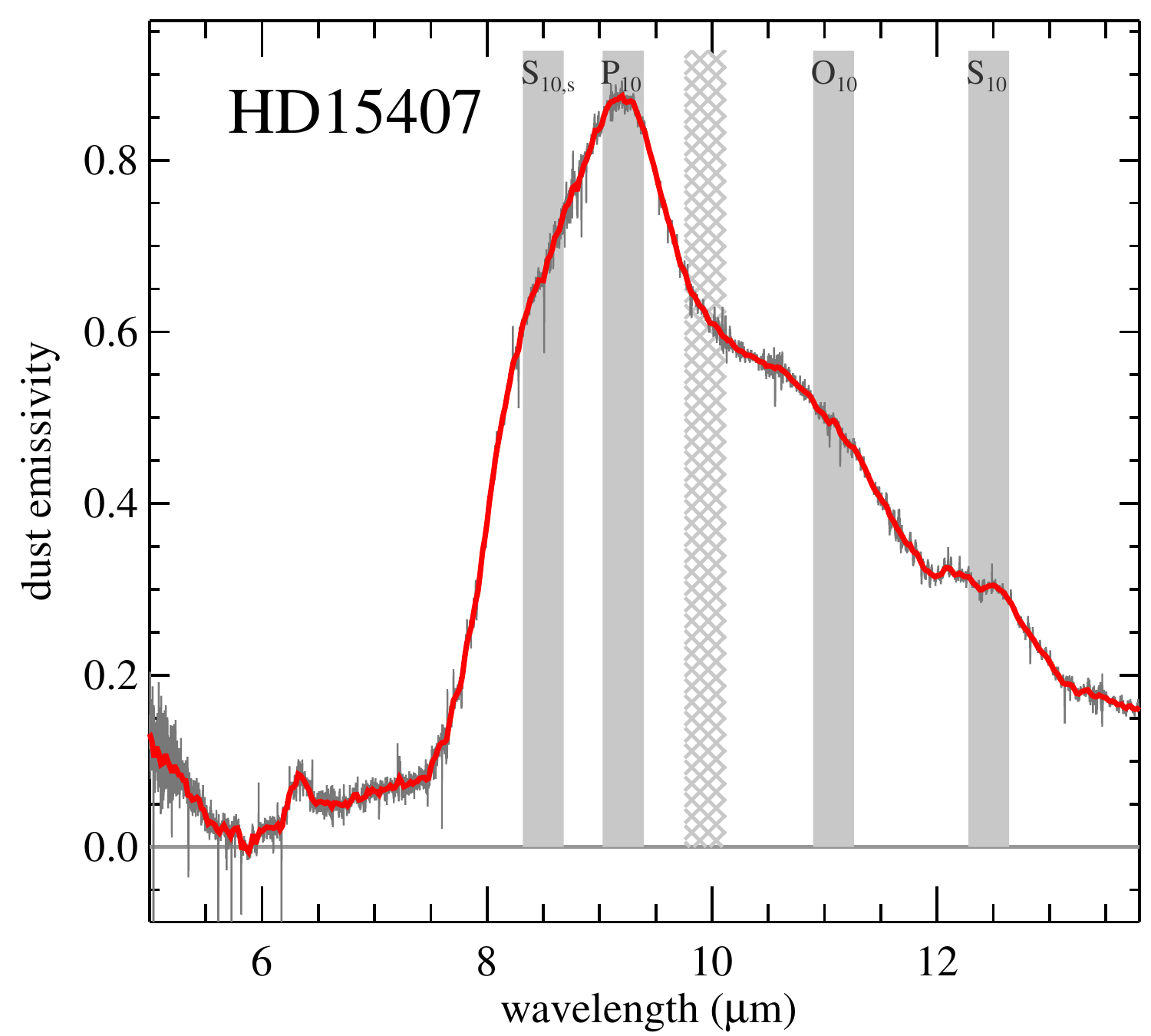}
    \includegraphics[width=0.23\linewidth]{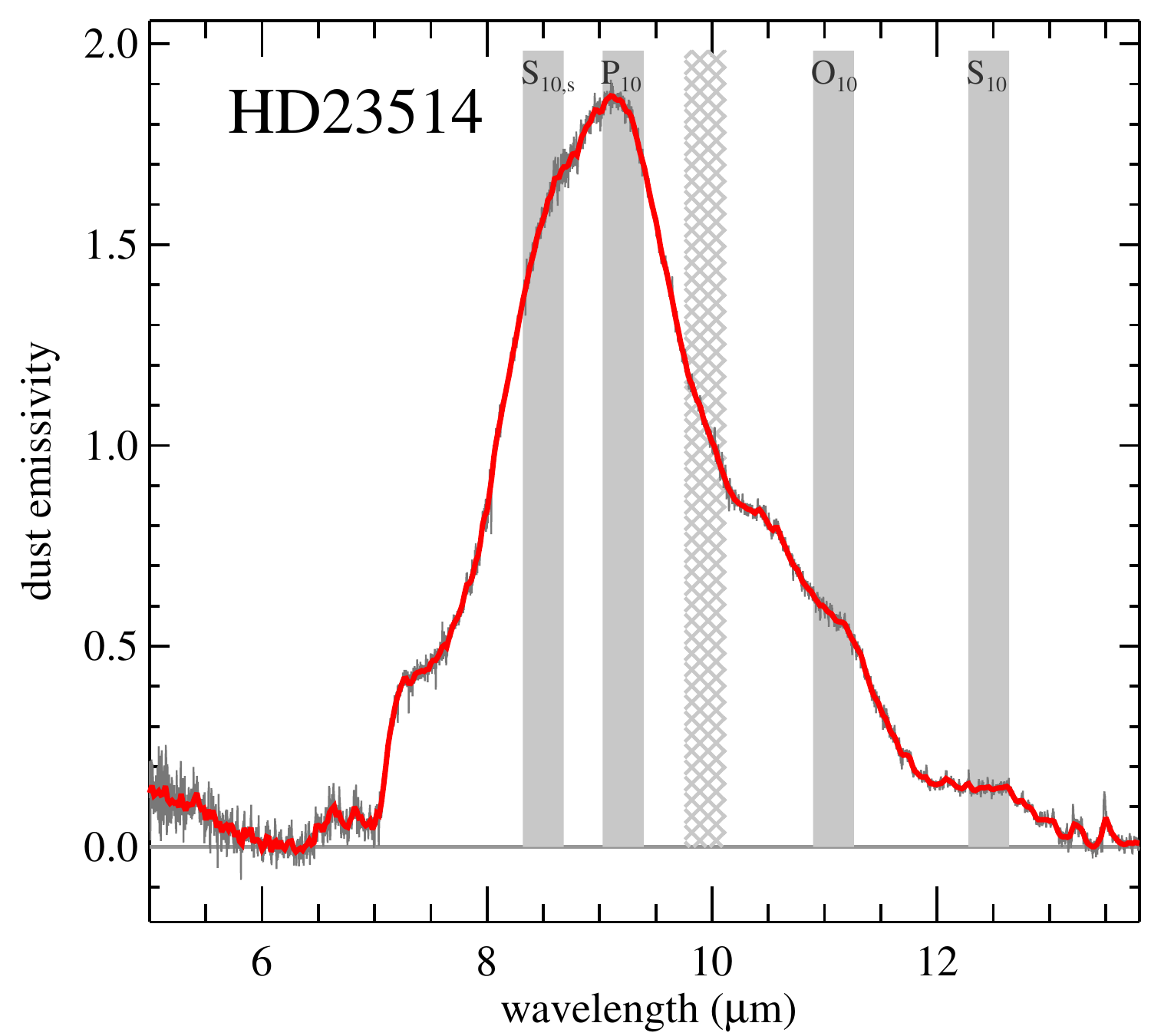}
    \includegraphics[width=0.23\linewidth]{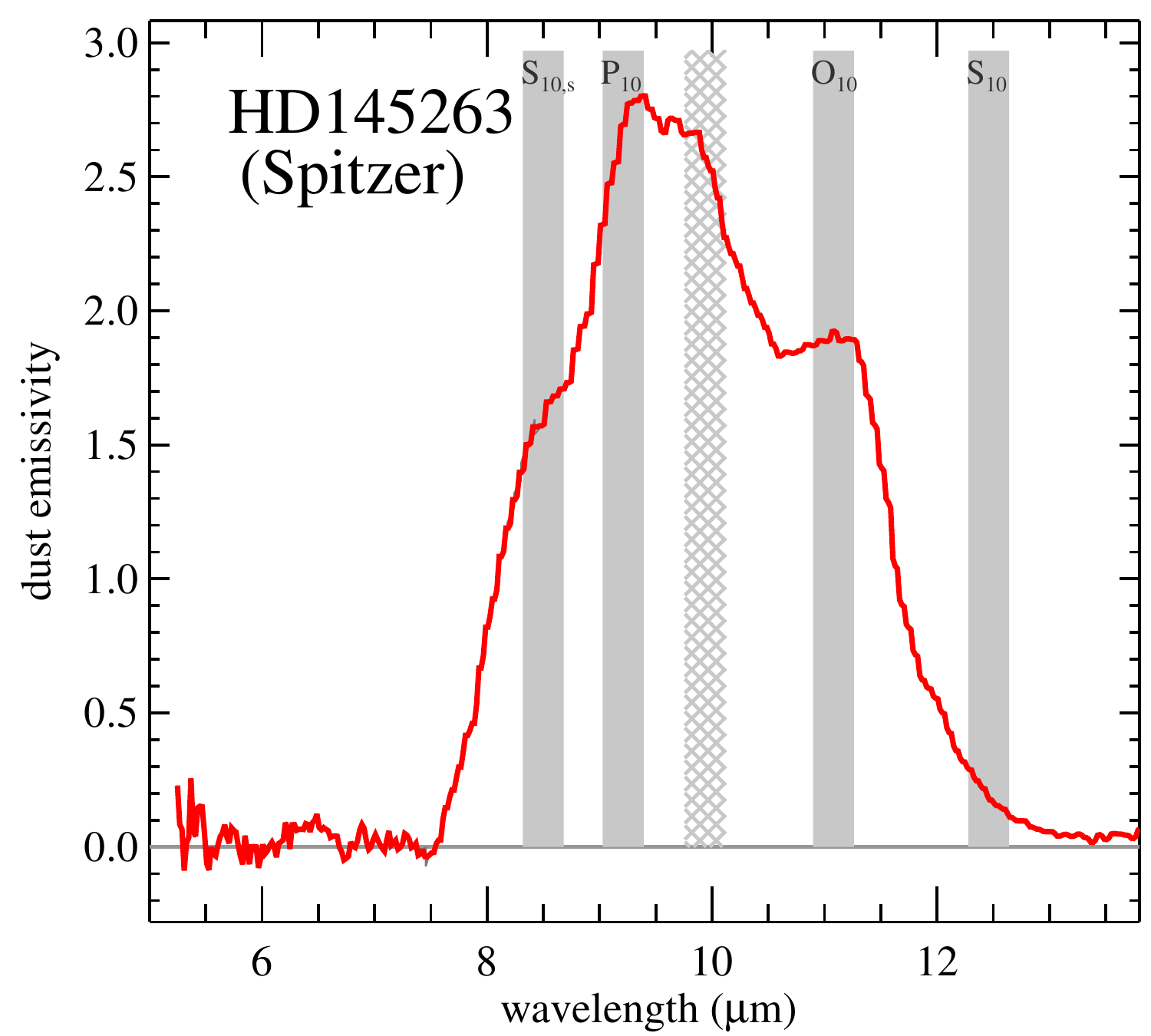}
    \includegraphics[width=0.23\linewidth]{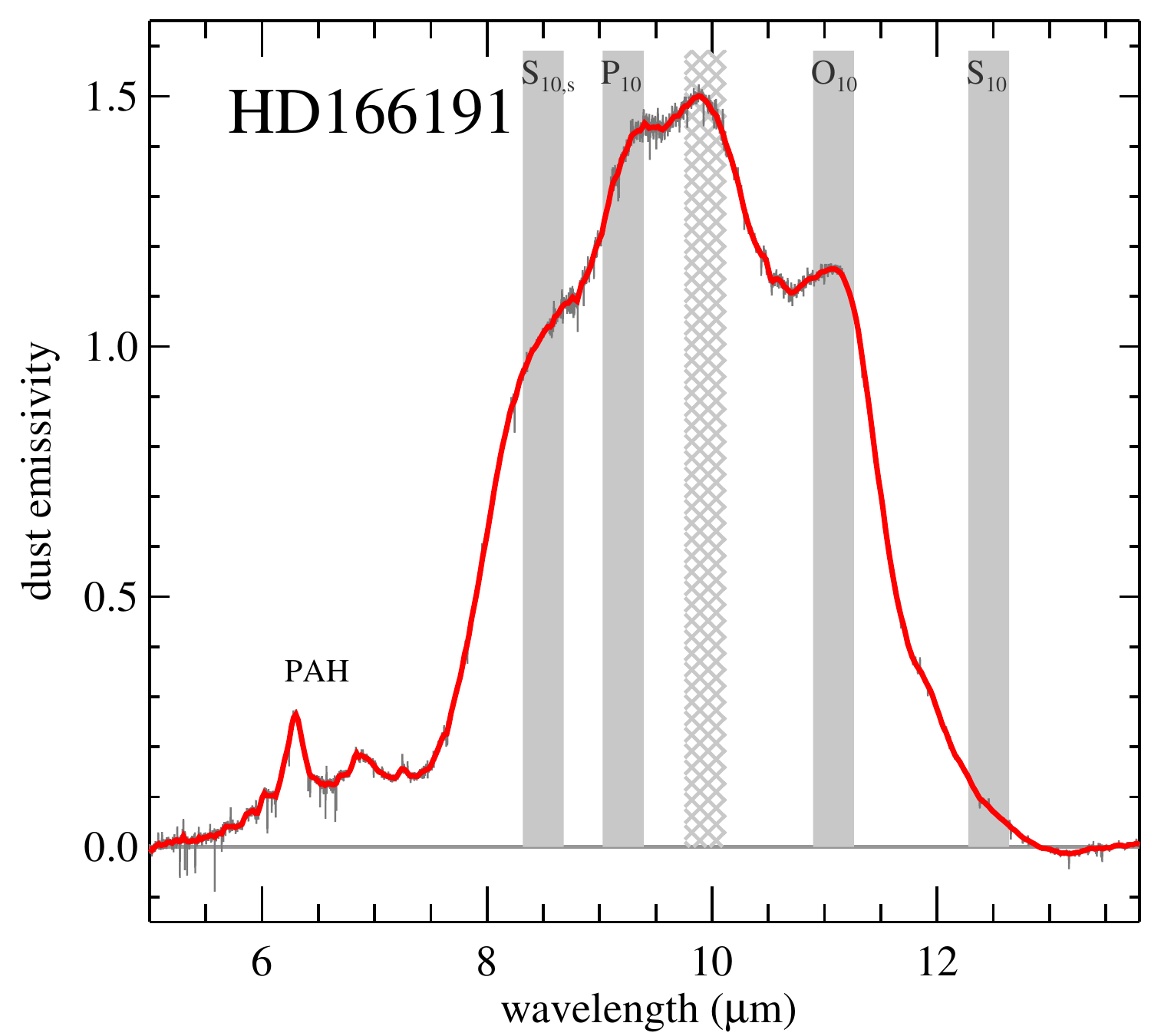}
    \includegraphics[width=0.23\linewidth]{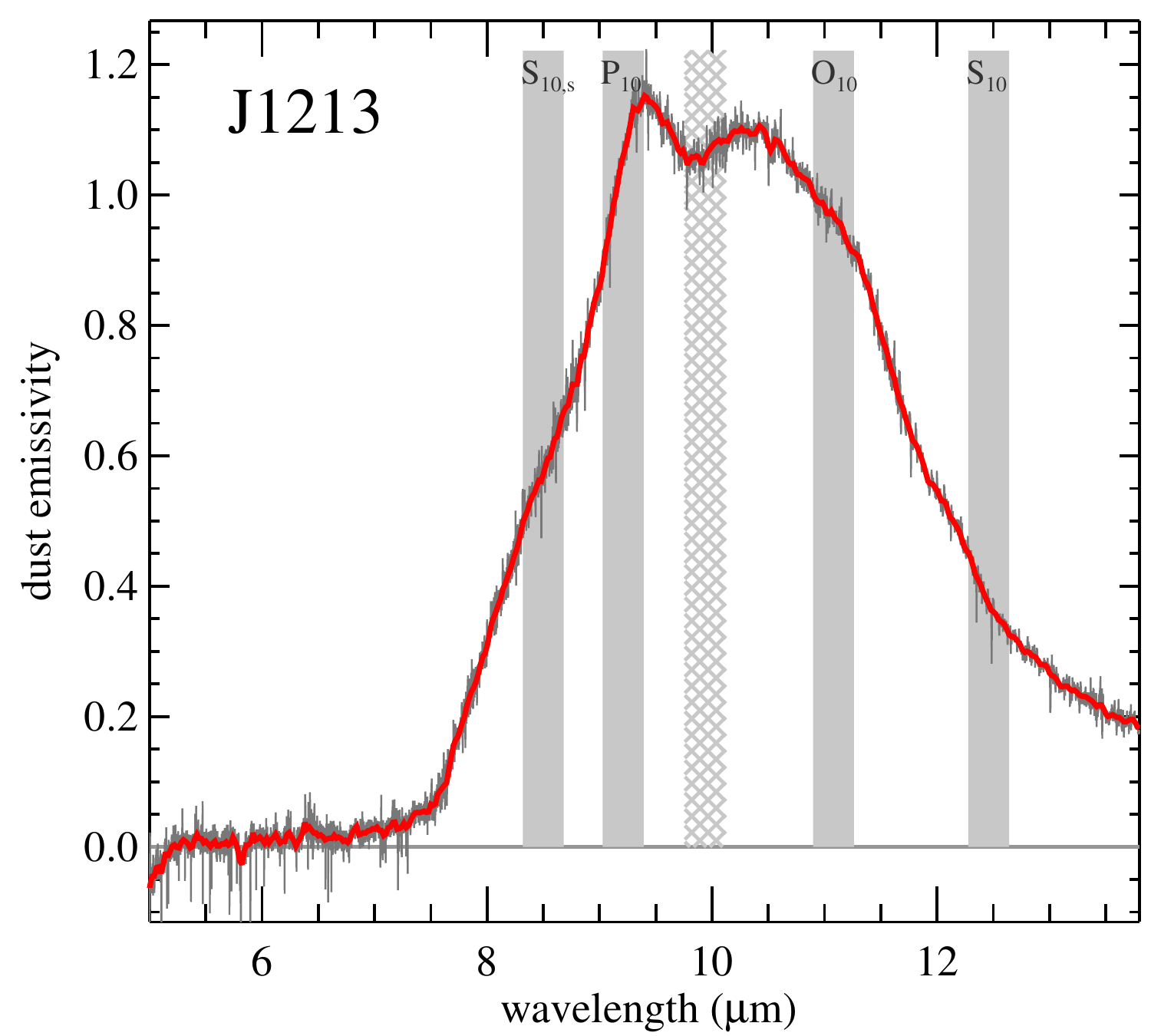}
    \includegraphics[width=0.23\linewidth]{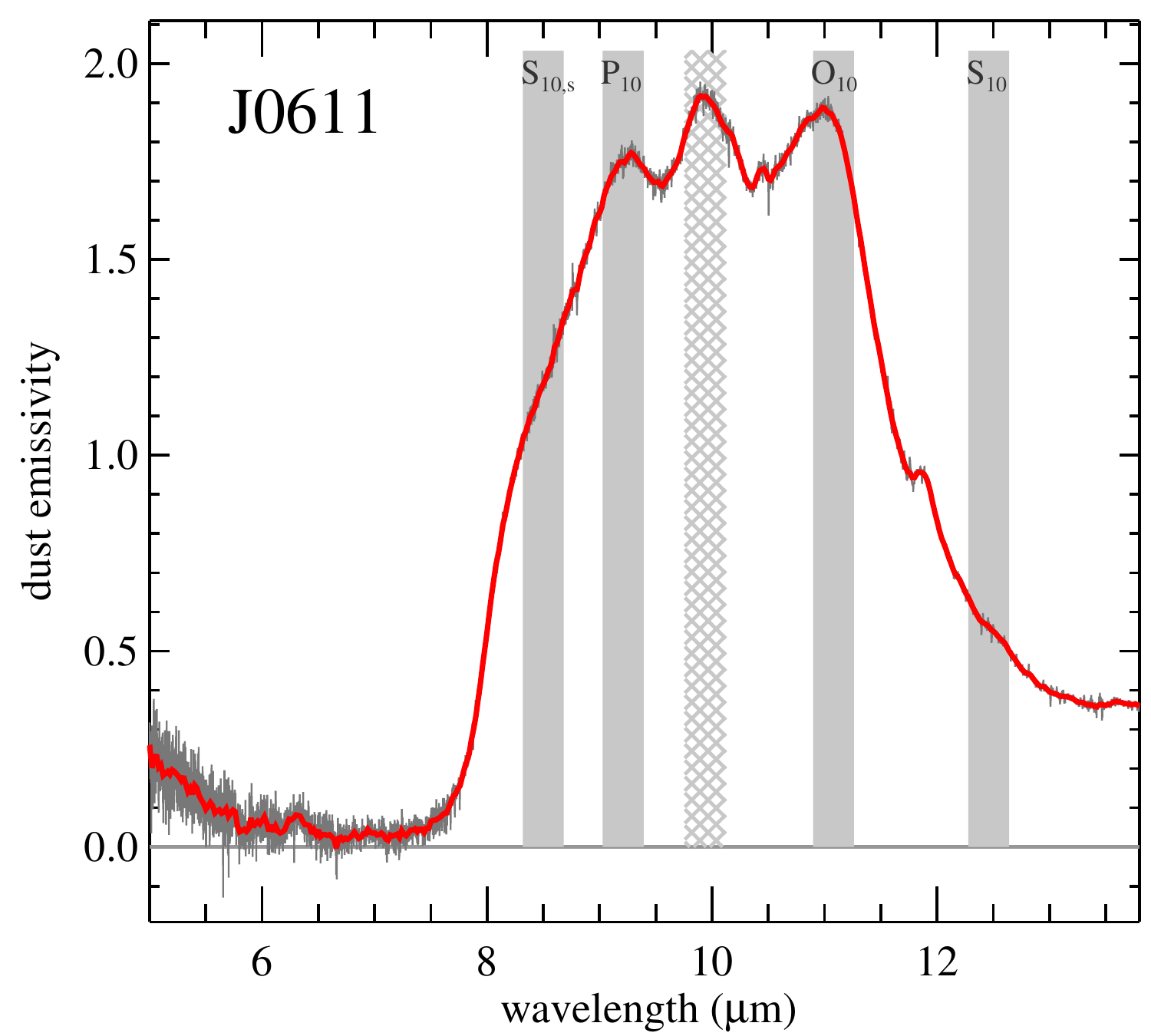}
    \includegraphics[width=0.23\linewidth]{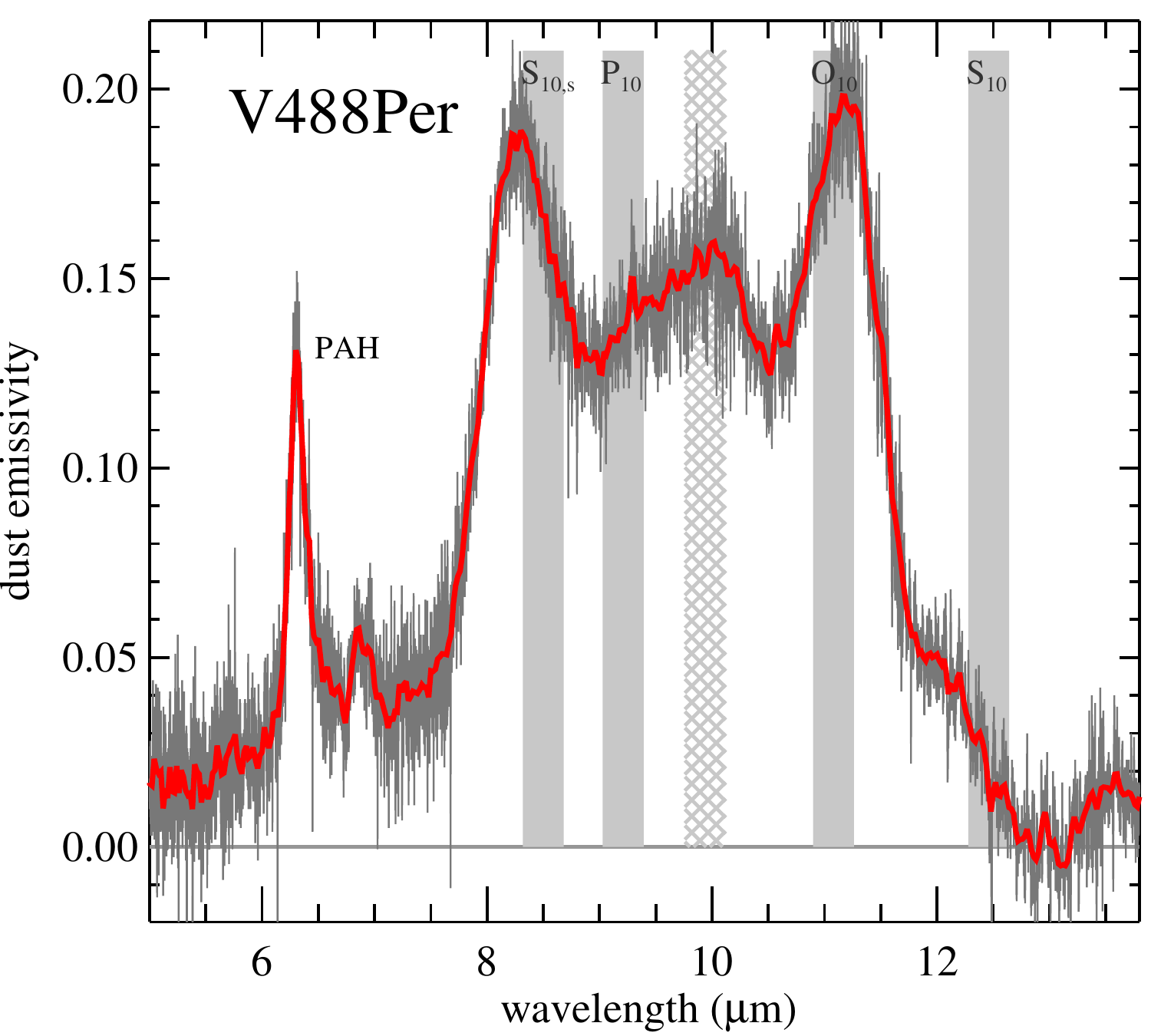}
    \includegraphics[width=0.23\linewidth]{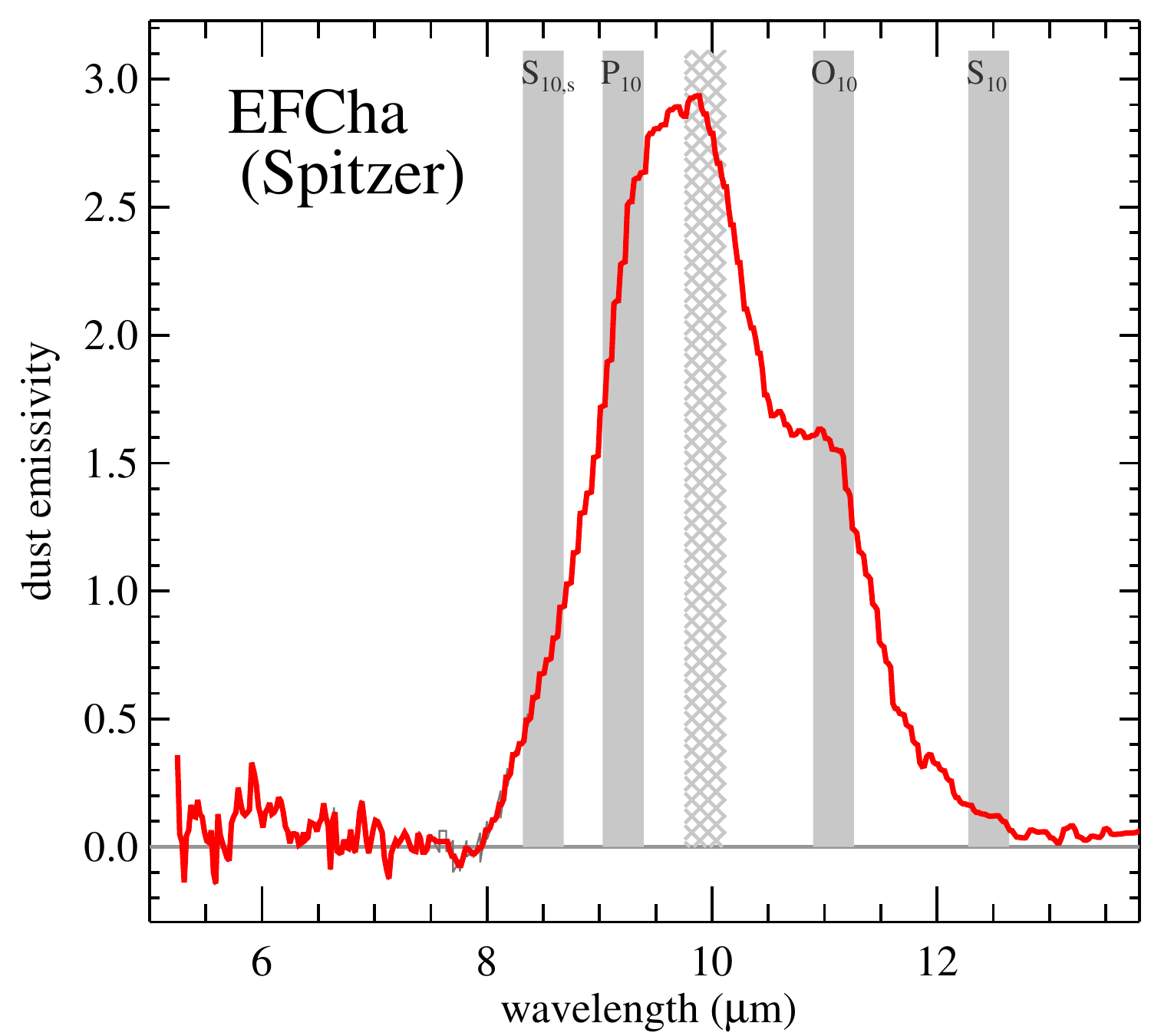}
    \includegraphics[width=0.23\linewidth]{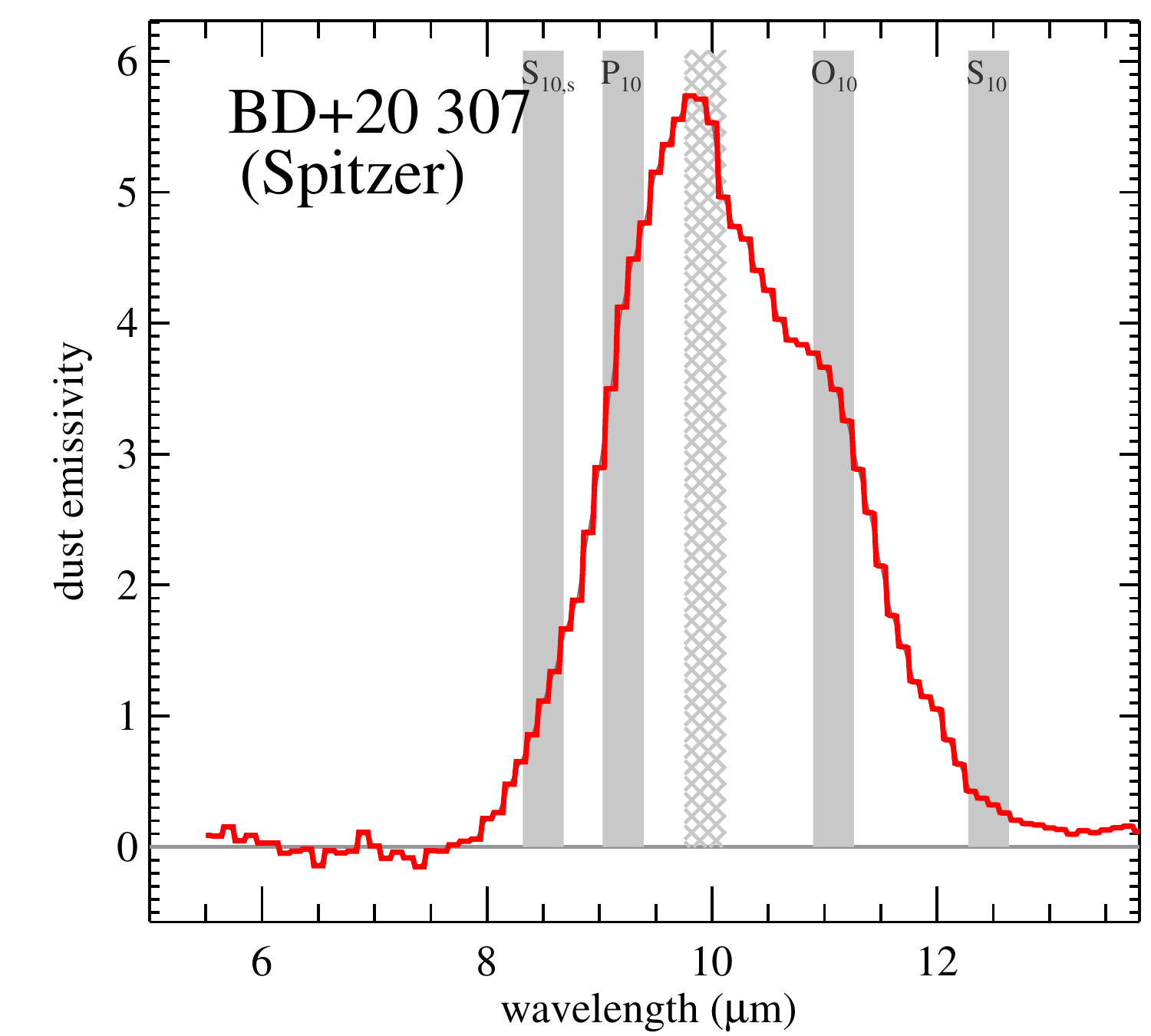}
    \includegraphics[width=0.23\linewidth]{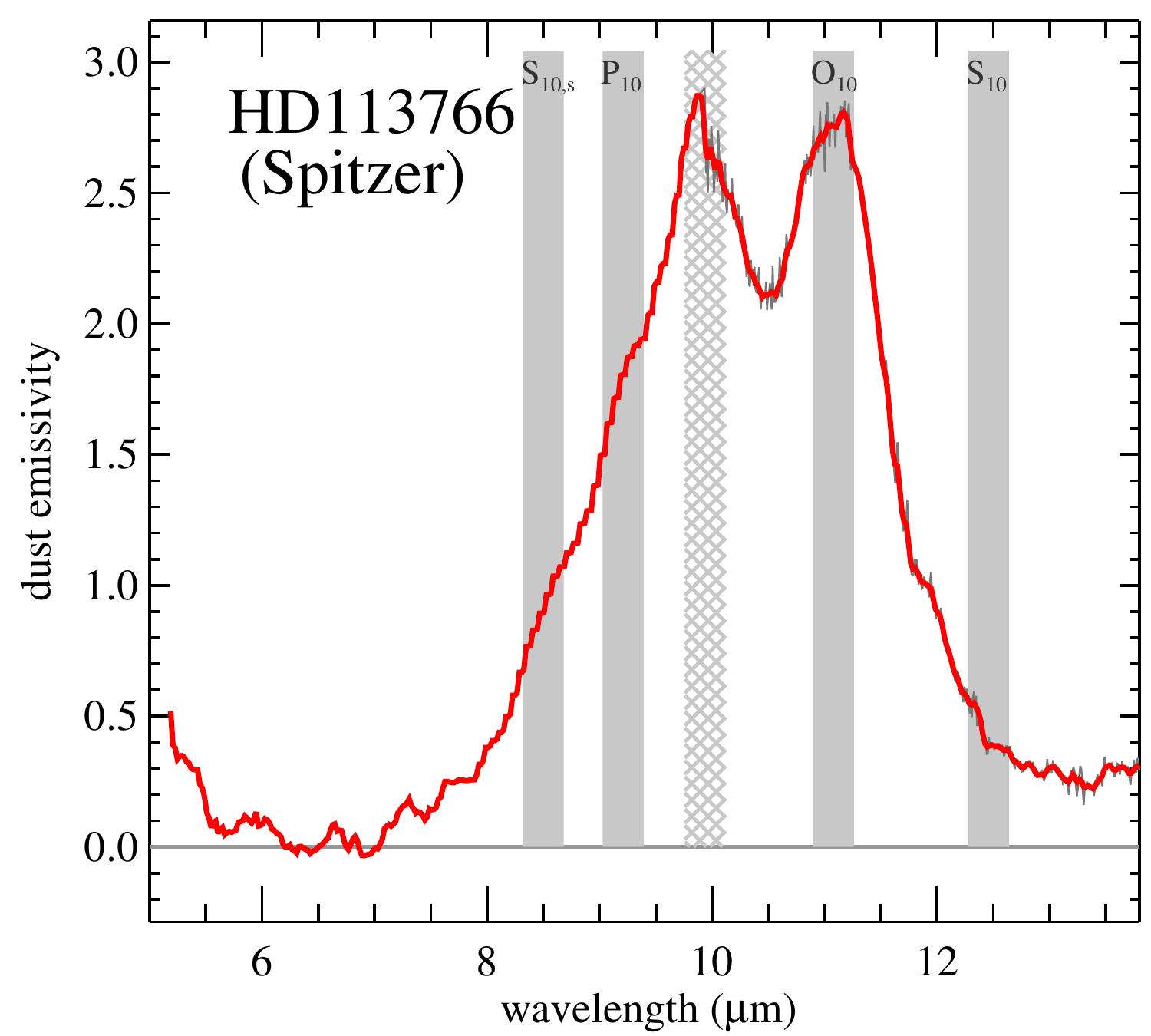}
   \includegraphics[width=0.23\linewidth]{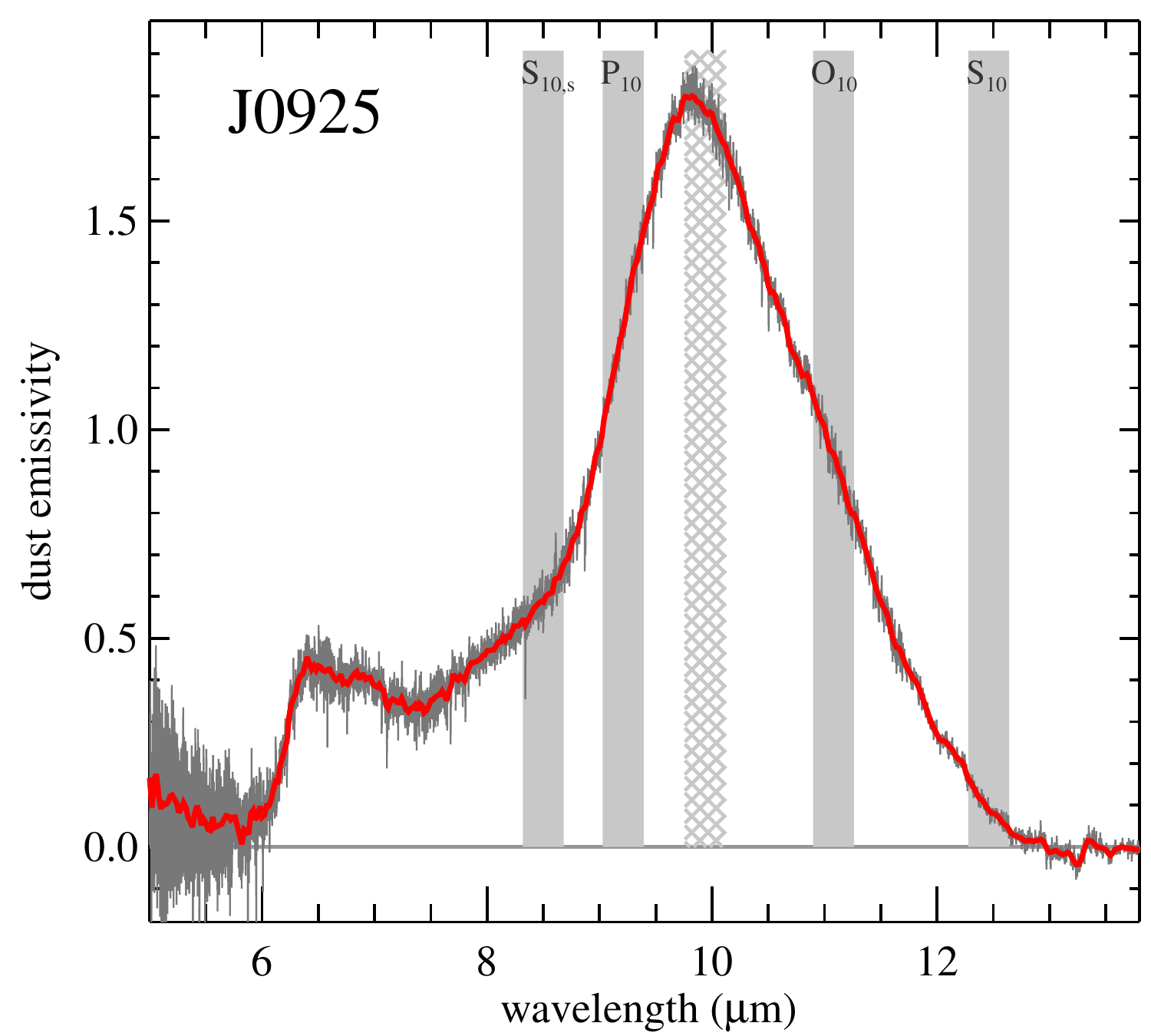}
    \includegraphics[width=0.23\linewidth]{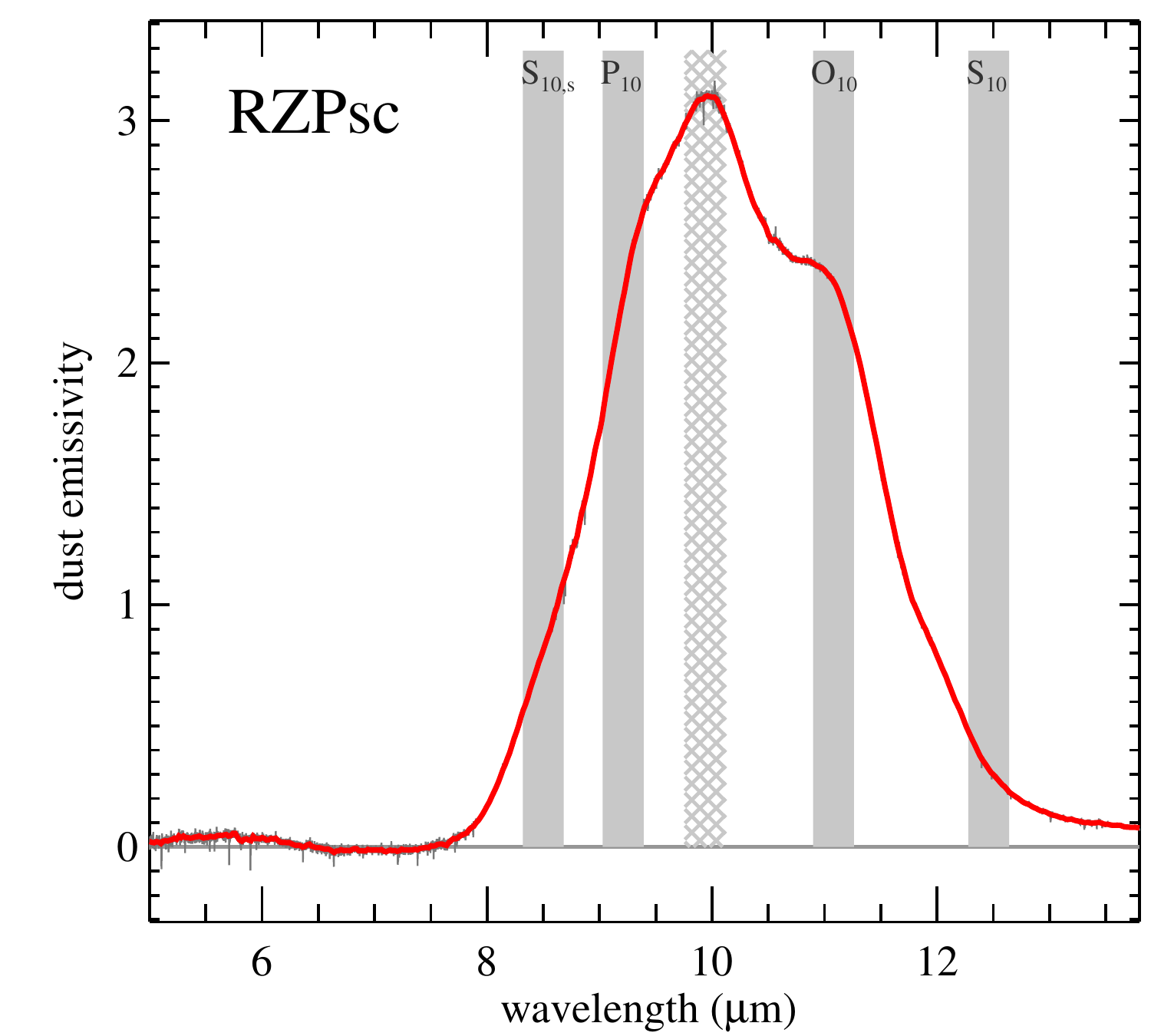}
    \includegraphics[width=0.23\linewidth]{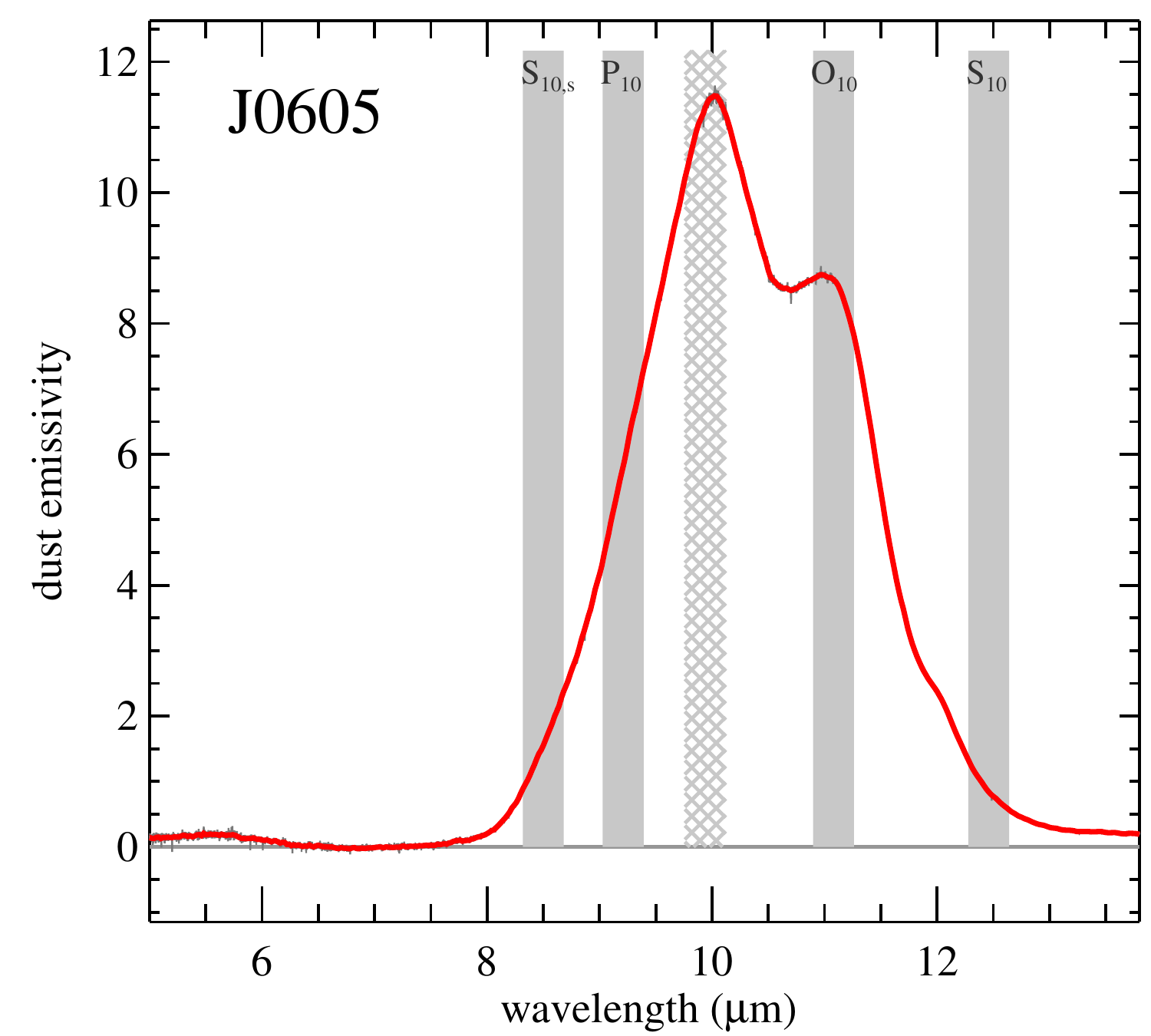}
    \includegraphics[width=0.23\linewidth]{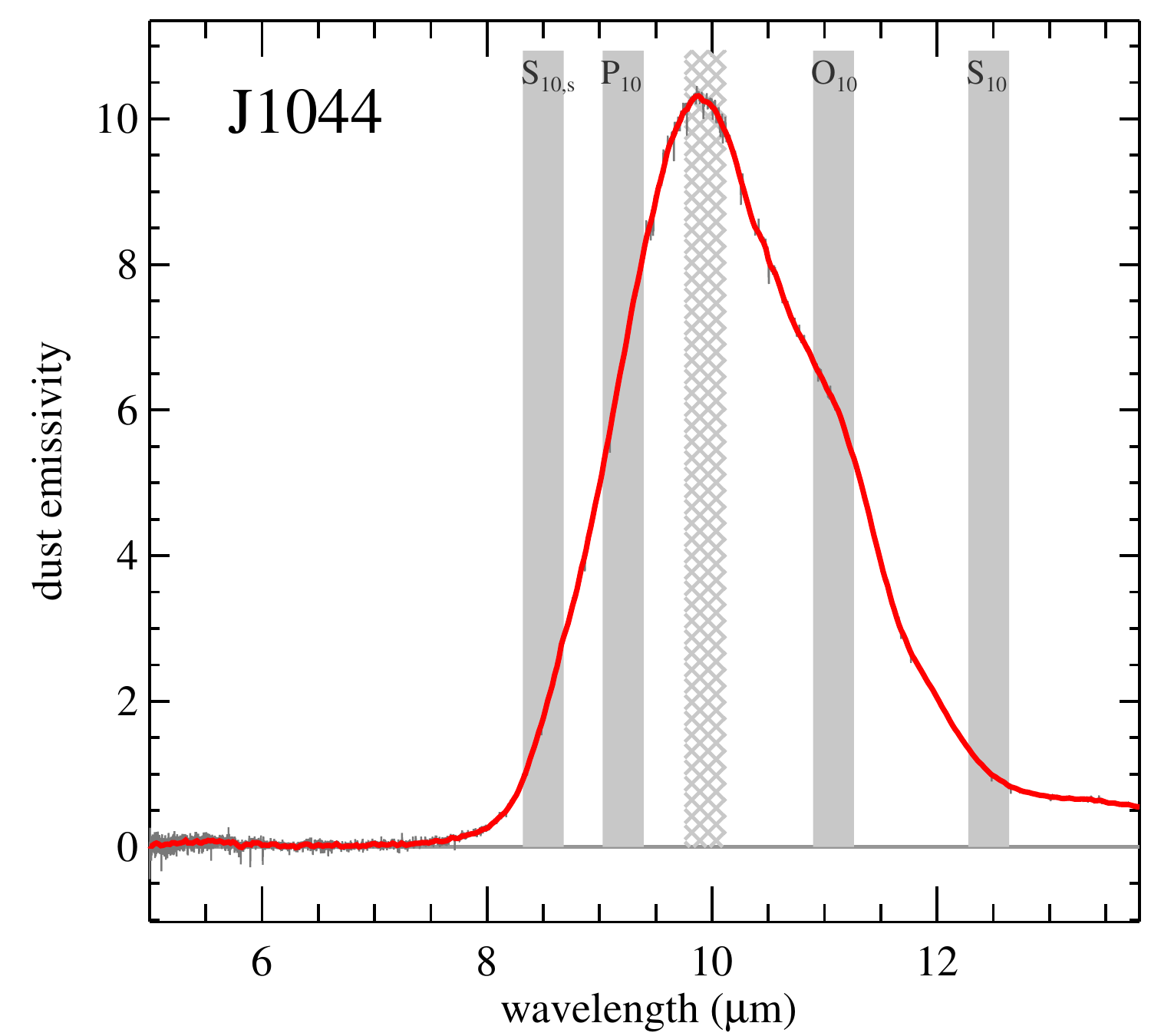}
    \includegraphics[width=0.23\linewidth]{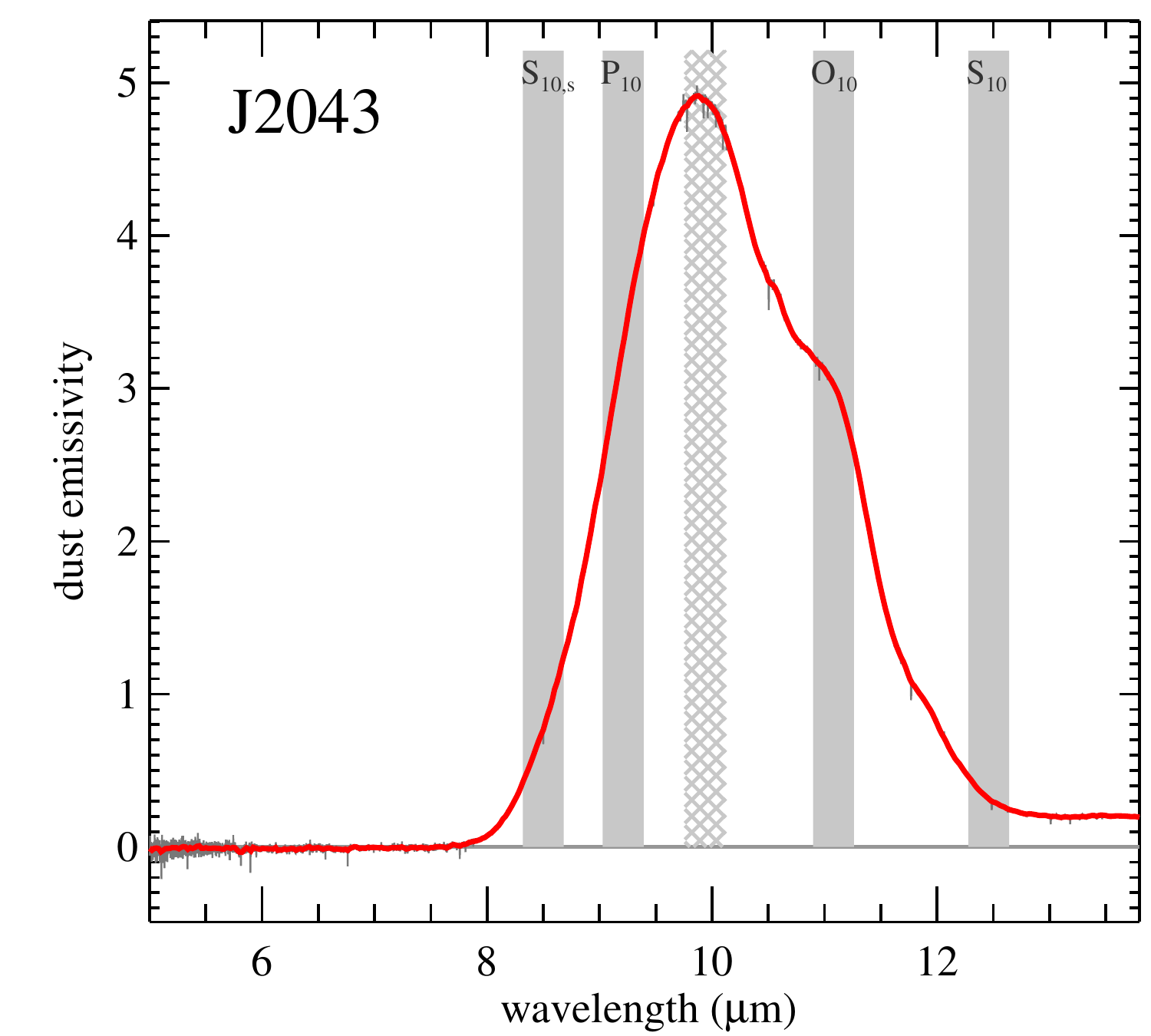}
    \includegraphics[width=0.23\linewidth]{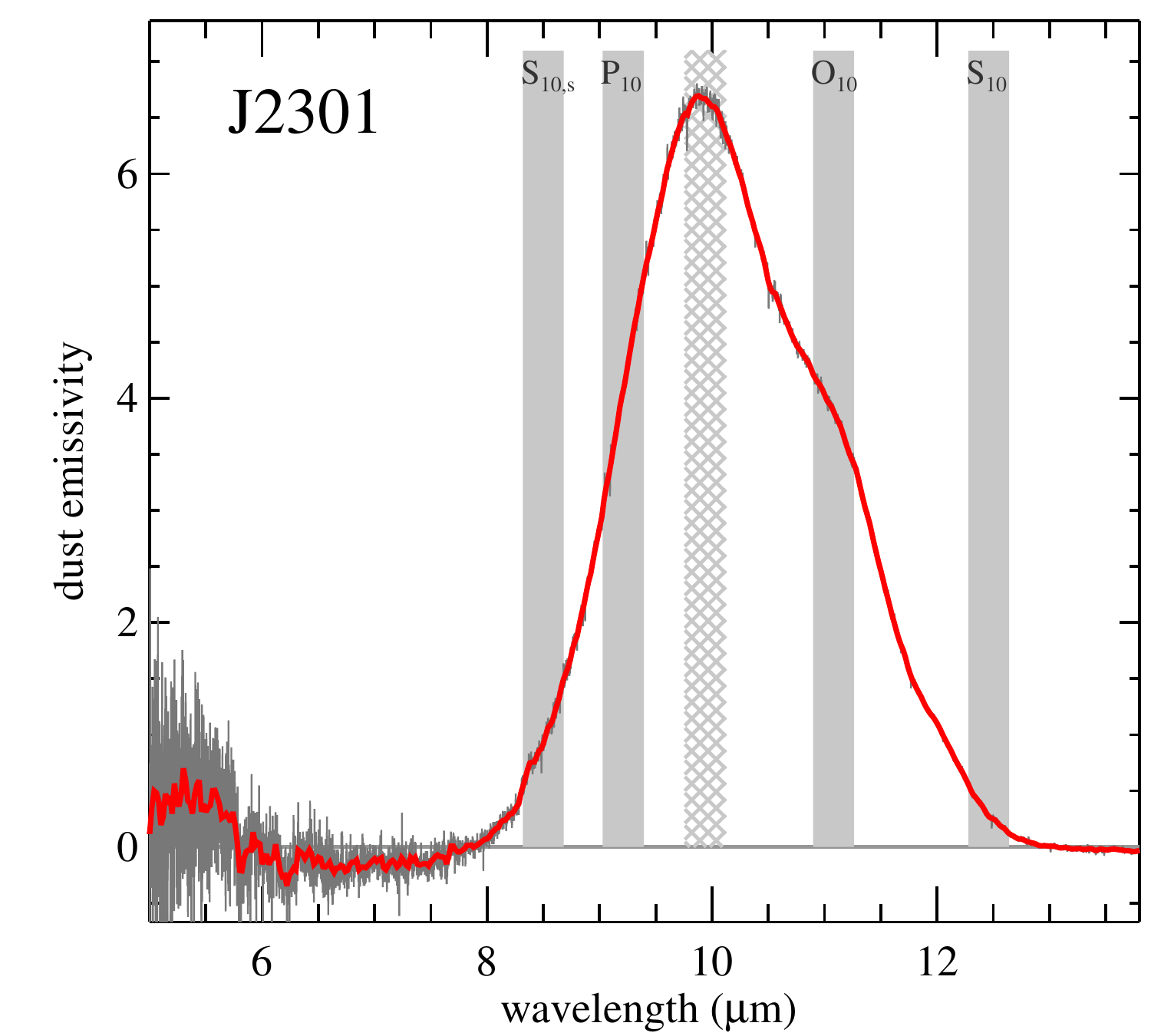}
    \includegraphics[width=0.23\linewidth]{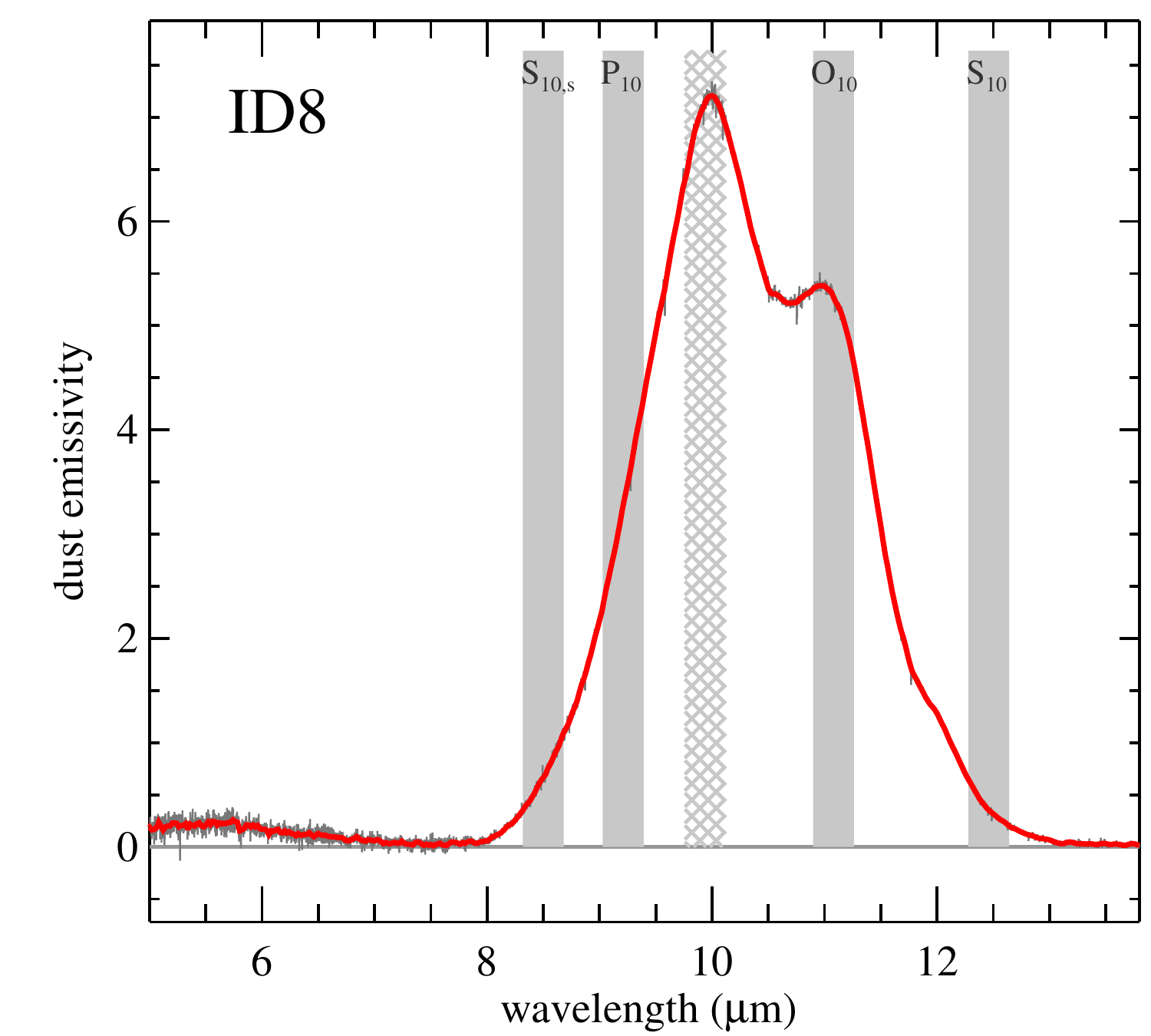}
    \includegraphics[width=0.23\linewidth]{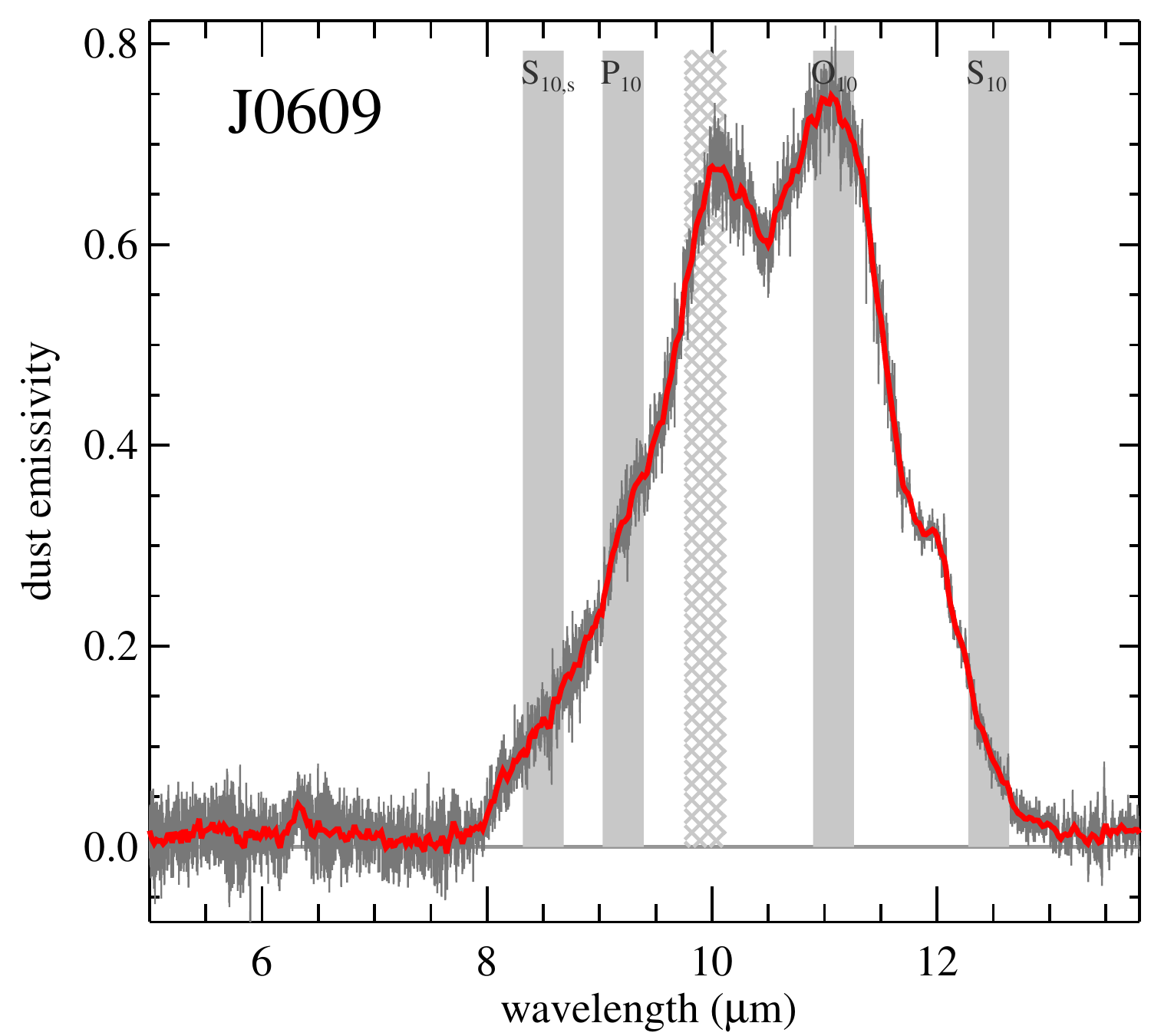}
    \includegraphics[width=0.23\linewidth]{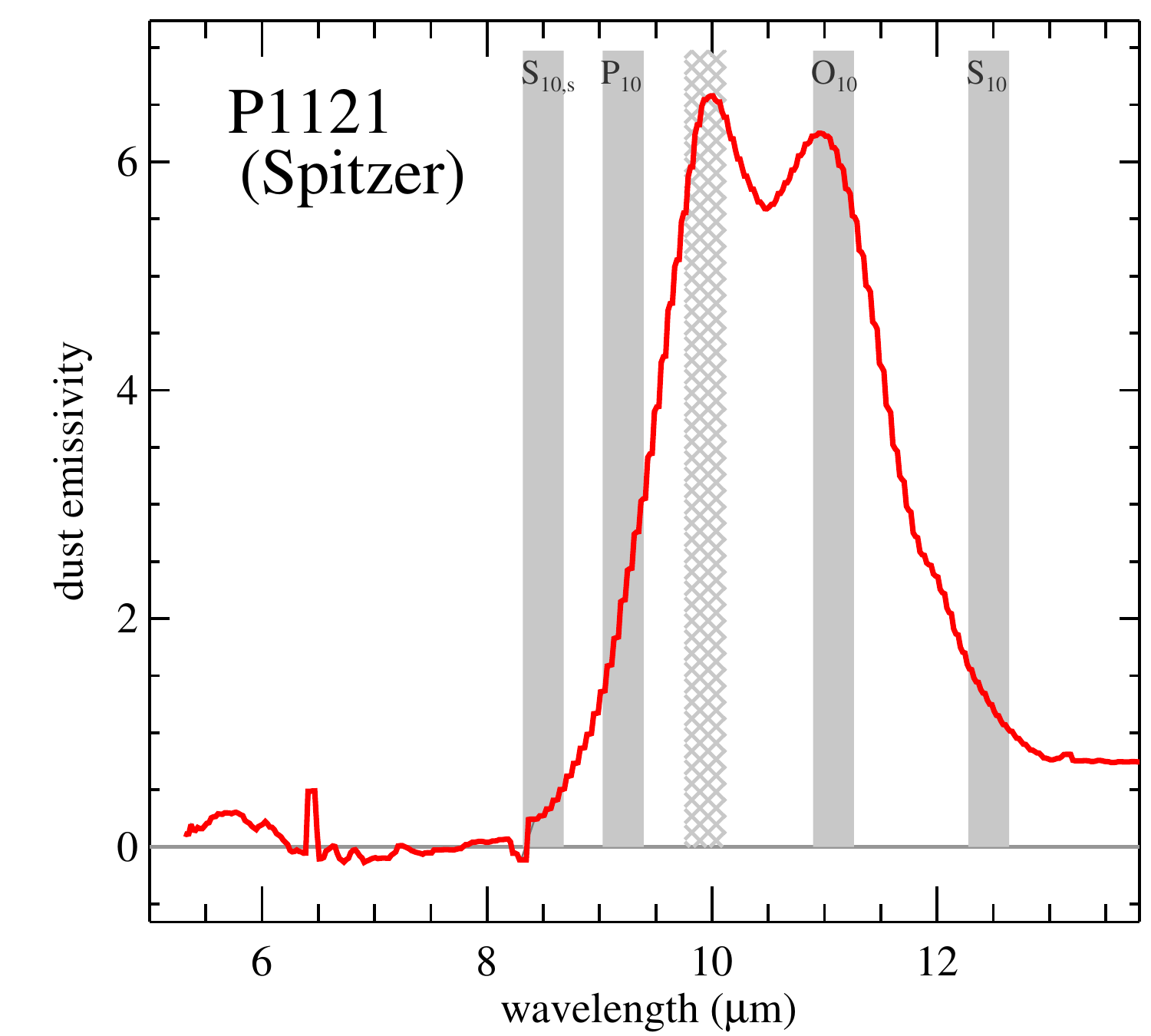}
    \includegraphics[width=0.23\linewidth]{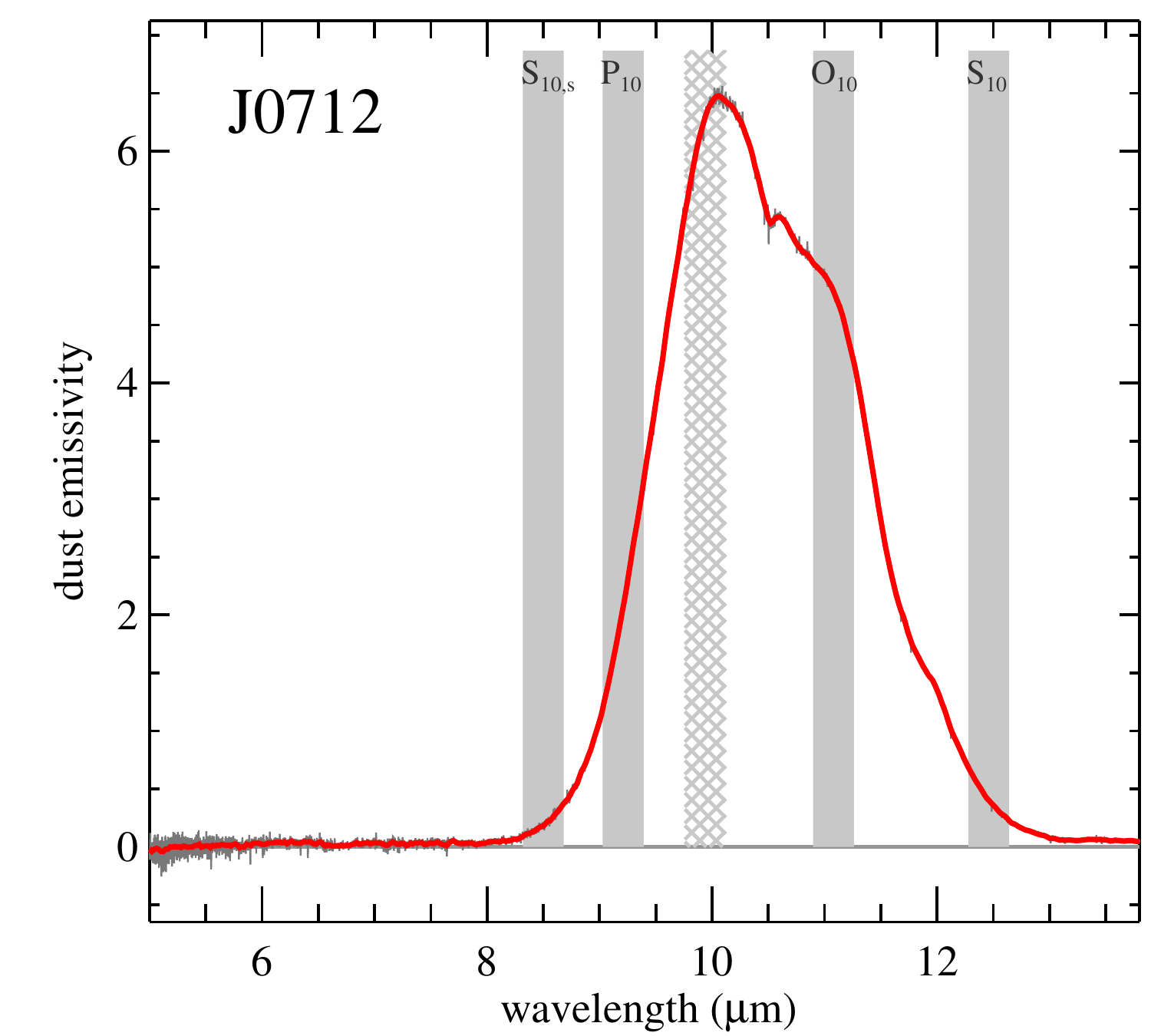}
    \includegraphics[width=0.23\linewidth]{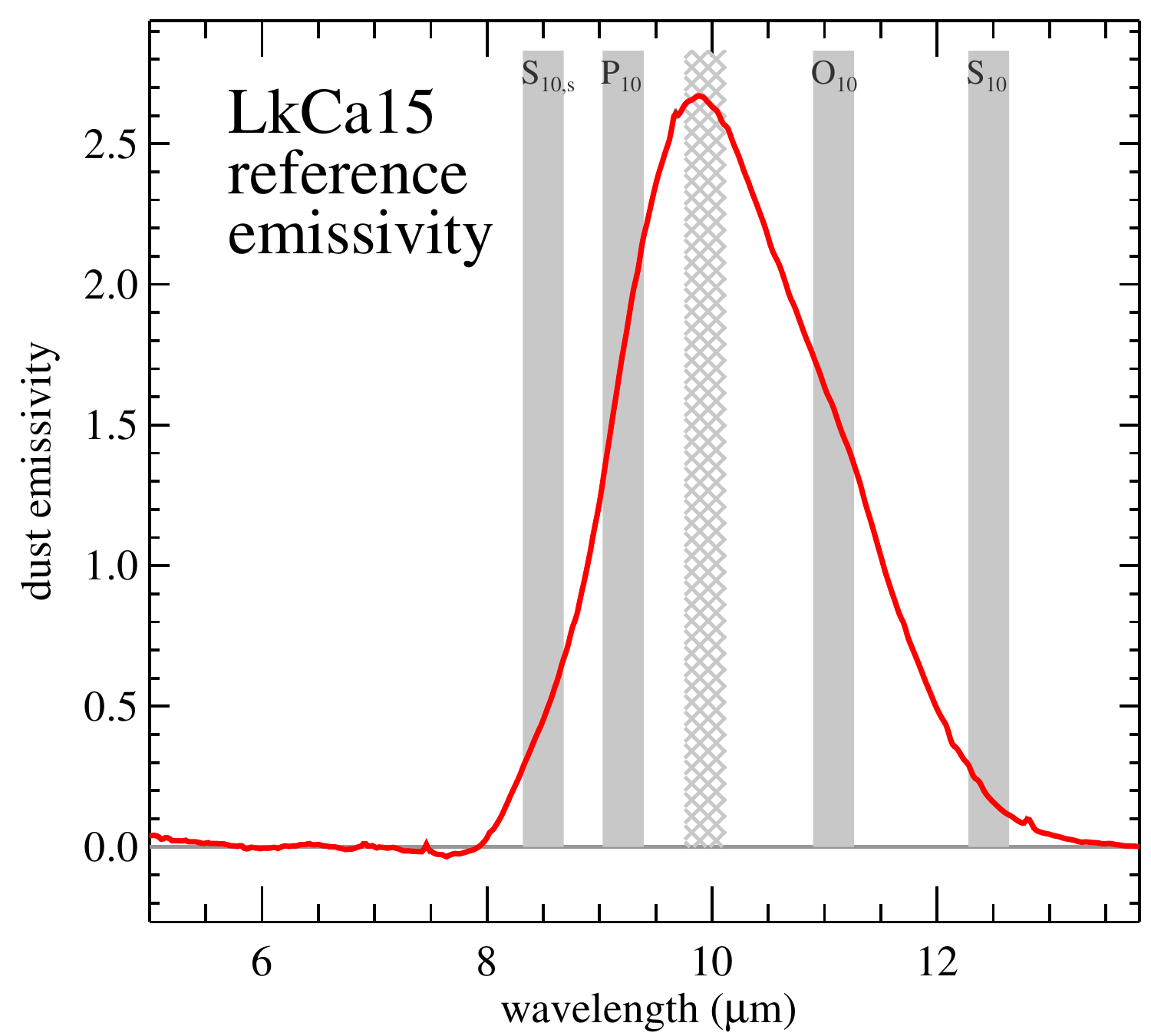}
    \includegraphics[width=0.23\linewidth]{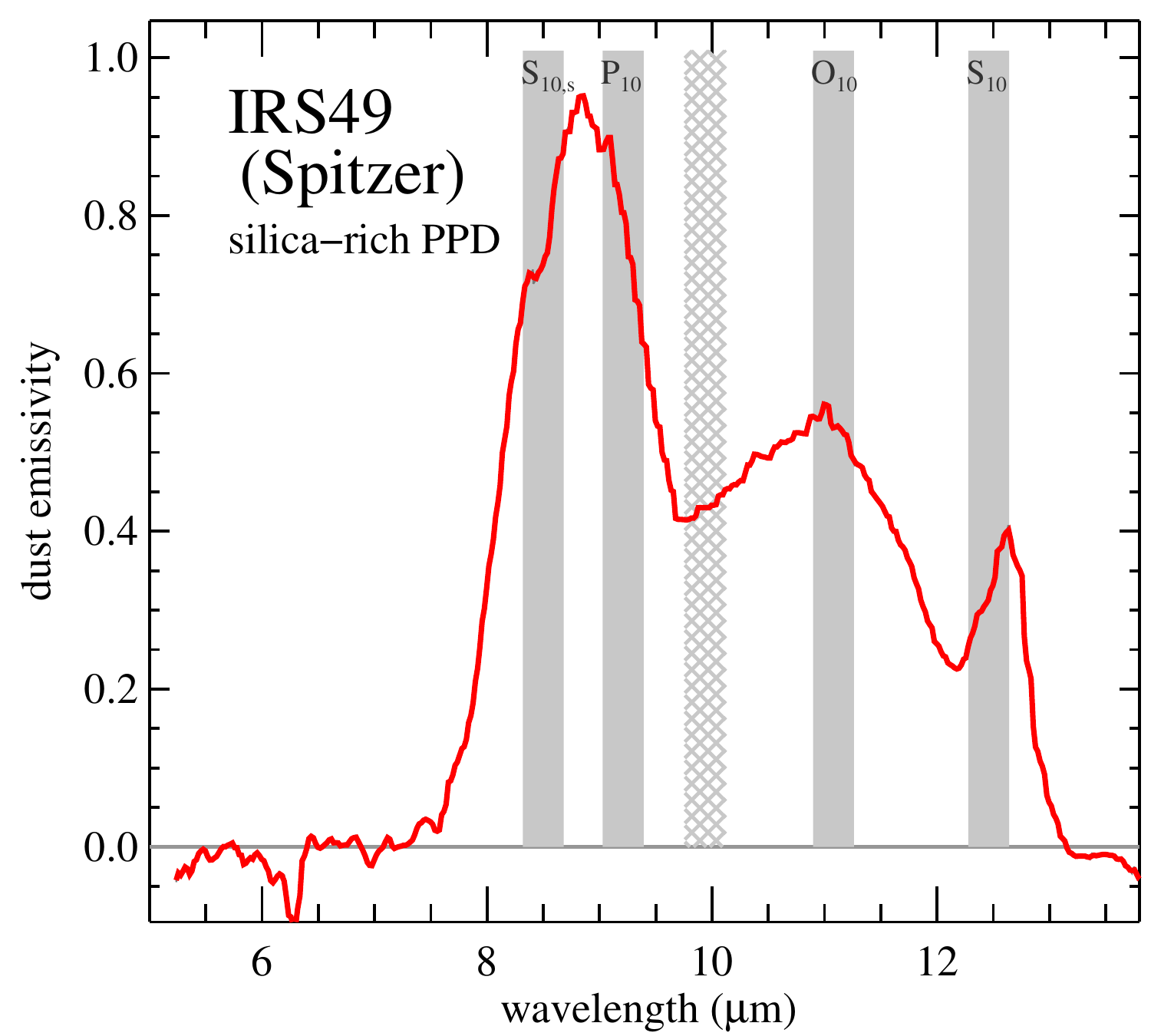}    
    \includegraphics[width=0.23\linewidth]{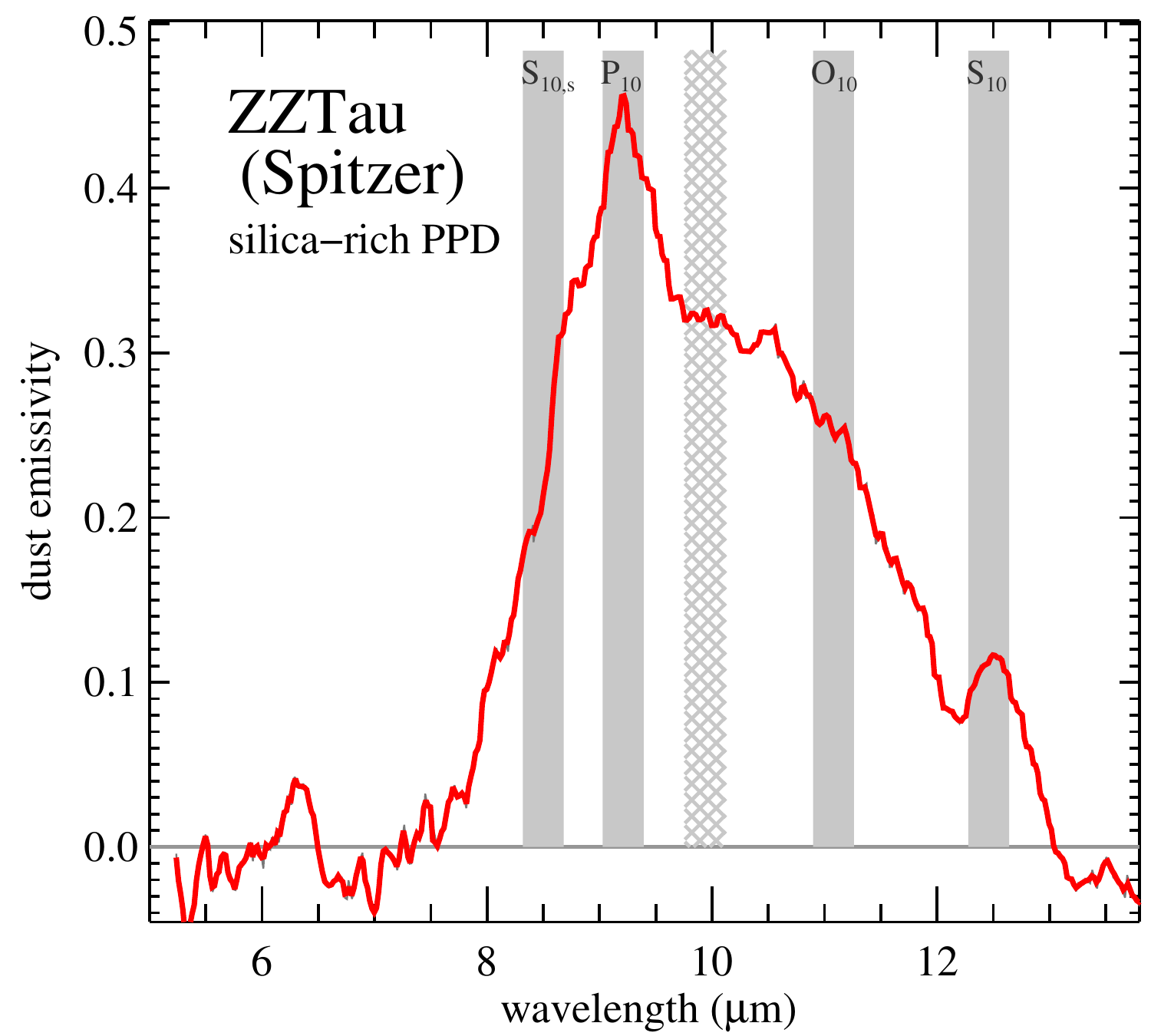}
    \caption{Dust emissivity of EDDs in the 10 \micron\ region with the pseudocontinuum modeled as a combination of blackbodies.  The top row shows the silica-rich system identified by Spitzer, while the second row shows the new silica-rich system identified by this work. The bottom row's last three panels show the reference emissivity from LkCa\,15 and two silica-rich PPDs (IRS\,49 and ZZ\,Tau). The vertical gray bars mark the wavelength regions for computing the dust indices (defining $S_{10,s}$, $P_{10}$, nominal 10 \micron\ reference peak, $O_{10}$ and $S_{10}$, details see Appendix \ref{sec:mathforms}).}
    \label{fig:10um_emissivity}
\end{figure*}

In this section, we define the properties of the 10 \micron\ features used in this work and 
recap the mathematical forms of the dust indices from \citet{watson09}, with the modifications described below. The feature's FWHM, equivalent width ($W_{10}$), and dust indices are measured from the derived dust emissivity as shown in Figure \ref{fig:10um_emissivity}: the FWHM and $W_{10}$ probe the dominant small‑grain sizes and the total mass of small grains, while the dust indices quantify the amount of thermally processed small grains in a system. We note that although FWHM is calculated assuming a Gaussian‑like profile, real features are often non‑Gaussian; FWHM can therefore be artificially small when a single sharp component dominates (e.g., the $\sim$9 \micron\ silica peak in HD\,172555). Increased crystallinity also broadens the 10 \micron\ feature by extending its red and blue wings through contributions from crystalline olivine and pyroxene. Although we propose using $W_{10}$ as a proxy for the dust mass of optically-thin, small grains, this quantity is subject to several uncertainties. The pseudocontinuum anchoring ignores the low‑level continuum contributed by small grains, causing a modest underestimation of true emission, and variations in grain‑size distribution complicate mass comparisons across different systems. Consequently, these values should be treated with caution and used primarily for relative comparisons across a sample whose values are derived consistently. 

We used the JWST MIRI/MRS spectrum of LkCa\,15 as the ``pristine-dust" reference, $W_{\lambda,0}$ near the peak wavelength $\lambda_R=$ 9.94 \micron\ (Figure \ref{fig:10um_emissivity}), following the work of \citet{su25_hd23514}. Note that the reference spectrum used in \citet{watson09} is a combined Spitzer spectrum of LkCa\,15 and UY\,Tau; UY\,Tau was not observed with JWST. We will discuss the suitability of using LkCa\,15 as the reference in Appendix \ref{sec:labindices}. The equivalent width and dust indices at 10 \micron\ are calculated as integrals over a wavelength range $d\lambda$ as: 
\[  W_{10} = \int W_\lambda \ d\lambda,  \]
\[    X_{10} = \frac{\int^{\lambda_X+\Delta\lambda}_{\lambda_X-\Delta\lambda} W_\lambda \ d\lambda}{\int^{\lambda_R+\Delta\lambda}_{\lambda_R-\Delta\lambda} W_\lambda \ d\lambda}  \frac{\int^{\lambda_R+\Delta\lambda}_{\lambda_R-\Delta\lambda} W_{\lambda,0} \ d\lambda}{\int^{\lambda_X+\Delta\lambda}_{\lambda_X-\Delta\lambda} W_{\lambda,0} \ d\lambda}, \]
where $X = P$, $O$, and $S$ for the peak wavelengths of crystalline pyroxene and olivine and silica, located at $\lambda_P$= 9.21 \micron, $\lambda_O$= 11.08 \micron, and $\lambda_S$ = 12.46 \micron. The procedure so far has followed the methodology from \citet{watson09}, where the reasoning behind the choices is well documented. We added a new dust index, $S_{10,s}$, centered at $\lambda_{S_{10,s}}=$ 8.50 \micron, to better capture the blue side of the silica feature \citep{koike13}, and used both $S_{10}$ and $S_{10,s}$ to characterize the silica dust. To minimize the overlapping wavelength between crystalline pyroxene and silica, we also reduced the integration range by a third compared to the value used by \citet{watson09}, i.e., $\Delta\lambda=$ 0.182 \micron. For $W_{10}$, the integration was performed between 7.2 and 13.0 \micron. 

We initially estimated uncertainties in the derived 10 \micron\ properties using a bootstrap: resampling the emissivity 500 times with propagated errors and taking the standard deviation of the resulting properties, following \citet{watson09}. The bootstrap performs well when measurement noise dominates but yields very small uncertainties for high S/N JWST spectra (typically $<$1\%). Therefore, we adopt two different pseudocontinuum forms to quantify the systematic uncertainty in the derived 10 \micron\ properties as given in Table \ref{tab:measuredindices}. The differences are typically $<$1\% for FWHM and $<$6\% for $W_{10}$. Uncertainties are similar for systems with JWST- and Spitzer-derived emissivities, although Spitzer ones include additional anchor points beyond 30 \micron. Typical uncertainties are 2\% for both $P_{10}$ and $O_{10}$ (up to 5\% in rare cases). For the silica indices, the typical uncertainty is 8\% for both $S_{10,s}$ and $S_{10}$ but exceeds 20\% in rare cases. 

To ensure that our modifications yield consistent results within uncertainties, we compare the indices calculated using both the old and new methods for 13 PPDs, where Spitzer IRS spectra are available in CASSIS \citep{cassis_ref}, including the five silica-rich PPDs studied by \citet{sargen09_silia}. For crystalline silicate indices ($P_{10}$ and $O_{10}$), the maximum difference between the old and new definitions is 5\%. For silica index ($S_{10}$), the difference is three times larger, with the new values mostly being higher -- likely because the narrower integration wavelength range increases the feature contrast. When plotting the old index values from \citet{watson09} on the new-index axes (e.g., Figure \ref{fig:indices}), differences for $P_{10}$ and $O_{10}$ are negligible, but $S_{10}$ shows a noticeable upward offset (indicated by arrows). We note that the systematic offset between the new and old index definitions may affect exact, system‑by‑system comparisons, but the overall distribution (i.e., spread) remains unchanged for comparative studies. Additionally, we excluded PPDs with host masses $<$0.5\,$M_\odot$, yielding a sample of 34 PPDs from \citet{watson09} that matches the host‑mass range of the EDDs for comparisons in Figure \ref{fig:indices}. For the new $S_{10,s}$ indices, in additional to the previous Spitzer PPD spectra, we also derived 10 \micron\ dust properties for additional 16 PPDs from archival JWST MIRI/MRS observations, all using a polynomial pseudocontinuum.

\subsection{Dust Indices for Various Materials}
\label{sec:labindices}

\begin{figure*}
    \includegraphics[width=\linewidth]{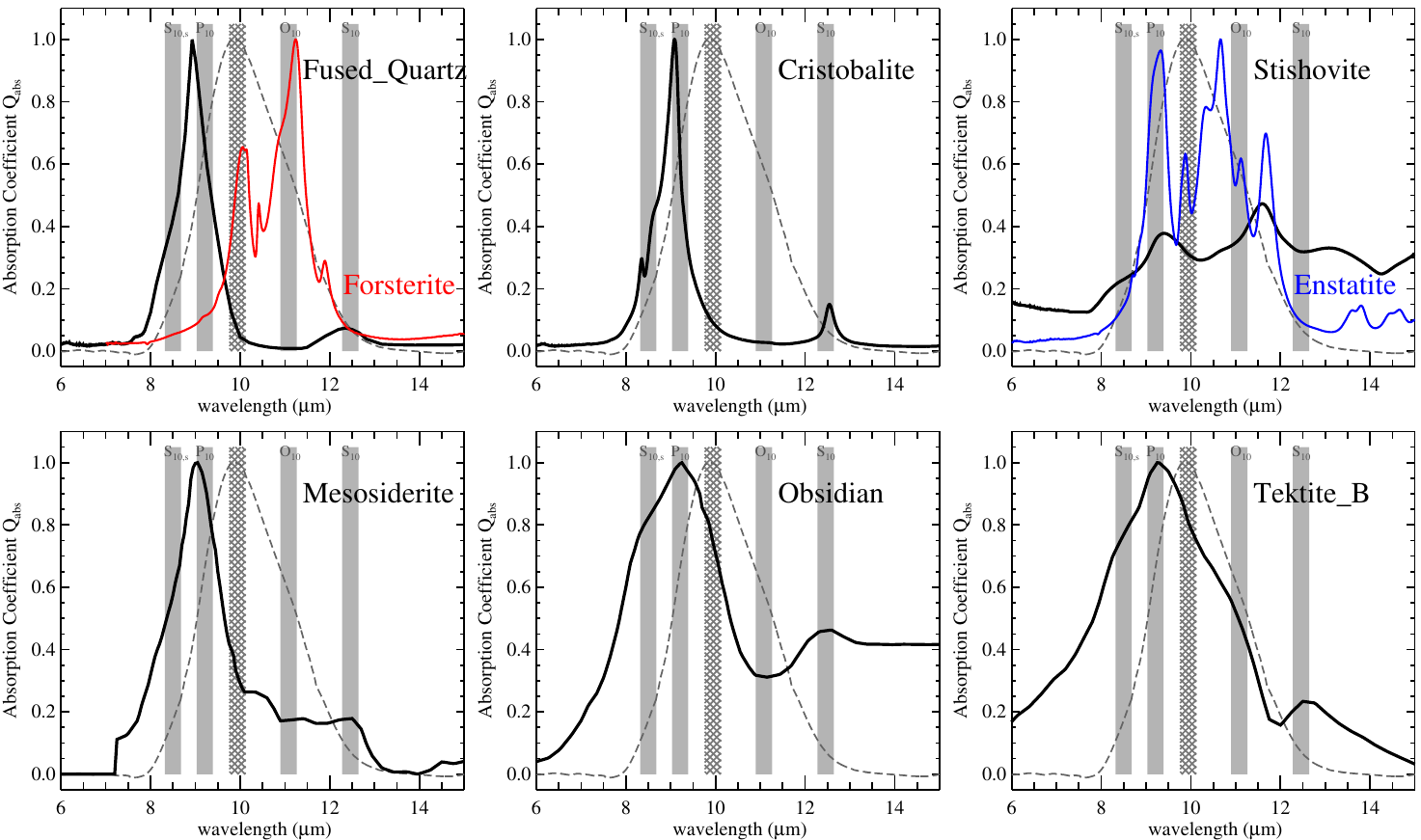}
    \caption{A sample of silica dust absorption coefficients (thick black lines, measured in the laboratory) in the 10 \micron\ region compared to pristine, ISM-like dust emissivity (gray dashed line). The top row shows the synthesized, silica materials from \citet{koike13}, along with synthesized forsterite (red, Mg-rich olivine) from \citet{koike03} and clinoenstatite (blue, Mg-rich pyroxene) from \citet{chihara02}. The bottom row shows the three of the natural impact samples from \citet{morlok14}. The vertical gray bars mark the wavelength regions for computing the dust indices as in Figure \ref{fig:10um_emissivity}.}
    \label{fig:silica}
\end{figure*}

The dust indices represent the contrast between the observed and pristine spectra. Specifically, the indices should be close to 1 if the two profiles are nearly identical, while a departure from that value indicates the prominence of a specific signature, assuming that the grain sizes are similar. To provide better perspective, we computed the 10 \micron\ properties (peak, FWHM, and dust indices) for a wide variety of compositions, ranging from synthesized materials to theoretically calculated absorption coefficients. The results are listed in Table \ref{tab:lab_indices}, with some shown in Figure \ref{fig:silica}. We particularly focus on the silica-like composition because different polymorphs of silica are formed under varying pressures and temperatures, making them valuable as a thermo-pressure gauge for their formation environment.

\begin{figure}
    \centering
    \includegraphics[width=\linewidth]{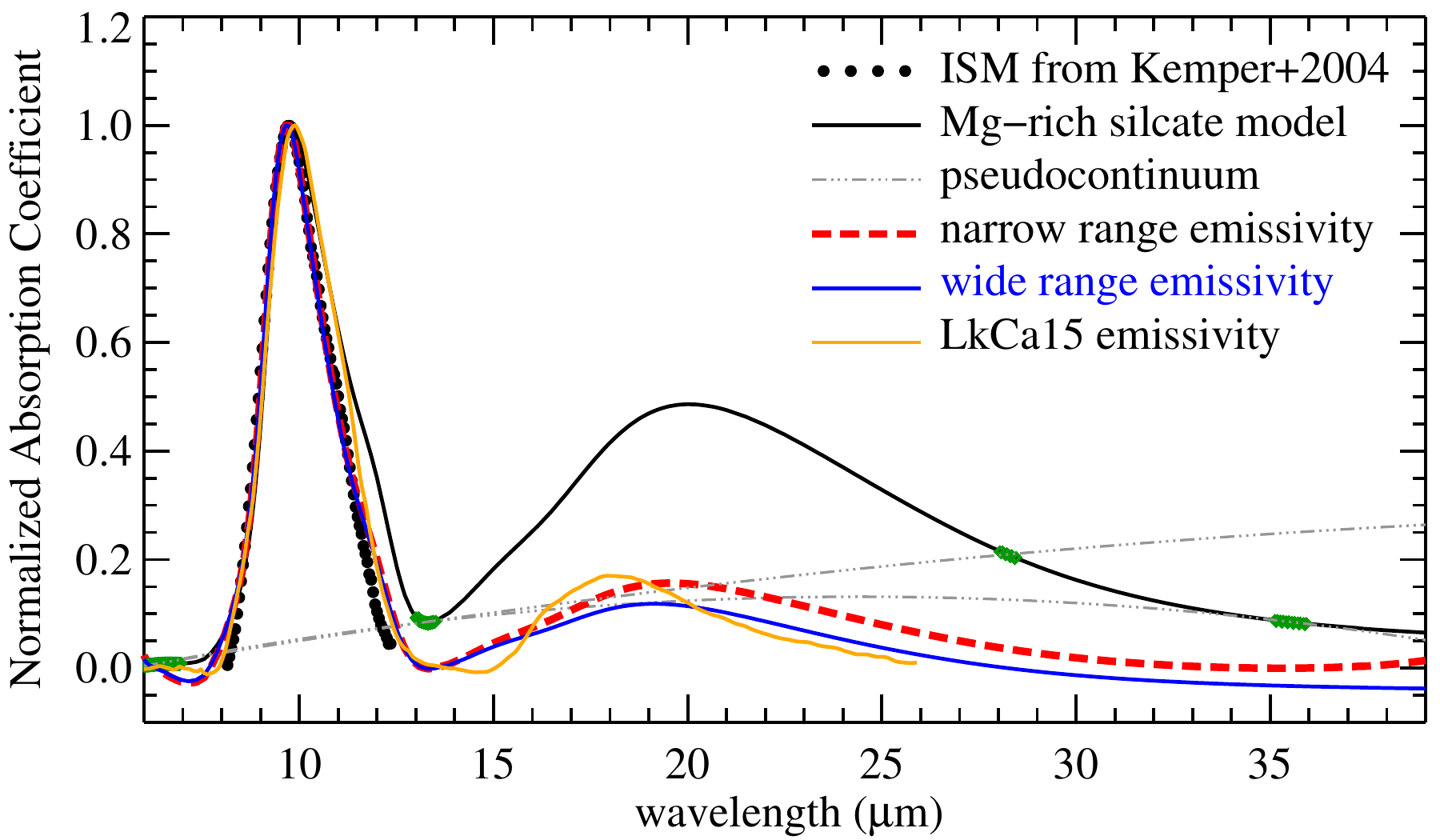}
    \caption{Comparison in the ISM-like dust emissivity. Dots: the observed absorption coefficient derived from the optical depth from \citet{kemper04}. Orange line: the derived dust emissivity of LkCa\,15 using the JWST spectrum. Blue and red-dashed lines are the model dust emissivities computed from a model absorption coefficient for 0.1 \micron, Mg-rich silicates (black) with two different pseudocontinua (narrow and wide anchoring ranges). The 10 \micron\ profiles are all very similar (relatively insensitive to exact composition) while the 20 \micron\ profiles are more uncertain. }
    \label{fig:ism_emi}
\end{figure}

We first assess how similar the LkCa\,15 10 \micron\ emissivity is to ISM-like dust, the presumed source material for protoplanetary disks. Comparing this metric with actual observations and theoretical ISM opacities (dominated by $\sim$sub-\micron\ size) clarifies the composition effects and associated uncertainties. The observational reference is \citet{kemper04}, who used ISO spectra toward the Galactic Center to derive the interstellar dust optical depth across the 10 \micron\ region. We adopt the precomputed absorption coefficient for Mg-rich, 0.1 \micron\ amorphous silicates from the \texttt{DuCKLing} package \citep{Kaeufer24_DuCKLinG}, mimic the effect of pseudocontinuum anchoring, and compute the resulting emissivity both with and without the last point beyond 30 \micron. Figure \ref{fig:ism_emi} compares the results with the LkCa\,15 dust emissivity. The 10-\micron\ profiles are very similar, while differences are larger in the 20-\micron\ region, consistent with our earlier conclusion that the derived 20-\micron\ emissivity is more uncertain. Although the profiles resemble each other, the computed dust indices are not exactly unity (see Table \ref{tab:lab_indices}, which also lists three theoretical ISM opacities from \citet{min2007}), indicating subtle differences. We define an empirical boundary based on these ISM-like dust indices. Deviations from that boundary are unlikely to arise solely from pseudocontinuum removal or from the calculation method and instead most likely indicate an increased crystalline contribution (thermal processing), since ISM dust is known to have low crystallinity \citep{kemper04,li08_low_crystallinity_ISM}. \citet{min2007} confirmed the absence or scarcity of crystalline silicates (and silica) in the ISM by fitting the observed 10 and 20 \micron\ optical‑depth profiles using three computational approaches that account for grain porosity. Their study also found interstellar silicates to be highly Mg‑rich and to include a $\sim$3\% silicon‑carbide (SiC) component, which broadens the overall 10‑\micron\ feature. For $O_{10}$, ISM dust spans 0.73--0.88, and for $P_{10}$ it spans 1.17--1.50. Although the equivalent ranges for silica indices are broad -- $S_{10}$: 1.19--2.57 and $S_{10,s}$: 0.98--2.61, likely reflecting compositional effects -- these values remain below the lowest values measured for laboratory silica-rich materials. These comparisons suggest that LkCa\,15's dust composition may differ slightly from the ISM -- a difference also noted for the ``pristine" composition of Solar System comets \citep{wooden_2002_review_CometGrains}. In practice, the precise definition of ISM-like dust composition is not critical since the indices reflect relative measurements as long as the reference remains consistent. This is why we adopt LkCa\,15 as the dust emissivity reference, allowing for the consistent integration of earlier studies with current work.

\begin{deluxetable}{rrrrrrrc}
    \tablewidth{0pt}
    \footnotesize     
    \tablecaption{10 $\mu$m Properties for Dust Material}
    \label{tab:lab_indices}
\tablehead{
   \colhead{Sample Name}  &  \colhead{Peak}  &   \colhead{FWHM}  &  \colhead{$P_{10}$}  &  \colhead{$O_{10}$}  &  \colhead{$S_{10}$}  &  \colhead{$S_{10,s}$} & \colhead{Note} \\  
     \colhead{}          &\colhead{($\mu$m)}&\colhead{($\mu$m)} &  \colhead{}       &  \colhead{}       &  \colhead{}       &  \colhead{}       &  \colhead{}   
}
\startdata
\multicolumn{8}{l}{silica-like composition} \\
         Quartz  &  9.15  &  0.50  &  13.67  &  0.35  &  19.39  &  21.09   & [1] \\  
   Fused Quartz  &  8.93  &  0.74  &  13.23  &  0.19  &  12.15  &  32.36   & [1] \\  
   Cristobalite  &  9.09  &  0.51  &  10.61  &  0.49  &  14.25  &  20.57   & [1] \\  
        Coesite  &  9.17  &  0.69  &  8.77   &  0.36  &  5.73  &  16.03   & [1] \\  
     Stishovite  &  11.58  &  4.42  &  1.68  &  1.95  &  13.72  &  4.19   & [1] \\  
   Mesosiderite  &  9.05  &  1.20  &  4.39  &  0.91  &  7.19  &  10.33   & [2] \\  
       Obsidian  &  9.25  &  2.40  &  2.03  &  0.72  &  8.19  &  6.35   & [2] \\  
Shocked Shergott &  9.35  &  3.00  &  1.51  &  1.31  &  4.08  &  2.98   & [2] \\  
  SiO$_2$ Glass  &  9.00  &  1.55  &  9.94  &  0.98  &  14.11  &  30.65   & [2] \\  
      Tektite A  &  9.20  &  3.85  &  1.88  &  1.09  &  6.08  &  6.21   & [2] \\  
     Tektite B  &  9.25  &  3.20  &  1.84  &  1.05  &  3.90  &  5.43   & [2] \\  
       \hline
\multicolumn{8}{l}{processed material} \\
   Unshocked$^\dagger$  &  10.15  &  1.60  &  0.64  &  0.74  &  0.38  &  0.42   & [3] \\   Medium Shock &  9.95  &  2.00  &  0.72  &  0.95  &  0.55  &  0.53   & [3] \\  
Average Shock  &  10.05  &  2.30  &  0.77  &  1.40  &  0.91  &  0.68   & [3] \\  
   High Shock  &  11.05  &  2.70  &  1.10  &  1.75  &  1.29  &  1.22   & [3] \\  
Maximum Shock  &  11.15  &  1.80  &  0.46  &  2.56  &  1.95  &  0.71   & [3] \\  
   \hline
\multicolumn{8}{l}{synthesized material} \\
     Forsterite  &  11.24  &  1.58  &  0.31  &  2.68  &  1.75  &  0.56   & [4] \\  
  C. Enstatite   &  10.66  &  2.83  &  2.51  &  1.81  &  2.48  &  1.63   & [5] \\  
     Diopside     &  9.28  &  2.56  &  2.77  &  1.90  &  2.02  &  1.16   & [5] \\  
  Enstatite Gel  &  10.00  &  1.27  &  0.74  &  0.51  &  1.97  &  0.97   & [5] \\  
 Diopside Glass  &  10.10  &  2.84  &  1.16  &  1.41  &  3.10  &  1.86   & [5] \\  
       \hline
\multicolumn{8}{l}{ISM-like composition} \\
      Diff. ISM  &  9.75  &  2.00  &  1.18  &  0.84  &  \nodata  &  0.98   & [6] \\  
      Mg-rich silicates$^\ddagger$ & 9.66 & 1.86 & 1.19 & 0.78 & 1.19 & 1.11 & [8] \\ 
     coatedGRF$^\ddagger$  &  9.60  &  2.30  &  1.50  &  0.88  &  2.57  &  2.61   & [7] \\  
     dhsfmax0p7$^\ddagger$  &  9.55  &  2.15  &  1.49  &  0.84  &  2.20  &  2.60   & [7] \\    
      porousGRF$^\ddagger$   &  9.55  &  2.15  &  1.47  &  0.87  &  2.41  &  2.26   & [7] \\  
      spherical$^\ddagger$   &  9.55  &  1.95  &  1.45  &  0.73  &  1.52  &  1.93   & [7] \\  
\enddata
    \tablecomments{$^\dagger$CM type carbonaceous chondrites \citep{morlok10_shockedCM}. $^\ddagger$Models of ISM composition with removal of narrow anchoring pseudocontinuum. References: [1] \citealt{koike13}, [2] \citealt{morlok14}, [3] \citealt{morlok10_shockedCM}, [4] \citealt{koike03}, [5] \citealt{chihara02}, ]6] \citealt{kemper04}, [7] \citealt{min2007}, [8] \citealt{Kaeufer24_DuCKLinG}}
\end{deluxetable}  

We use two other sources of laboratory materials to further validate the dust indices: processed materials (same initial composition subjected to different shock pressures) and synthesized materials with varied compositions. Processed materials are represented by mid‑infrared spectra of carbonaceous chondrites (CM) experimentally shocked at different pressures, as measured by \citet{morlok10_shockedCM}. CM chondrites are thought to be some of the most primitive materials in our solar system, having undergone minimal alteration since their formation. As shown in Table \ref{tab:lab_indices}, $O_{10}$ dust indices are good tracers for experimental shock pressure; as the pressure increases, the forsterite peak becomes increasingly prominent. As discussed by \citet{morlok10_shockedCM}, the strongly shock-melted sample (above 36 GPa) exhibits features of olivine that have recrystallized from melted material, closely resembling those found in the disks HD\,113766 and HD\,69830. Interestingly, while the indices for the maximally shocked material are similar to those of forsterite dust, they do not precisely match those of a pure synthesized forsterite sample. This further reinforces the idea that varying degrees of impurity in the material, along with the specific Mg/Fe ratio in the shocked, forsterite-rich material, influence the resulting indices. We also note that these melted, shocked forsterite‑rich materials have silica indices well below those of silica‑rich samples, making misidentification between the two unlikely.

\begin{deluxetable}{rrrrrrrrc}
    \tablewidth{0pt}
    \footnotesize     
    \tablecaption{10 $\mu$m properties of silica-rich PPDs}
    \label{tab:indices_silica_ppds}
\tablehead{
   \colhead{Name}  &  \colhead{Peak}  &   \colhead{FWHM}  &  \colhead{$W_{10}$} &  \colhead{$P_{10}$}  &  \colhead{$O_{10}$}  &  \colhead{$S_{10}$}  &  \colhead{$S_{10,s}$} & \colhead{Origin} \\  
     \colhead{}          &\colhead{($\mu$m)}&\colhead{($\mu$m)} &  \colhead{}  &  \colhead{}       &  \colhead{}       &  \colhead{}       &  \colhead{}       &  \colhead{}   
}
    \startdata
          IRS\,49 &   8.85 &   3.19 &   2.59 &   2.79 &   2.15 &  10.69 &  10.35 &    [1]  \\ 
          FZ\,Tau &  11.04 &   5.16 &   0.91 &   1.62 &   2.09 &  10.76 &   6.61 &        [2]  \\ 
    1RXS\,J1614   &   9.08 &   3.31 &   0.75 &   2.05 &   1.58 &   4.15 &   5.73 &    [1]  \\ 
            T\,51 &   9.15 &   3.02 &   4.00 &   1.96 &   1.50 &   2.34 &   4.97 &    [1]  \\ 
          ZZ\,Tau &   9.22 &   2.73 &   1.21 &   2.02 &   1.35 &   4.61 &   4.12 &    [1]  \\ 
        ROXs\,42C &   9.15 &   2.66 &   1.81 &   1.98 &   1.50 &   2.41 &   3.86 &      [1]  \\ 
    \enddata
    \tablecomments{[1] Spitzer data from \citealt{sargen09_silia}, [2] JWST data from \citealt{pontoppidan24_jdisc}}
\end{deluxetable}

Using dust indices from PPDs and previously analyzed EDDs/DDs (from spectral‑decomposition studies), we perform a comparative mineralogical analysis emphasizing the degree of thermal processing. The index distribution in PPDs lies mostly outside the low-crystallinity ISM-like boundary.  The ranges of indices in $P_{10}$ and $O_{10}$ reflect silicate crystallinity (the top row of Figure \ref{fig:indices}), varying from low (a few percent) to high (several tens of percent), calibrated from PPD studies \citep{oliverira11_ppd_dust_mineralogy}. Similarly, EDDs (and DDs where small, warm grains are present) generally exhibit crystallinity levels comparable to those in PPDs. Two well-studied forsterite-rich debris systems, $\beta$ Pic and HD\,69830, showcase this association; high-quality Spitzer/IRS spectra indicate crystallinities of approximately 3\% for $\beta$ Pic \citep{chen07} and 30--40\% for HD\,69830 \citep{olofsson12}, aligning with their positions in the $O_{10}$ index distribution. An additional consistency check comes from well-known forsterite-rich EDDs, where detailed dust mineralogy has been derived using Spitzer/IRS spectra. Both HD\,113766 and BD+20\,307 exhibit high crystallinity ($\gtrsim$30\%) \citep{olofsson12}. In the case of HD\,113766, there is a greater contribution from forsterite (as traced by $O_{10}$) compared to enstatite (as traced by $P_{10}$), while BD+20\,307 shows roughly equal contributions from both minerals. This is consistent with the index values presented in Table \ref{tab:measuredindices}. 

All EDDs observed with JWST show clear departures from the “pristine” ISM‑like profile represented by the LkCa\,15, evident in both the disk spectra (Figure \ref{fig:disksed_linearly}) and the 10 \micron\ emissivity profiles (Figure \ref{fig:10um_emissivity}) even without detailed spectral decomposition. Forsterite‑rich systems are readily identified by a double‑peaked 10 \micron\ profile (at $\sim$10.05 and 11.23 \micron) with some also showing Fe-rich, crystalline olivine peaks at $\sim$16.3, 19.5, and 24.0 \micron\ \citep{koike03}, consistent with earlier spectral‑decomposition results \citep{olofsson12}. Similar peaks appear in many JWST spectra as well, but with shifted peak positions or markedly different profiles, indicating a range of mineral mixtures. Compared with Spitzer‑identified silica‑rich EDDs (HD\,172555, HD\,23514, HD\,15407, HD\,145263), several JWST spectra also show peaks near $\sim$8.6, 9.1, 12.5, and 16.1 \micron\ indicative of silica‑rich minerals \citep{koike13}, although some of these peaks are muted or less pronounced in certain sources. This reflects the fact that, in a real astronomical environment, dust is a mixture of many minerals, and the dominant composition can only be identified by comparing profiles dominated by different species. We perform this comparison using dust indices and by grouping systems with similar index values to identify minerals of interest, specifically silica. 

As shown in Table \ref{tab:lab_indices}, all nearly pure silica compositions prepared in the laboratory ($\sim$sub-\micron\ size) have peak wavelengths in the 10-\micron\ region ranging from 8.93 to 9.25 \micron, except for stishovite, which peaks at 11.58 \micron. All silica-rich materials exhibit high $S_{10}$ and $S_{10,s}$ indices: the lowest $S_{10}$ is 3.90 (Tektite\,B) and the lowest $S_{10,s}$ is 2.98 (shocked shergottite). Most silica-rich samples also show high $P_{10}$ but relatively low $O_{10}$ values. We use the silica indices from previously identified silica-rich systems -- four EDDs (HD\,172555, HD\,23514, HD\,15407, and HD\,145263 from Table \ref{tab:measuredindices}) and six PPDs  (Table \ref{tab:indices_silica_ppds}) from \citet{sargen09_silia} and \citet{pontoppidan24_jdisc} -- to set thresholds for identifying silica-rich systems. All ten of these silica-rich systems have $S_{10,s}$ values larger than the lowest $S_{10,s}$ measured for silica-like laboratory material\footnote{HD\,172555 has $S_{10,s}$=2.95$\pm$0.23, which places it at the boundary of the lowest $S_{10,s}$ value within the errors.}, but only six meet the $S_{10}$ threshold (the four that do not are HD\,172555, HD\,145263, T\,51, and ROXs\,2C). This is not surprising, since the $S_{10}$ index probes the less pronounced silica feature\footnote{\citet{watson09} chose this wavelength, rather than the more prominent $\sim$9-\micron\ feature, to avoid confusion with the crystalline‑pyroxene feature near $\sim$9.21 \micron\ when using a wider integration index.}, which is therefore more susceptible to pseudocontinuum uncertainties. Note that the silica indices do not distinguish between crystalline and amorphous silica; a sharper 12.46 \micron\ feature (e.g., cristobalite, a crystalline polymorph; see Fig. \ref{fig:silica}) would increase the $S_{10}$ value if present in sufficient abundance. In real environments dust is expected to contain mixtures of materials -- different polymorphs of silica (crystalline or amorphous) and other minerals with varying impurities and porosities -- that likely reduce the contrast of the weaker $S_{10}$ feature. To be consistent with previous identifications, we therefore use $S_{10,s}>$2.98 as the threshold for identifying silica-rich systems. The newly identified silica-rich EDDs are, V488\,Per, J0611, J1213, HD\,166191; together with the four previously identified silica-rich EDDs from Spitzer, eight of 21 EDDs (38$^{+11}_{-9}$\%) are silica-rich. The labels ”silica-rich” and ”silica-poor” are relative to the sample and chosen threshold; including or excluding edge cases near the boundary does not affect our overall conclusions. 
By analogy, our derived $S_{10,s}$ indices yield an apparent silica‑rich fraction of 27$^{+9}_{-6}$\% among PPDs (eight of 30)\footnote{Two of the eight are newly identified silica‑rich PPDs according to our criterion and were confirmed by \citet{varga26} using the spectral decomposition technique; the other six were previously published shown in Table \ref{tab:indices_silica_ppds}.}, but that sample includes many preselected silica‑rich systems and thus likely overestimates the true fraction. Using the $\sim$80 Class II PPDs analyzed by the Spitzer/IRS GTO team gives a more representative silica‑rich PPD fraction of $\sim$6$^{+4}_{-2}$\% \citep{watson09,sargen09_silia} using Spitzer data.

\begin{figure}
    \centering
    \includegraphics[width=\linewidth]{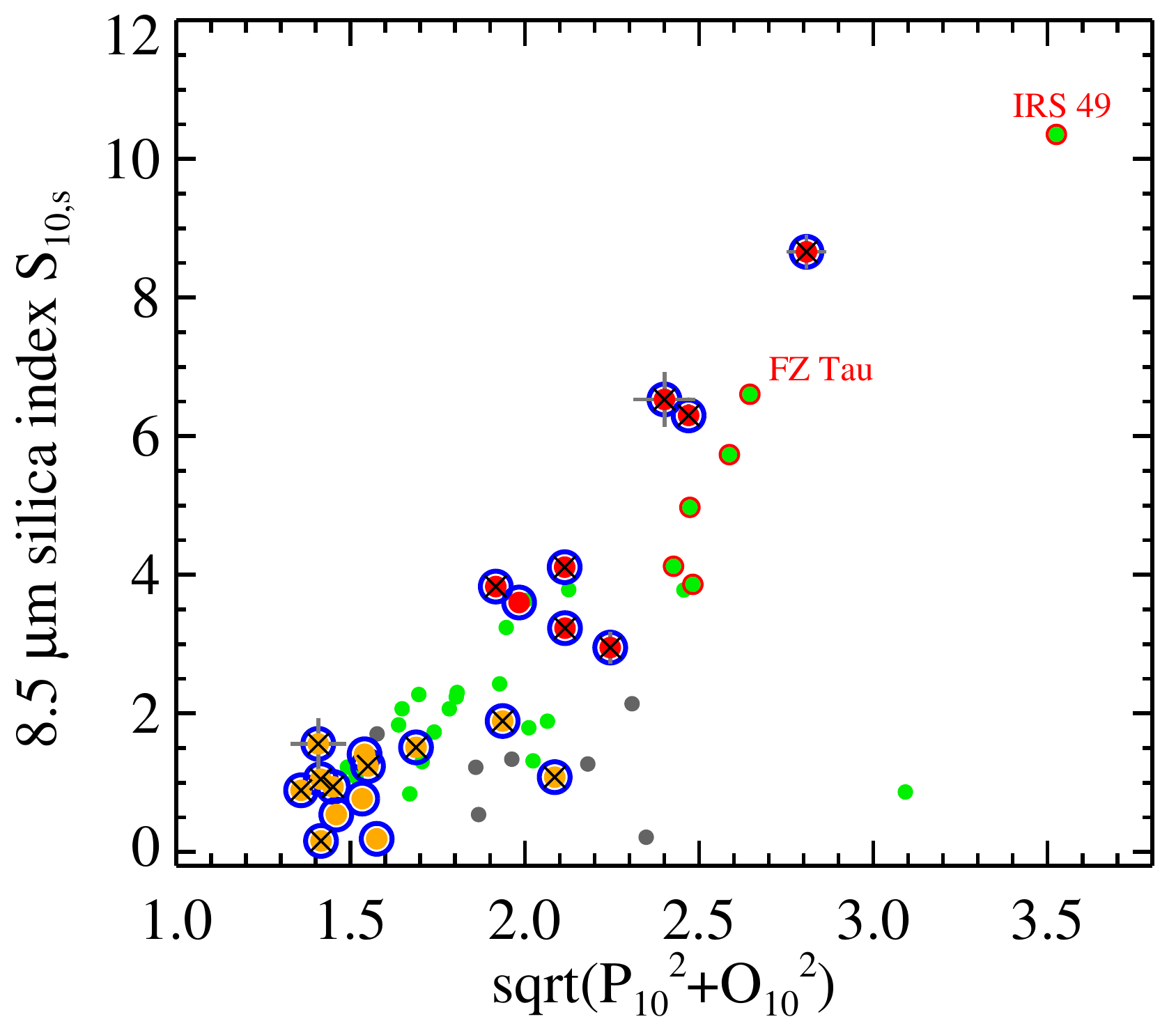}
    \caption{The distribution between the $S_{10,s}$ silica index and the combined crystalline silicate index ($\sqrt{P_{10}^2 + O_{10}^2}$), labeled as crystallinity. These indices suggest a positive correlation between the amounts of silica and crystalline silicates.}
    \label{fig:silica_vs_crystallinity}
\end{figure}

As discussed in the main text, while we adopt a dichotomous terminology to describe EDD dust mineralogy (silica-rich vs.\ silica-poor), this does not imply that silica-rich systems lack crystalline silicates or that silica-poor systems contain no silica at all. The study by \citet{olofsson12} demonstrated that $\sim$7\% of the dust in HD\,69830 originates from $\beta$-cristobalite (a form of silica), while the crystalline silicate features in BD+20\,307 are best fitted with equal amounts of forsterite and enstatite, both of which are classified as silica-poor systems in our framework. We also note that crystalline silicates from both the olivine and pyroxene families are present in nearly all previously identified silica EDDs and PPDs. For silica-rich EDDs HD\,172555 and HD\,145263, 11\% and 22\% of the emission, respectively, are attributed to Mg‑rich forsterite \citep{lisse09,badlisse2020}, while crystalline silicates account for a few percent up to $\sim$30\% of the mass in the five silica‑rich PPDs \citep{sargen09_silia}. Among the silica-poor systems, the last nine panels in Figure \ref{fig:disksed_linearly} show pronounced double peaks near 16 and 19 \micron\ -- indicative of Fe-rich forsterite -- which previously would have led to classification as forsterite-rich.  Silica dust also has a feature at 16.1 \micron, so determining whether the 16 \micron\ feature in some silica-rich systems arises from a mixture with forsterite or from pure silica requires future detailed spectral decomposition.

To further explore potential trends between silica dust (represented by $S_{10,s}$ index) and crystallinity in EDDs, we derive a crystallinity measured by combining $P_{10}$ and $O_{10}$ as $\sqrt{P_{10}^2 + O_{10}^2}$. The results are illustrated in Figure \ref{fig:silica_vs_crystallinity}. These indices suggest a positive correlation between the amounts of silica and crystallinity, indicating that silica-rich systems are more frequently found in environments with high crystallinity, including PPDs.  The presence of a small amount of silica in HD\,69830, combined with its high crystallinity, is consistent with the observed correlation, reinforcing that both silica and crystalline silicates are highly processed materials.

\section{Infrared Variability}
\label{sec:irvariability}

Infrared variability resulting from circumstellar material often serves as supporting evidence for identification as genuine EDDs. In light of the decommissioning of the NEOWISE mission, we have collected and documented all WISE W1 and W2 photometry from IPAC/IRSA and determined their variability behavior over the span of 15 years, as detailed in Appendix \ref{sec:wise4.6um}. In addition to the decade-long, regular 6-month sampling of disk variability using the WISE W1/W2 data, we further investigate the stability of the 10 \micron\ feature by comparing the synthetic photometry derived from the JWST/MIRI/MRS spectrum to other existing 10--20 \micron\ photometry as detailed in Appendix \ref{sec:10_20umVar}.

\subsection{3--5 \micron\ Infrared Variety as traced by WISE W1/W2 data}
\label{sec:wise4.6um}

\begin{figure*}
    \centering
    \includegraphics[width=0.495\linewidth]{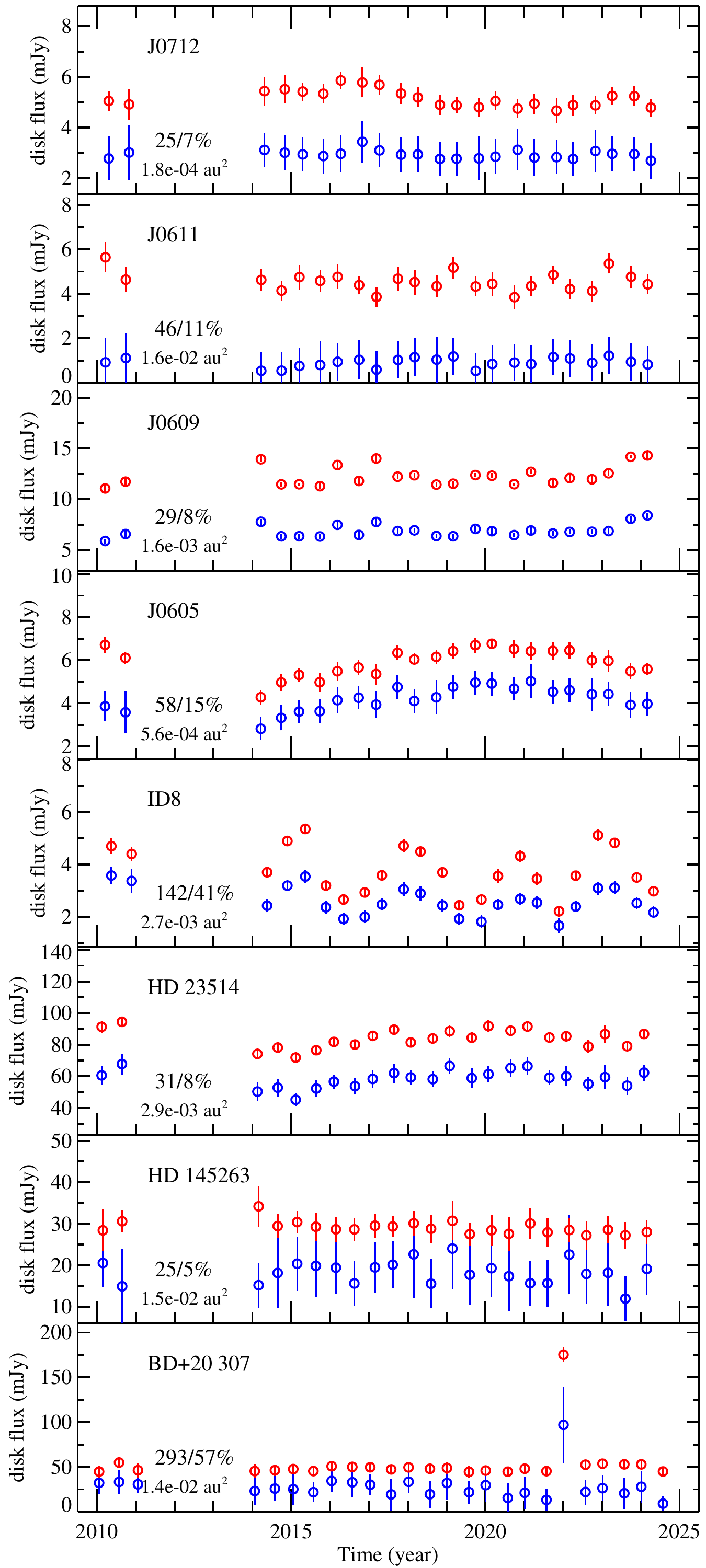}
    \includegraphics[width=0.495\linewidth]{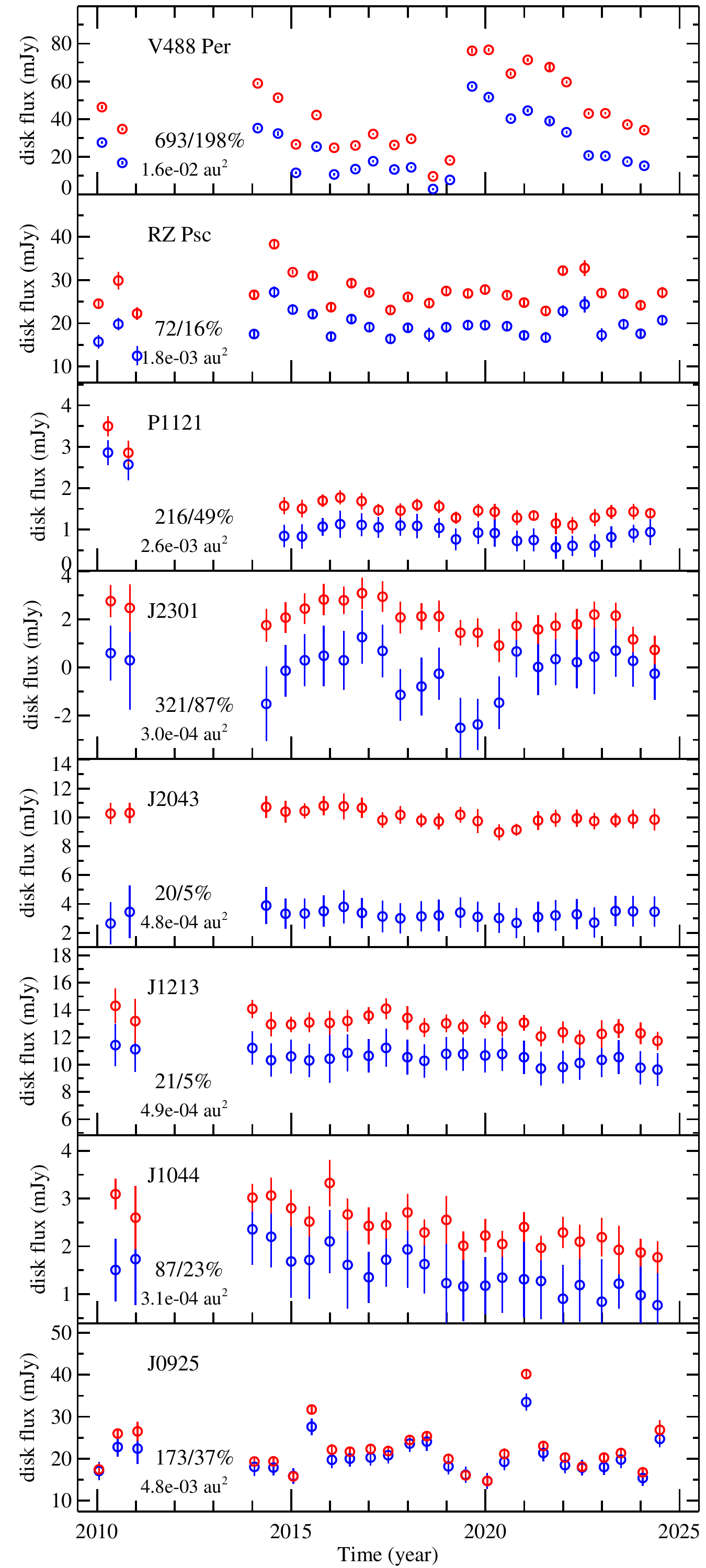} 
        \caption{Each subfigure shows the individual WISE W1 (blue) and W2 (red) light curves. The maximum and typical variability percentages relative to the minimum disk flux at 4.6 \micron\ are provided in each of the panel, along with the maximum change in the dust cross-section in units of au$^2$. }
    \label{fig:wiselc1}
\end{figure*}

Following the analysis steps outlined in \citet{moor21}, the W1/W2 measurements are calculated as an average of multiple single-exposure photometry taken every six months, and the disk contribution (i.e., excess flux) is obtained by subtracting the stellar contribution determined by atmospheric models (assuming a systematic uncertainty of 2\%), which is assumed to be non-variable. Although some optical variability is known to be associated with the host stars, the infrared variability is unlikely to arise from the star because the degree of infrared variability is often much larger than that coming from the stellar photosphere, particularly at longer wavelengths (i.e., 4.6 \micron\ photometry is a better tracer for disk variability than 3.4 \micron\ photometry). However, normalization of the stellar model does affect the resulting photospheric flux; particularly for systems with large interstellar extinction, which greatly influences the normalization using optical photometry.  As a result, the disk flux becomes negative, especially during low disk flux states, such as for J2301 (see Figure \ref{fig:wiselc1}).

The effect of stellar photospheric determination can be mitigated if we quantify the degree of variability in a relative sense by normalizing to a specific value. We used the minimum disk flux at 4.6 \micron\ to represent the nominal background flux level maintained by collisions among existing small planetesimals, along with two indicators to quantify infrared variability: the maximum and typical (rms) variations, both expressed as percentages relative to the minimum disk flux at 4.6 \micron. The typical variation reflects the stochastic nature of collisions, while the maximum variation captures the highest degree of change recorded in the data over the monitored period. Four out of the 21 EDDs are heavily saturated in W1/W2 photometry, making their single-exposure photometry unusable. HD\,23514, HD\,145263, and BD+20\,307 have single-exposure measurements that are partially affected by saturation. We employed the method described in the NEOWISE Explanatory Supplement\footnote{see \url{https://irsa.ipac.caltech.edu/data/WISE/docs/release/NEOWISE/expsup/sec2\_1civa.html}.} for saturation correction and found that the corrections are most significant at 3.4 \micron\ for the latter two systems but not significant for HD\,23514 or for the data at 4.6 \micron. Furthermore, we validate the short brightening event observed in the 2022 data point of BD+20\,307 by examining the relevant single exposure data of nearby stars. 
The final 16 disk light curves are presented in Figure \ref{fig:wiselc1}. Four of the saturated systems have warm Spitzer time-series data that can be used to supplement the WISE data. Published results for HD\,113766, HD\,172555, and HD\,166191 are sourced from \citet{su20,su22_hd166}, while we find HD\,15407 exhibiting stochastic variations of approximately 5--10\% using the archival data\footnote{Adopted values for the variability analysis are: 200/10 for HD\,166191, 15/5 for HD\,113766, and 10/5 for both HD\,172555 and HD\,15407}. 

We also computed the associated change in dust cross-section, $\Delta\Sigma = \Delta F d^2 / B_{\nu}(T_d)$, required to match the maximum flux change observed at 4.6 \micron\ ($\Delta F$) where $d$ is the distance and $B_{\nu}$ is the Planck function by assuming that (1) the emission is optically thin and (2) all dust is at the same temperature ($T_d$). Using the warmer temperature applied in the pseudocontinuum, we found that $\Delta \Sigma$ ranges from 1.85$\times10^{-4}$ au$^2$ to 1.60$\times10^{-2}$ au$^2$ with the largest values coming from V488\,Per and the smallest from J0712 (as illustrated in Figure \ref{fig:wiselc1}). These values correspond to 2.7 to 233 times that of the stellar surface (assuming a size comparable to the Sun). We emphasize that these values are true minima due to the prior assumptions, suggesting that 
the variation originates from the circumstellar environment rather than from the star or the emerging planet itself (if any). 

\begin{figure*}
    \centering
    \includegraphics[width=0.32\linewidth]{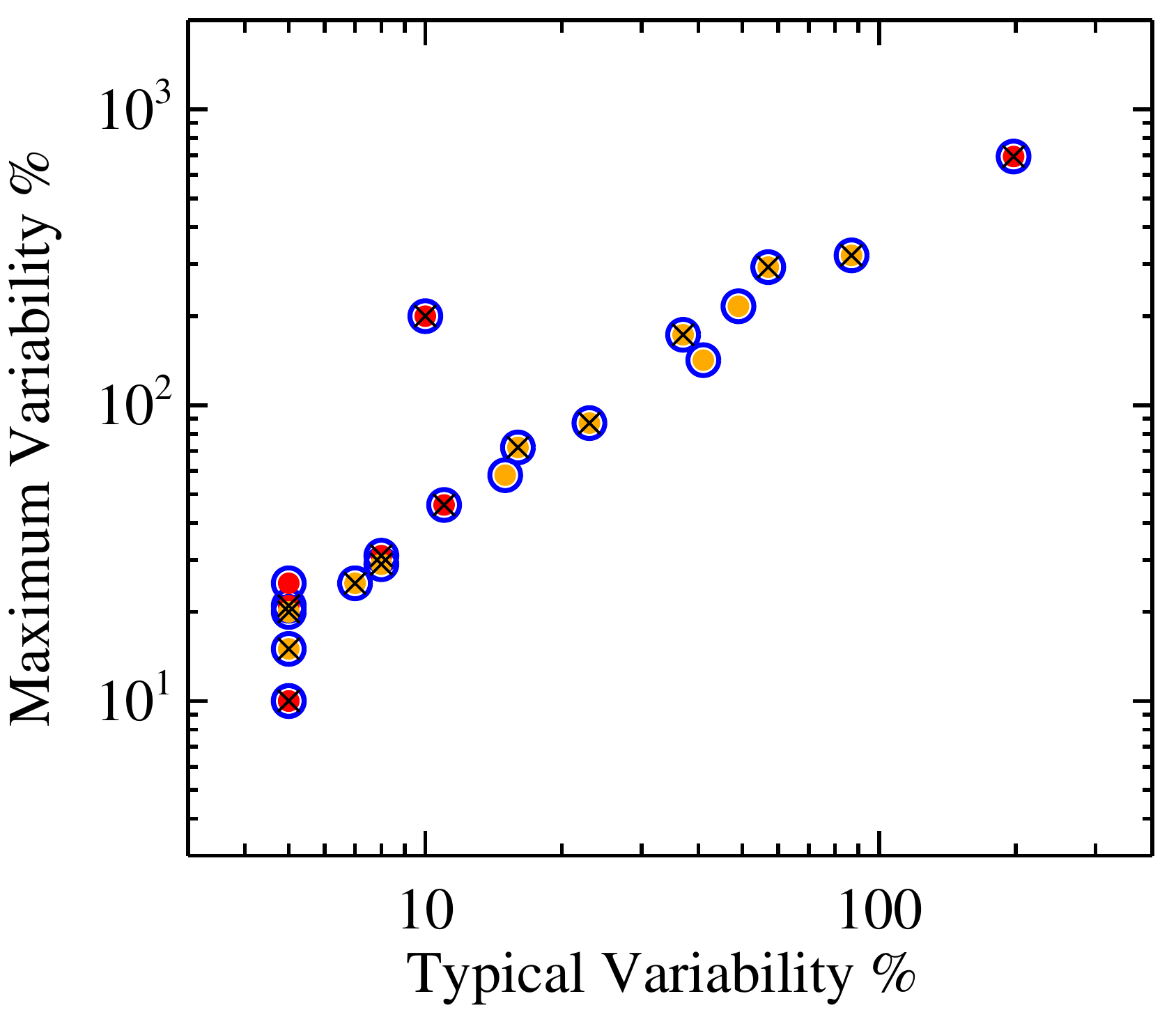}
    \includegraphics[width=0.32\linewidth]{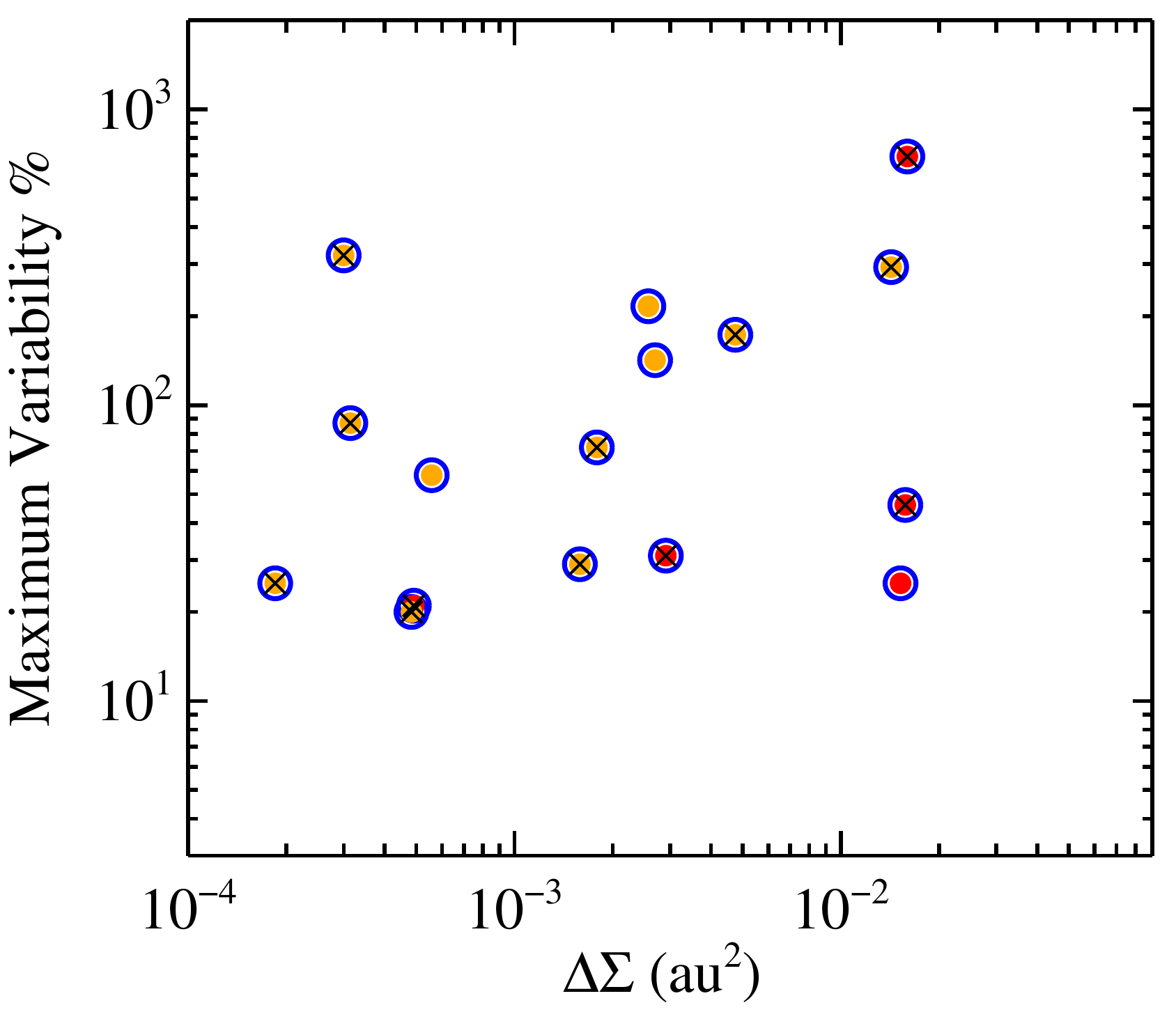}
    \includegraphics[width=0.32\linewidth]{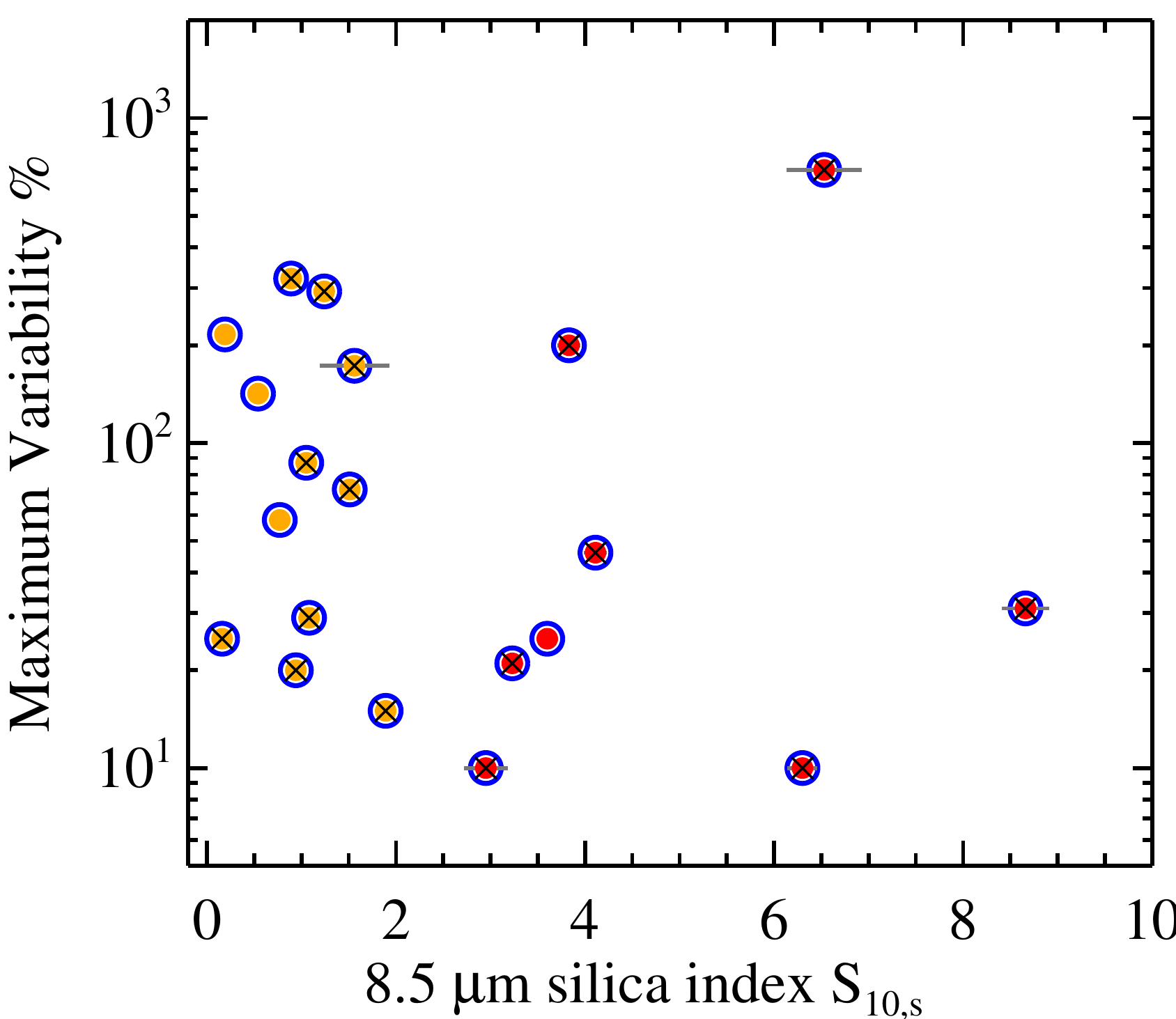}
    \includegraphics[width=0.32\linewidth]{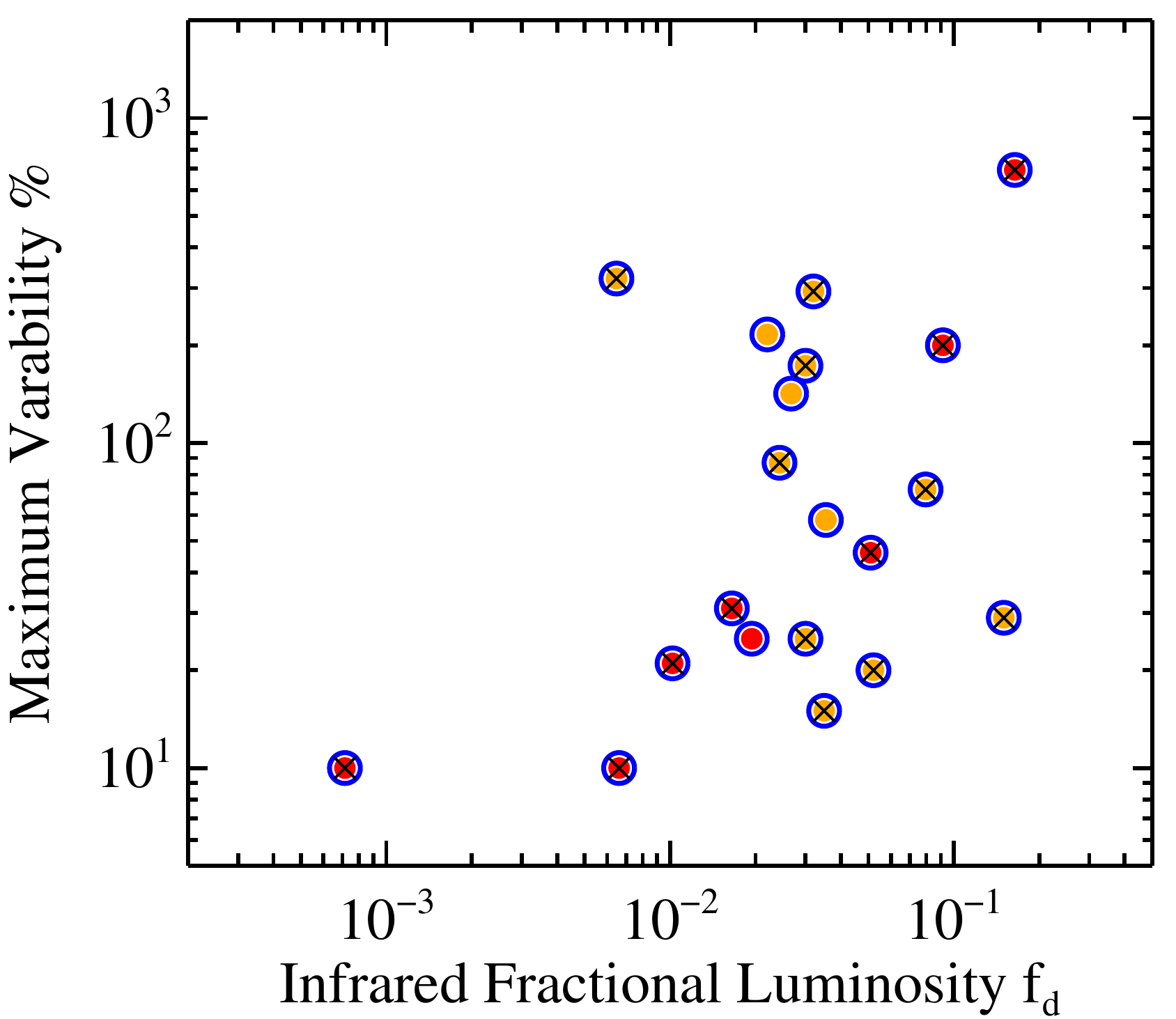}
    \includegraphics[width=0.32\linewidth]{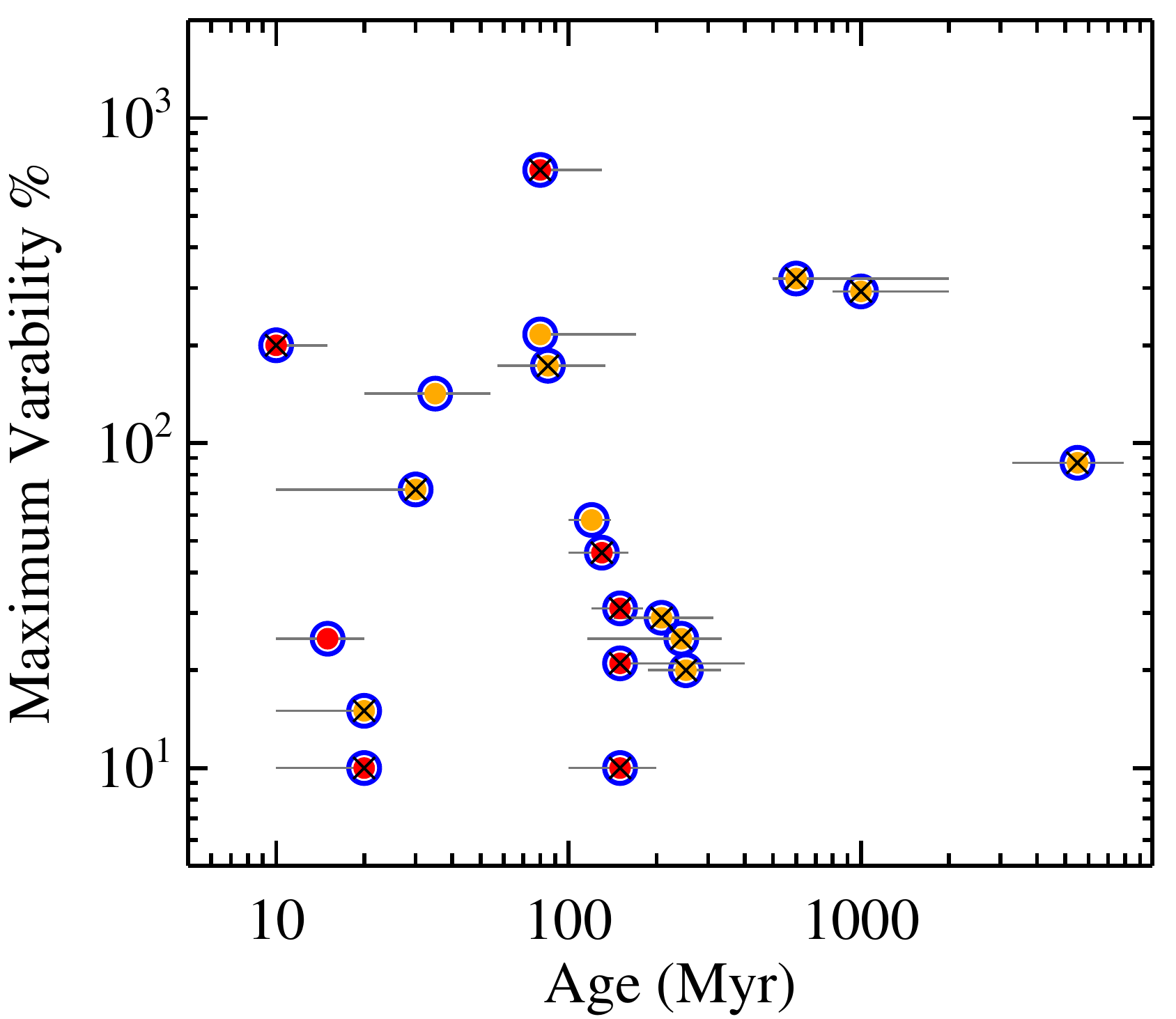}
    \includegraphics[width=0.32\linewidth]{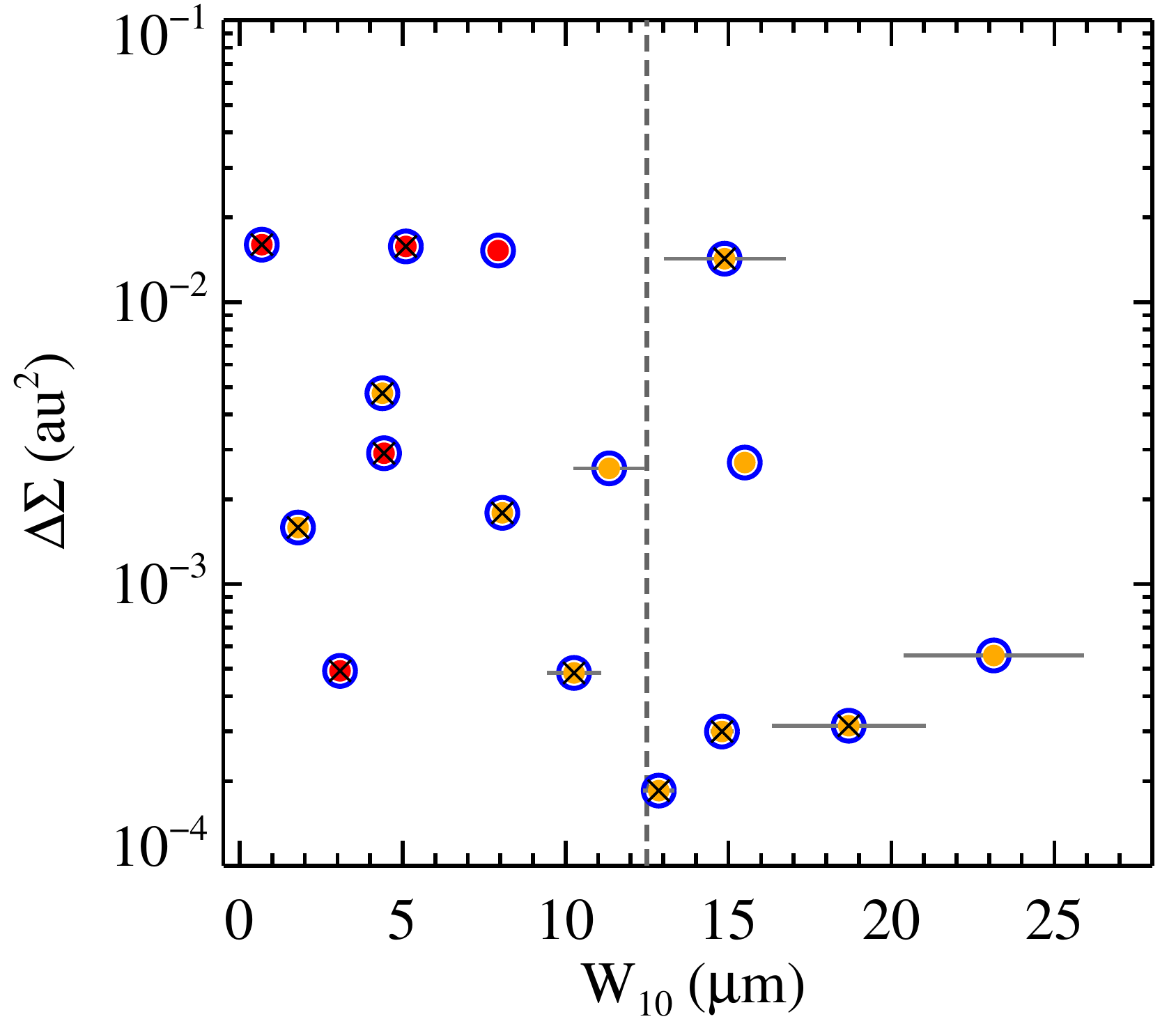}
    \caption{Relation between the observed 4.6 \micron\ disk variability and selected EDD properties. The symbols and colors are the same as in the previous figures (red: silica-rich, orange: silica-poor, cross: multiplicity). The top-left panel displays a comparison between the two variability percentage indicators showing a tight correlation that suggests both track the aftermath of a significant impact. However, their relation to the change in dust cross-section ($\Delta \Sigma$, on an absolute scale) is very weak, as illustrated in the upper-middle panel. The other panels show disk variability (using both the maximum variation percentage and $\Delta \Sigma$) vs.\ other EDD properties, such as silica indices $S_{10,s}$, infrared fractional luminosity ($f_d$), and age. None of these shows a strong correlation, including the derived dust mineralogy. The only potential trend is shown in Figure \ref{fig:var_w10} with implications discussed in Section \ref{sec:discussion}.}
    \label{fig:WISEvariability}
\end{figure*}

We compared the two variability percentage indicators (maximum and typical rms) and $\Delta\Sigma$ shown in the top two panels of Figure \ref{fig:WISEvariability}. The two variability percentage indicators are strongly correlated\footnote{To ensure that the correlation was not biased by the WISE cryogenic cooling effect, we also repeated the analysis using only NEOWISE data (after 2012), and the resulting correlation remained unchanged.}, suggesting they are likely driven by a common physical process such as the aftermath of a significant impact. Interestingly, there is no direct correspondence between the variability percentage (normalized to the minimum disk flux per system) and $\Delta\Sigma$, as illustrated in the upper-middle panel of Figure \ref{fig:WISEvariability}. 
While both quantities can be used to characterize the variable behaviors in EDDs, the former reflects the change normalized to the individual system, whereas the latter is anchored on an absolute scale, allowing for comparison among different stars. The uncertainty associated with both might be quite large in some cases due to the adopted assumptions, which transform the observed flux change to the variability levels. Both derivations assume that the observed flux change is solely due to changes in the emitting area, which is valid in the optically thin regime but should be treated as a lower limit otherwise. Additionally, $\Delta\Sigma$ has an additional assumption that the dust temperature is constant, which is unlikely to be the case.

We used both quantities to search for potential trends with other EDD properties, such as infrared fractional luminosity ($f_d$), age, 10 \micron\ feature properties and dust mineralogy. We note that EDDs are identified by large amounts of infrared excesses indicative of large amounts of circumstellar material with $f_d \gtrsim 10^{-2}-10^{-3}$. The only exception in our sample is HD\,172555 with $f_d\sim7\times10^{-4}$ whose classification is primarily based on its unique, silica-rich dust mineralogy \citep{lisse09}. We employed both Spearman and Kendall correlation coefficients, suited for smaller data sets, to search for the potential correlation among the derived properties with the $p$ values indicating the statistical significance of the results. The only pair of parameters with a potentially significant correlation is $P_{10}$ and $\Delta \Sigma$. Even in this case, however, the $p$-values are only slightly lower than 0.05. Combined with the fact that our sample is small (there are only three systems older than 300 Myr\footnote{The age of 150 Myr for J1213 is a lower limit \citep{moor21}, and it could be older than 300 Myr given the uncertainty.}), and the data are not normally distributed or contain outliers, these statistical tests reveal no strong correlations among the EDD properties.

The apparent lack of a strong correlation between the EDD properties and 3--5 \micron\ variability is not surprising, as a system could remain in a quiescent state for extended periods before exhibiting large variations following a significant impact \citep{su22_hd166}. Photometry at 3--5 \micron\ only traces the hottest part of the disk. 
The varying time since the giant impact that created the EDD phenomenon for different systems further complicates the situation and prevents a clear correlation. The only potential trend related to disk variability emerges when we differentiate EDDs in $W_{10}$ space, as illustrated in Figure \ref{fig:var_w10}. The median variability values for the large $W_{10}$ ($\gtrsim$12.5 \micron) systems, indicated by the horizontal dashed line, are $\sim$5 times higher than the median values for the small $W_{10}$ systems. Notably, all large $W_{10}$ systems are classified as silica-poor.  However, such a difference is washed out in the $\Delta \Sigma$ space, likely due to variation in the dust location (i.e., temperature).  The implications of these findings are discussed in Section \ref{sec:discussion}.

We searched for possible correlations between multiplicity and EDD properties, and no obvious trends were found (some examples are shown in Figure \ref{fig:WISEvariability}). As noted by \citet{moor24_edds_visir}, three of the old ($>$300 Myr) EDDs, which are now identified as having large $W_{10}$ values, are all in wide binary systems with high eccentricity. However, not all EDDs with large $W_{10}$ ($\gtrsim$12.5 \micron) values have stellar companions (see Figure \ref{fig:var_w10} and the lower-right panel of Figure \ref{fig:WISEvariability}). Given the small sample size, the exact role that multiplicity plays in creating EDD phenomenon is unclear. A larger sample, particularly one that extends to older ages, along with better determination of binary properties, is needed for future in-depth study.

\subsection{10/20 \micron\ Variability As Traced by Synthesized Photometry }
\label{sec:10_20umVar}

\begin{figure*}
    \centering
    \includegraphics[width=0.329\linewidth]{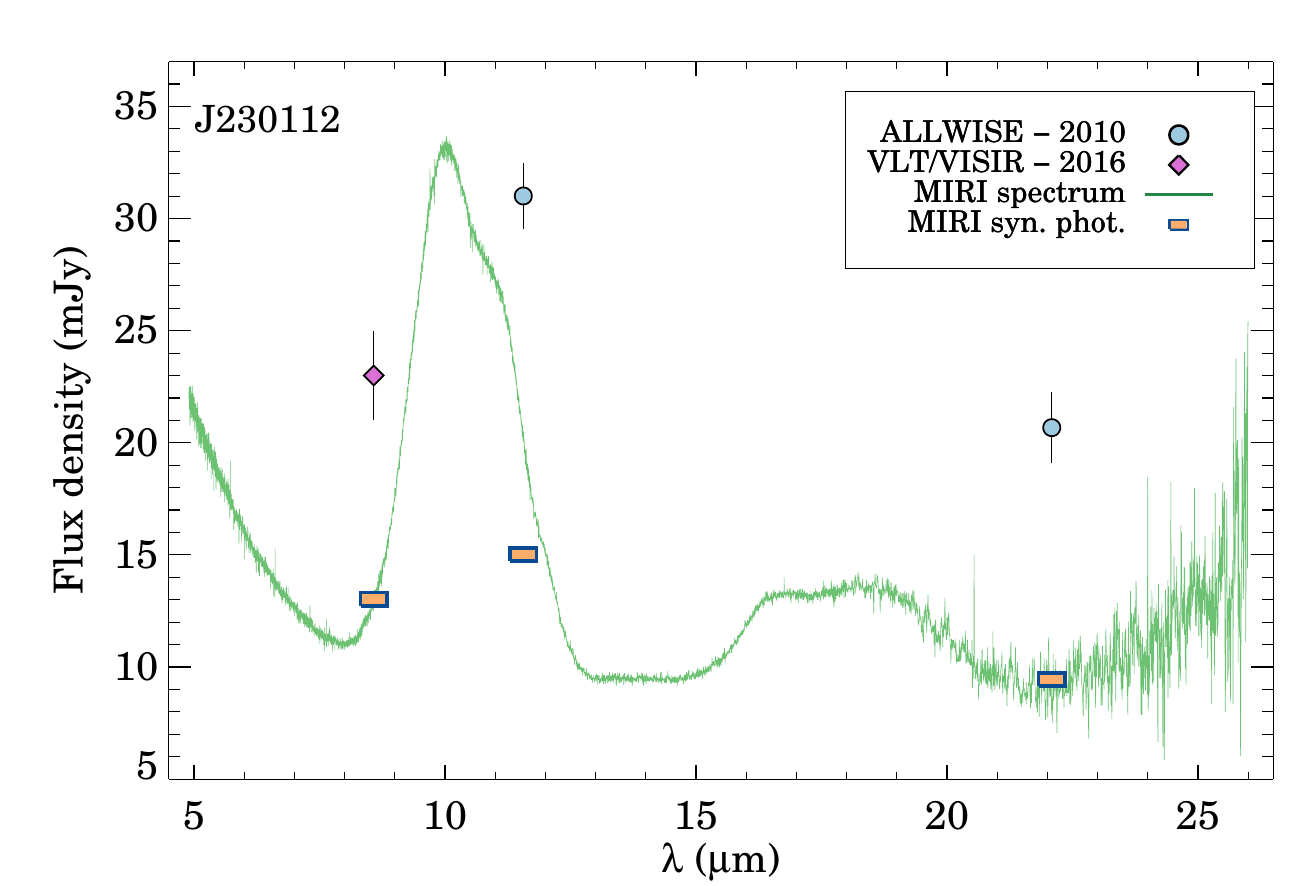}
    \includegraphics[width=0.329\linewidth]{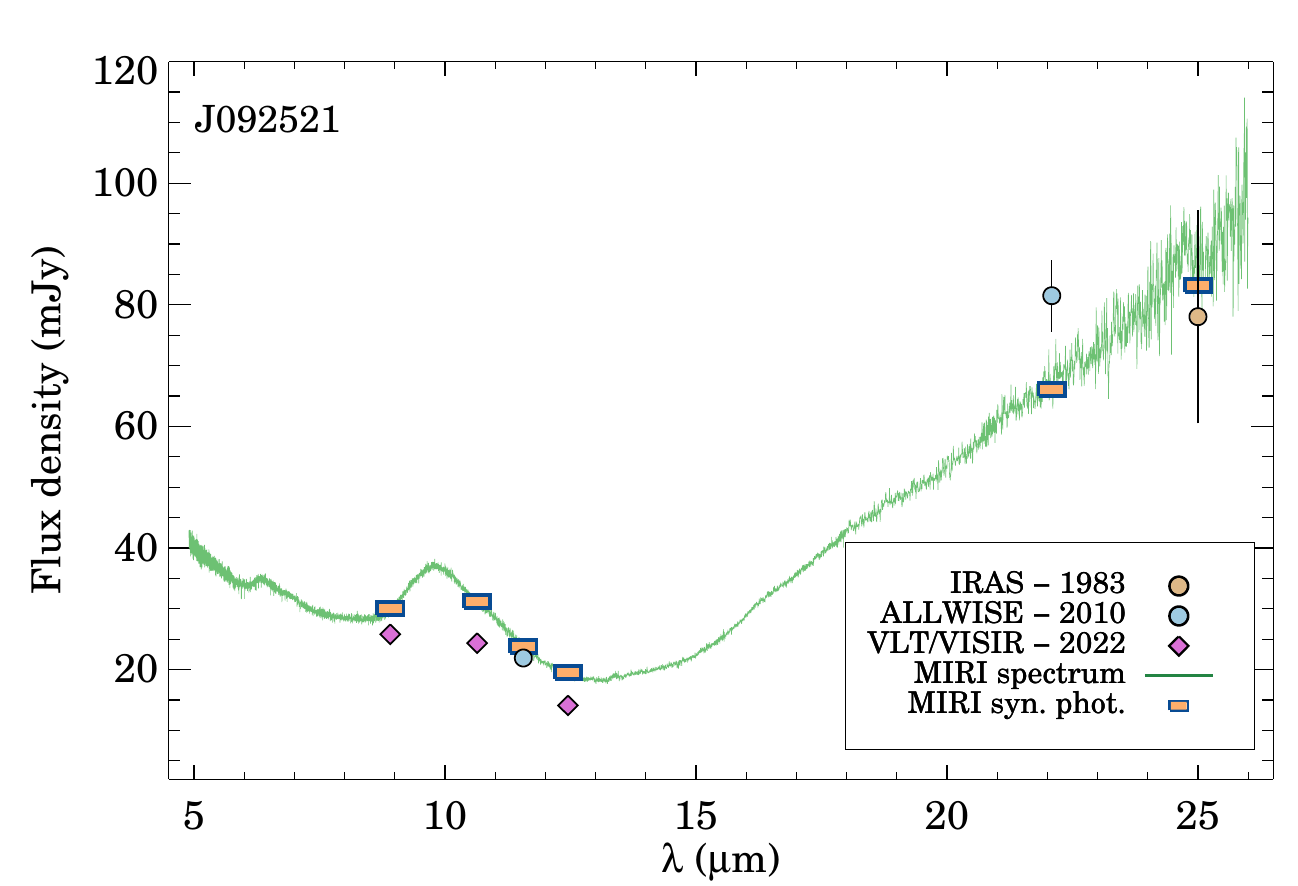}
    \includegraphics[width=0.329\linewidth]{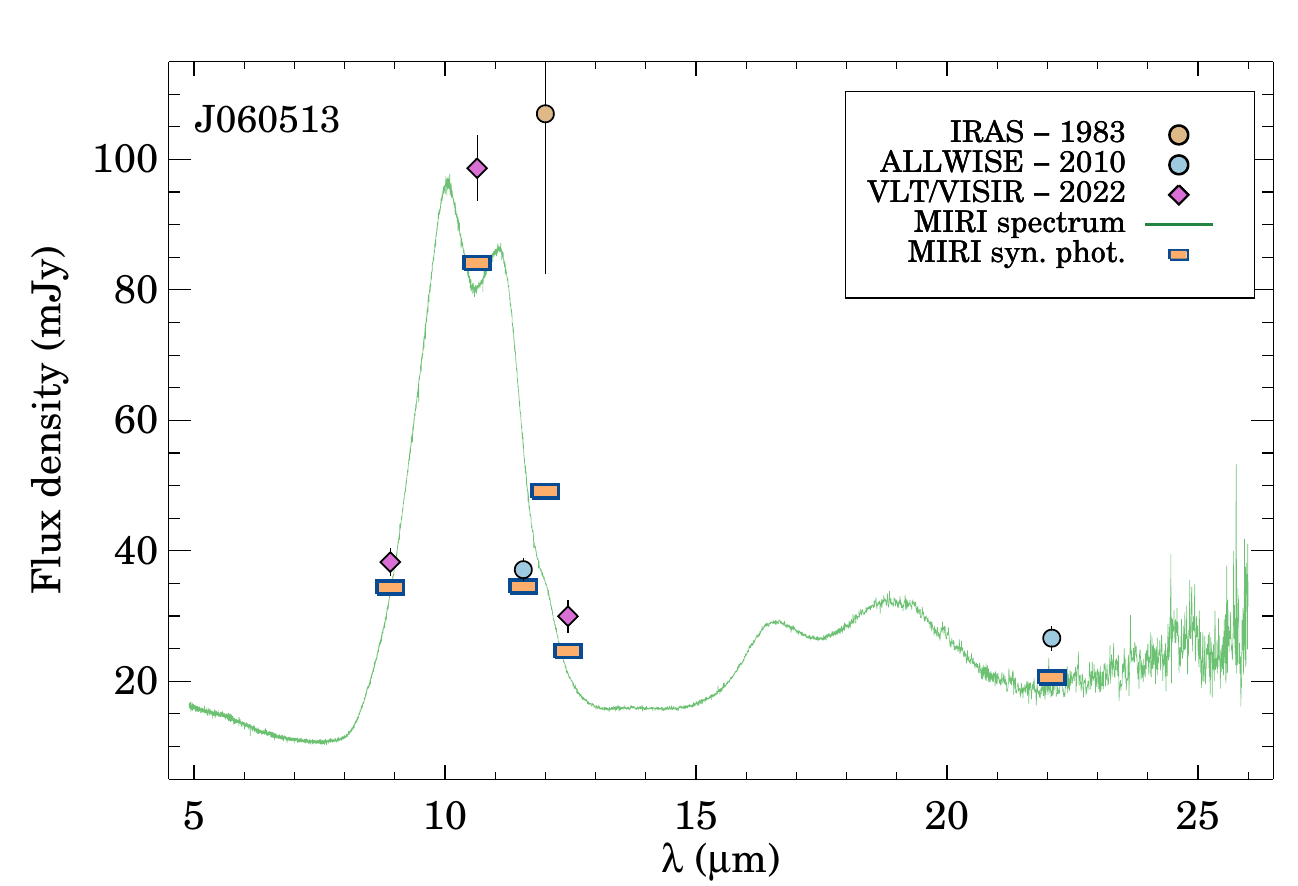}
    \includegraphics[width=0.329\linewidth]{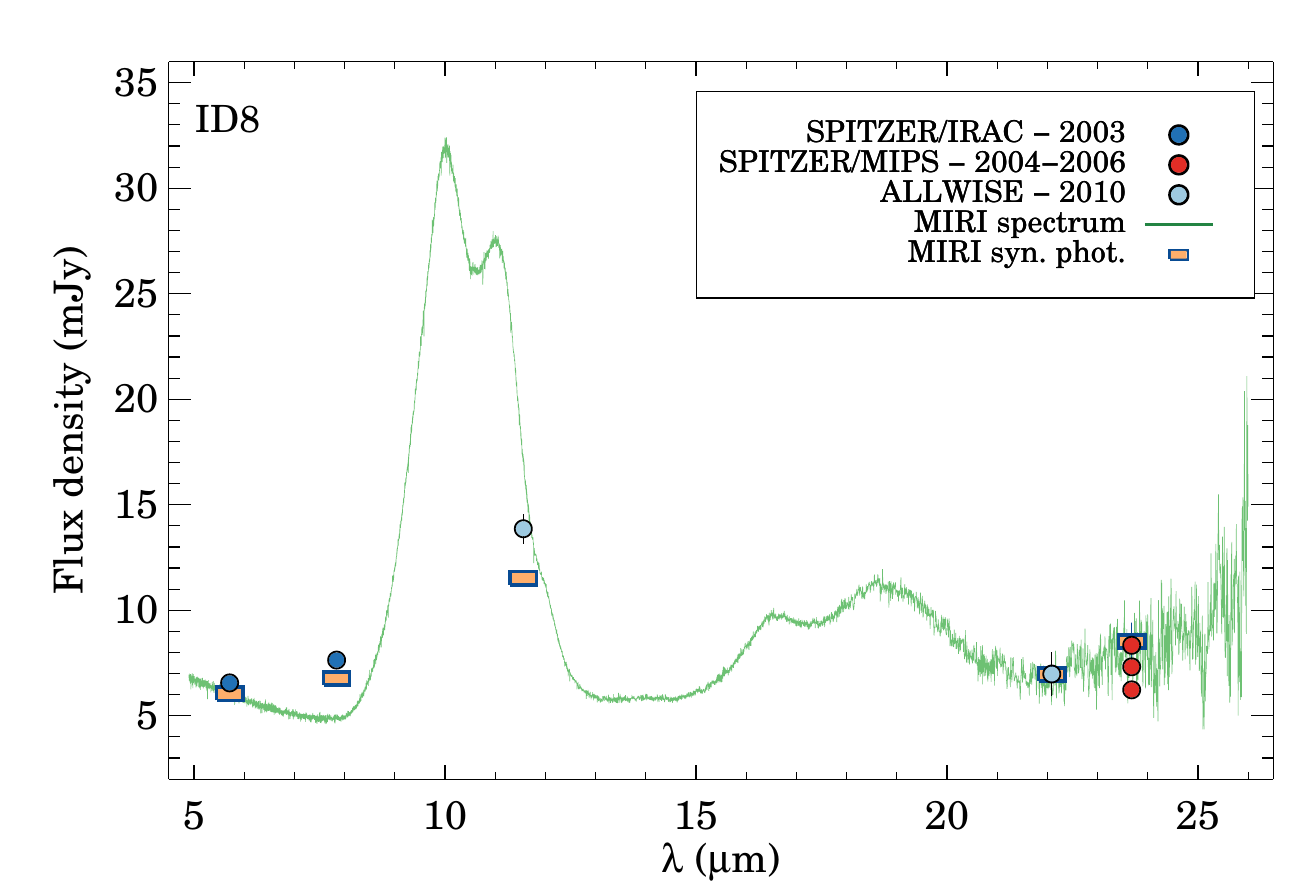}
    \includegraphics[width=0.329\linewidth]{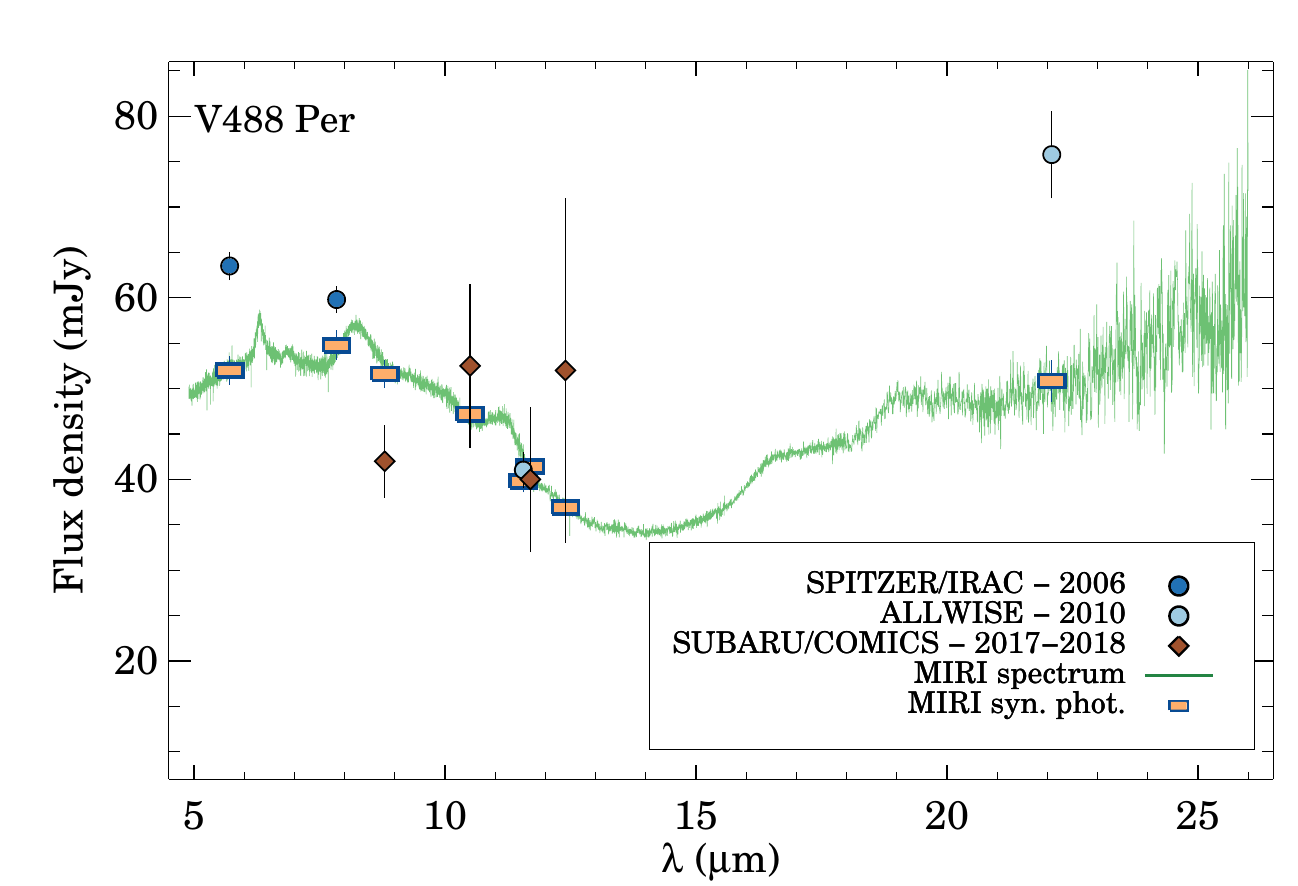}
    \includegraphics[width=0.329\linewidth]{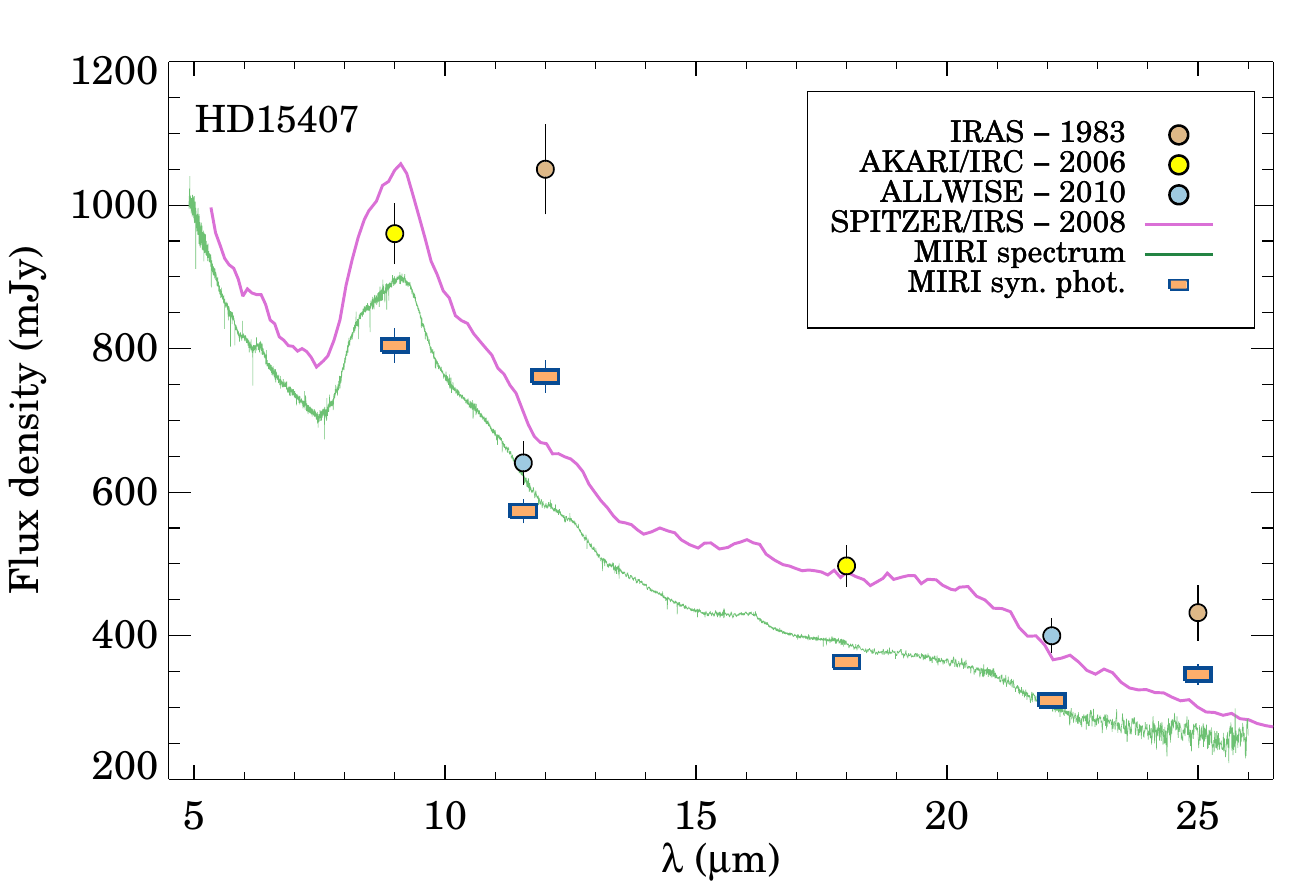}
    \includegraphics[width=0.329\linewidth]{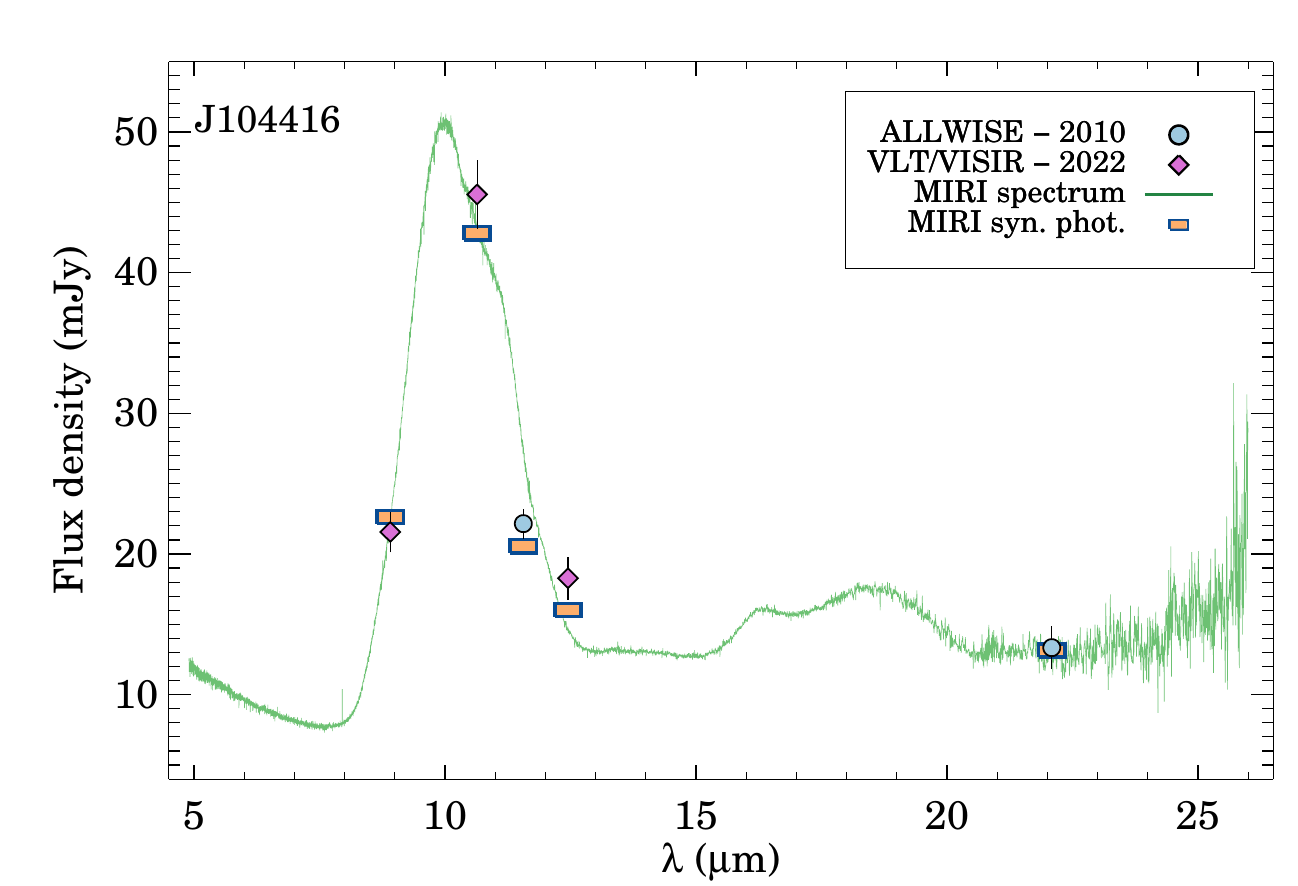}
    \includegraphics[width=0.329\linewidth]{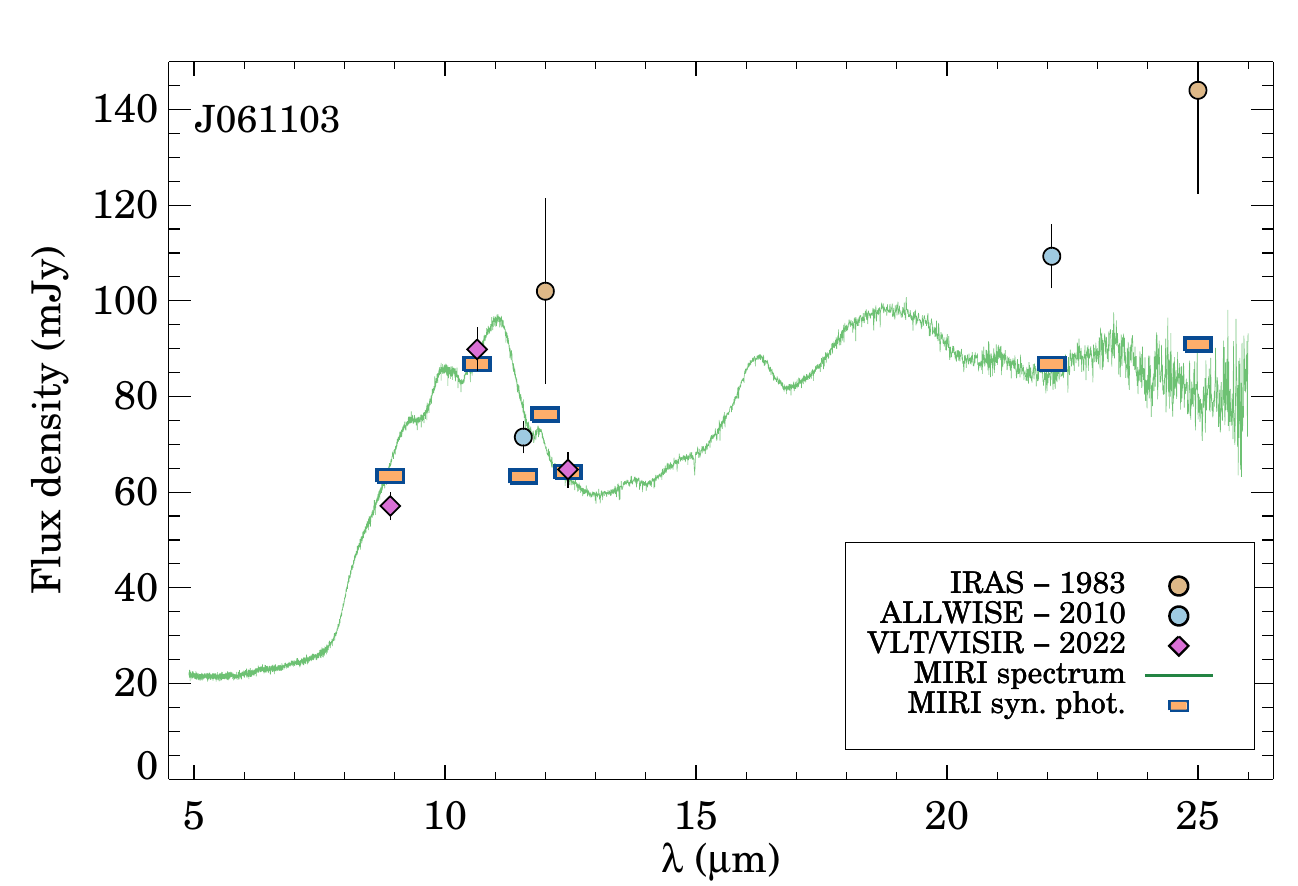}
    \includegraphics[width=0.329\linewidth]{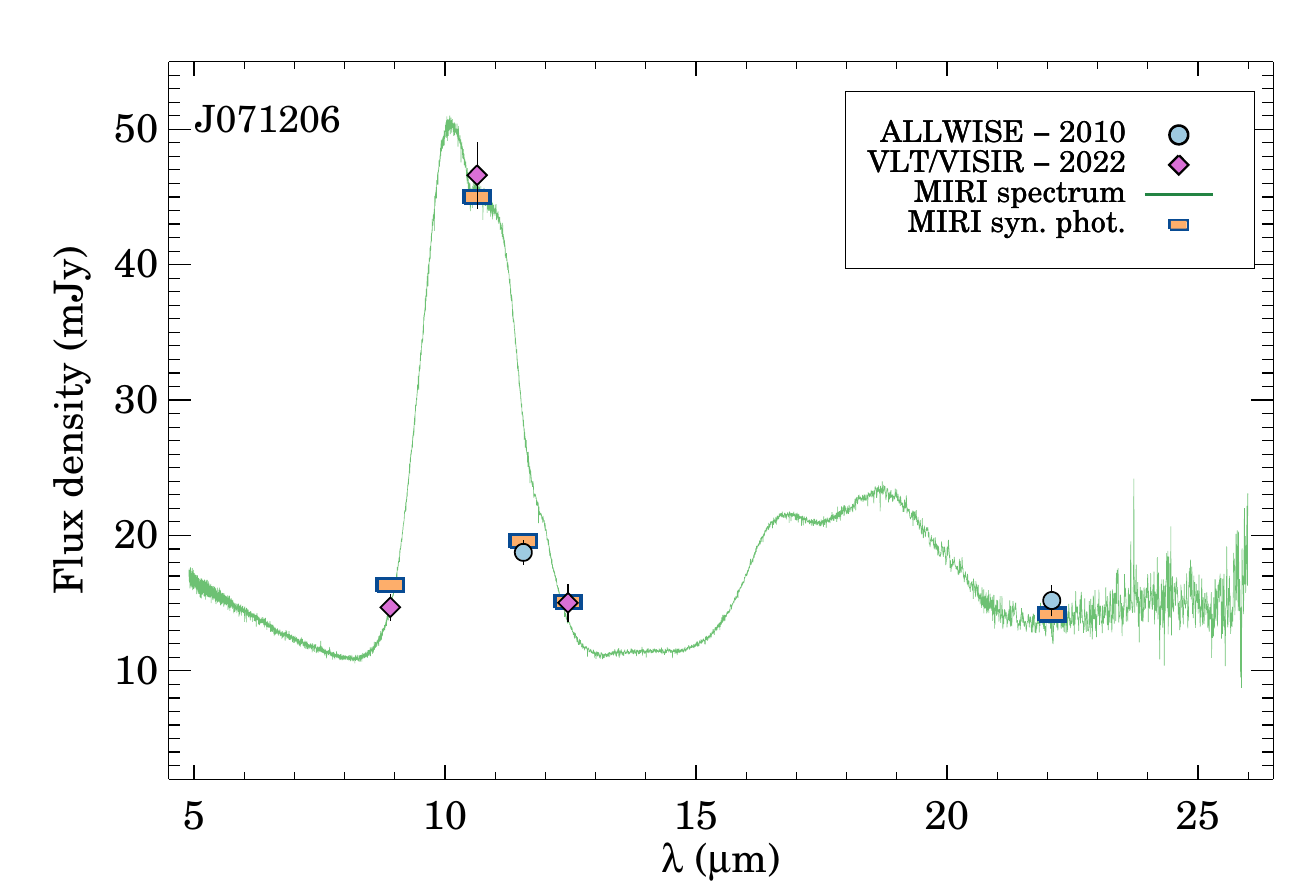}
    \includegraphics[width=0.329\linewidth]{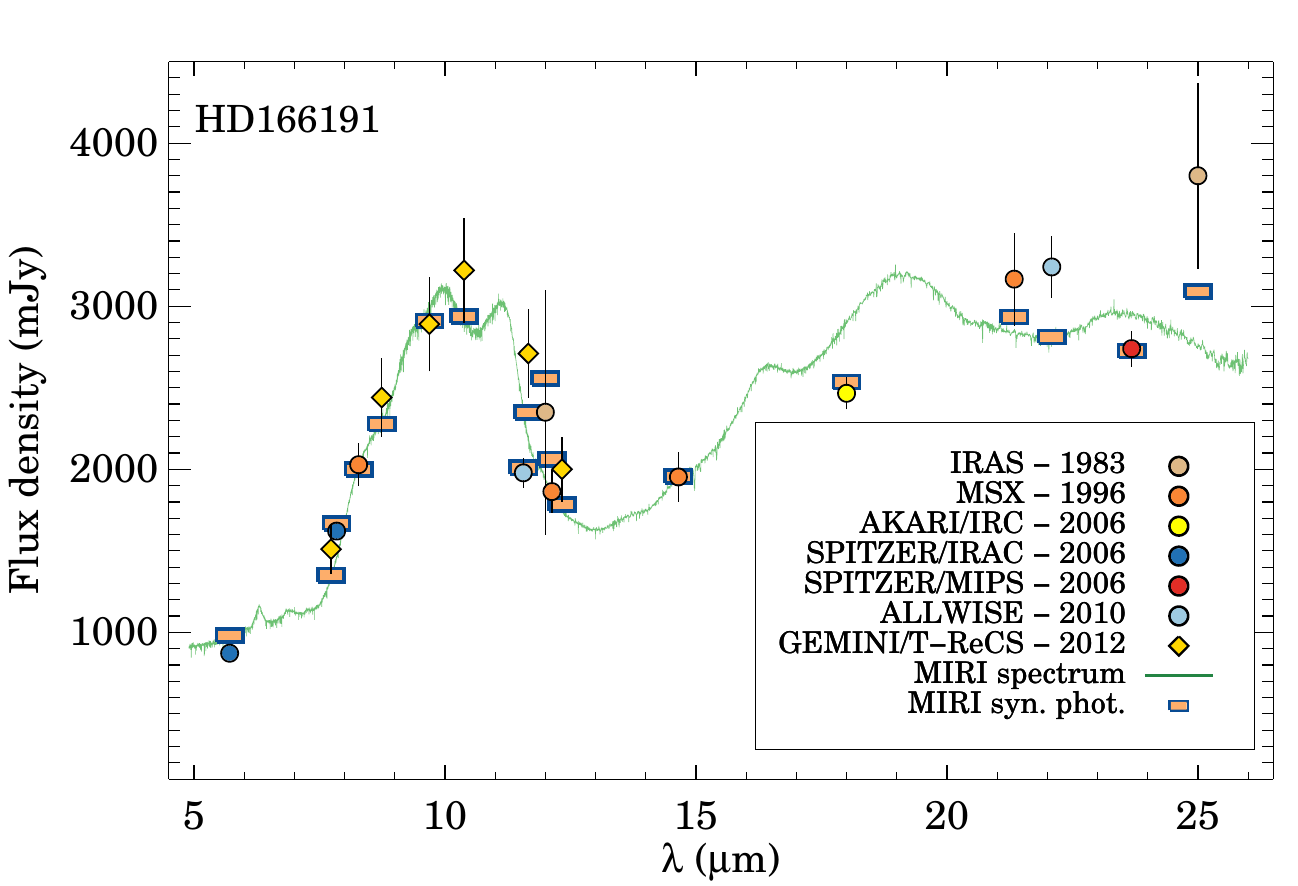}
    \includegraphics[width=0.329\linewidth]{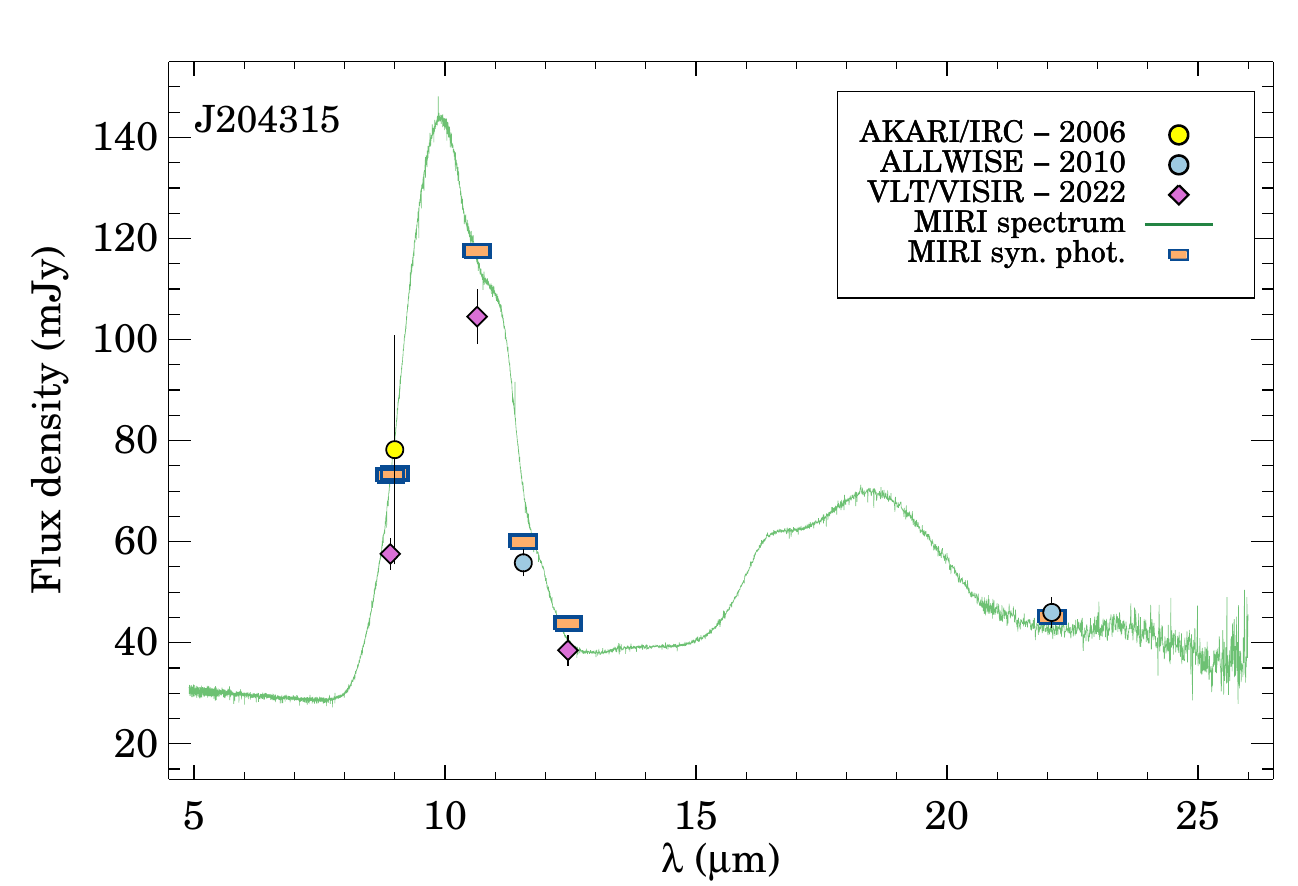}
    \includegraphics[width=0.329\linewidth]{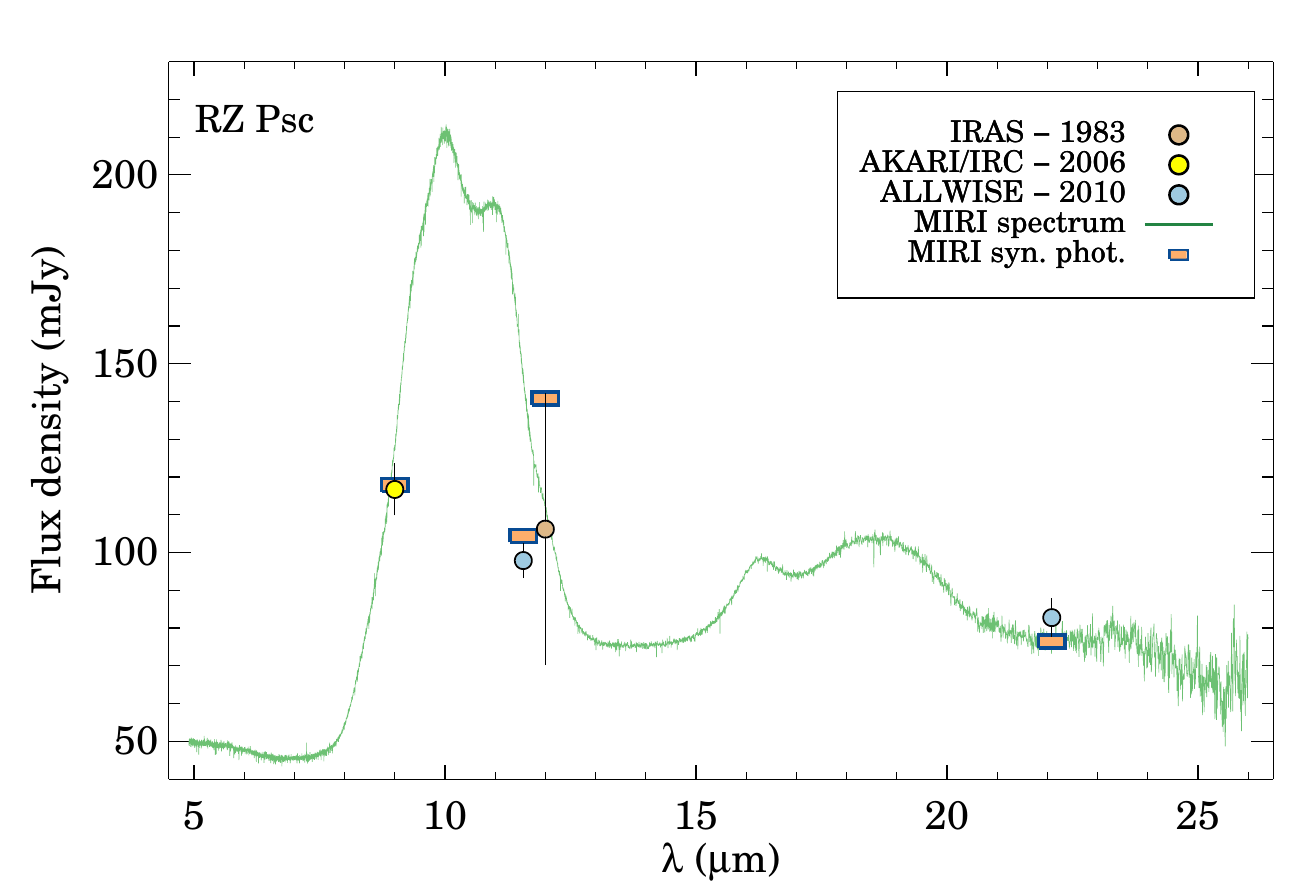}
    \includegraphics[width=0.329\linewidth]{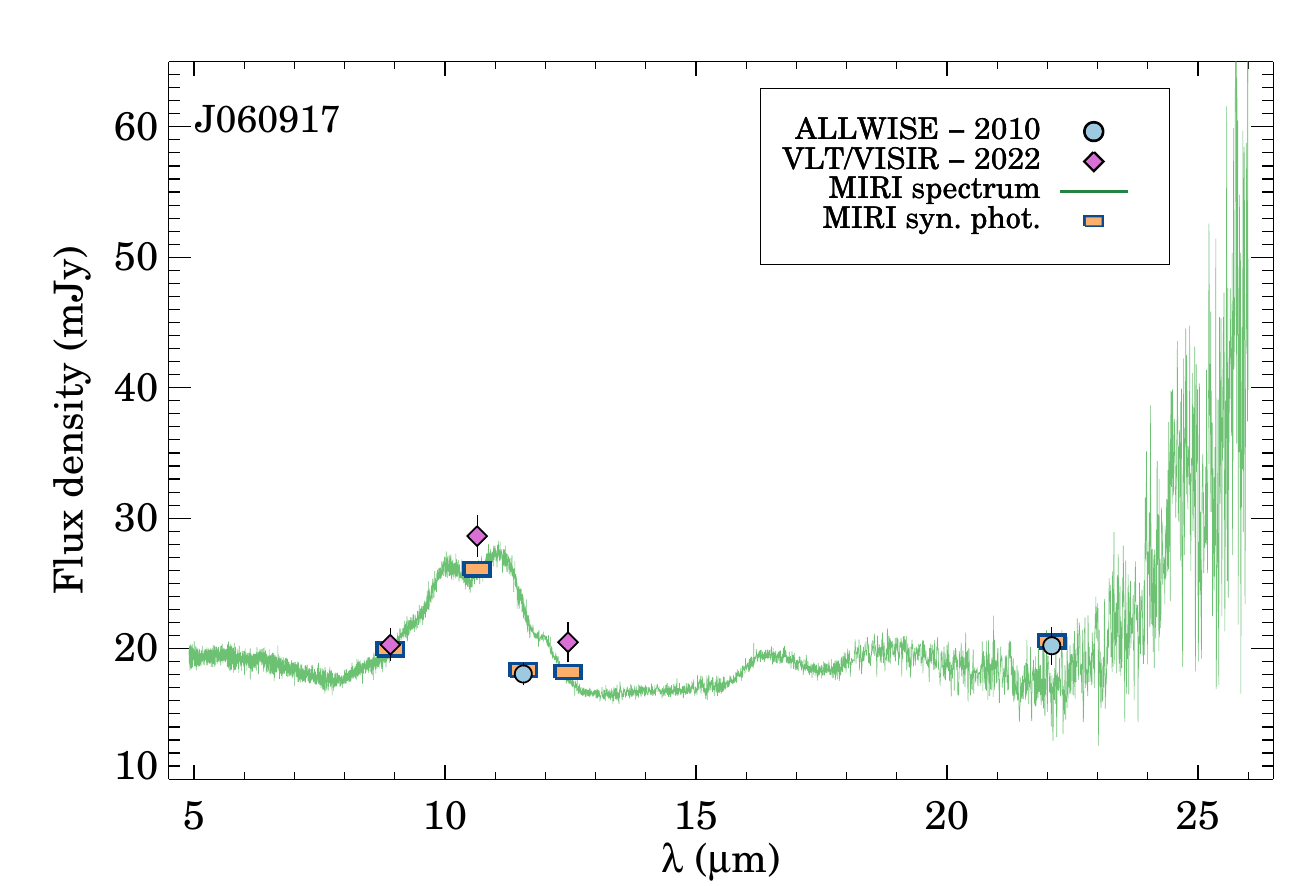}
     \includegraphics[width=0.329\linewidth]{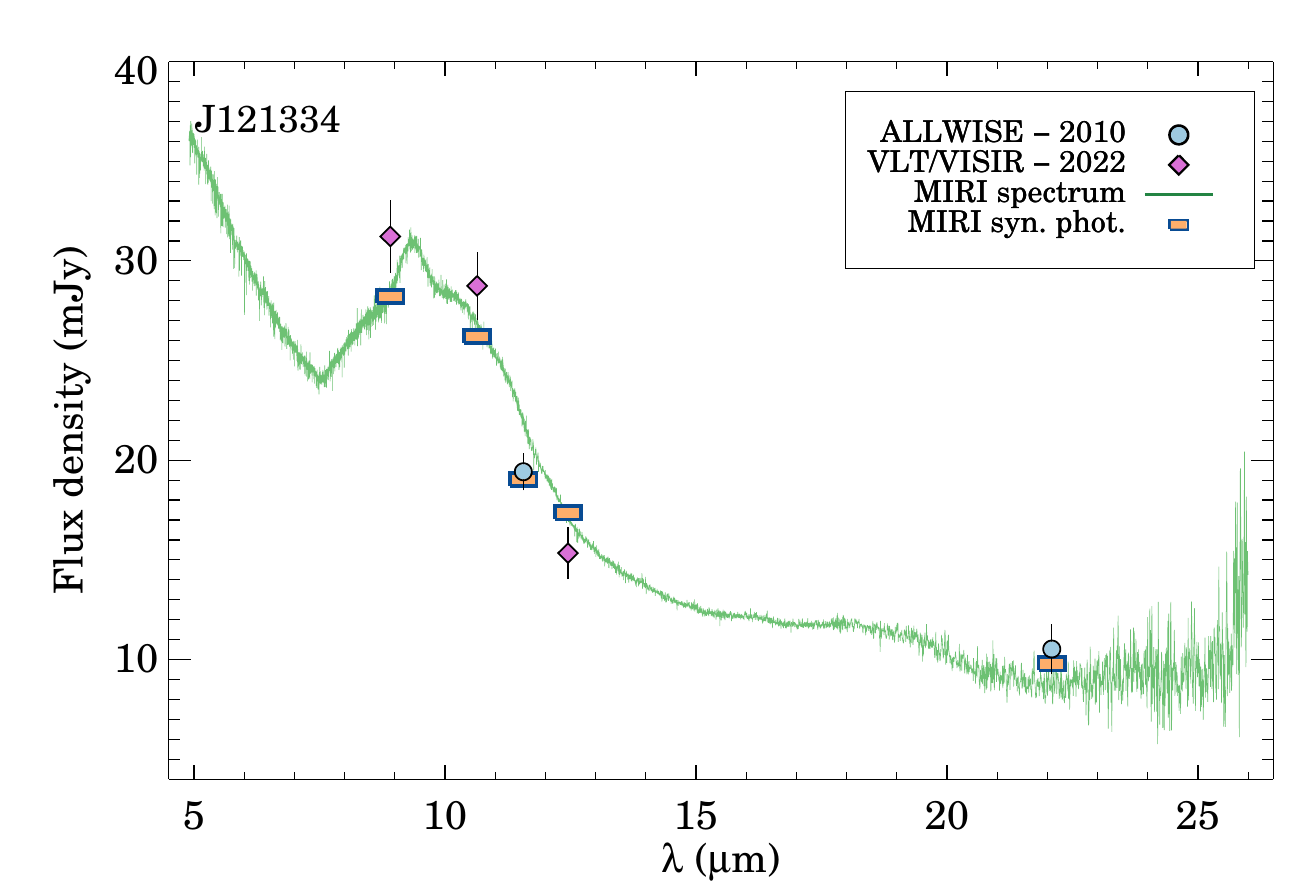}   
    \caption{Mid-infrared 10/20 \micron\ variability for 14 EDDs studied in this work by comparing previous photometry taken over the past few decades to the synthetic photometry derived from the 2024 JWST/MIRI/MRS spectra. The 10/20 \micron\ variation behaviors are summarized in Table \ref{tab:measuredindices} along with other published studies (see \citealt{su25_hd23514} for HD\,23514 and \citealt{samland25_hd172555} for HD\,172555). For HD\,15407, the Spitzer/IRS spectrum taken in 2008 is also shown, illustrating the fading behavior along with similar solid-state features -- commonly found in these EDDs -- regardless of whether they exhibit fading, brightening, or both behaviors. }
    \label{fig:midir_var}
\end{figure*}

\begin{figure*}
    \centering
    \includegraphics[width=\linewidth]{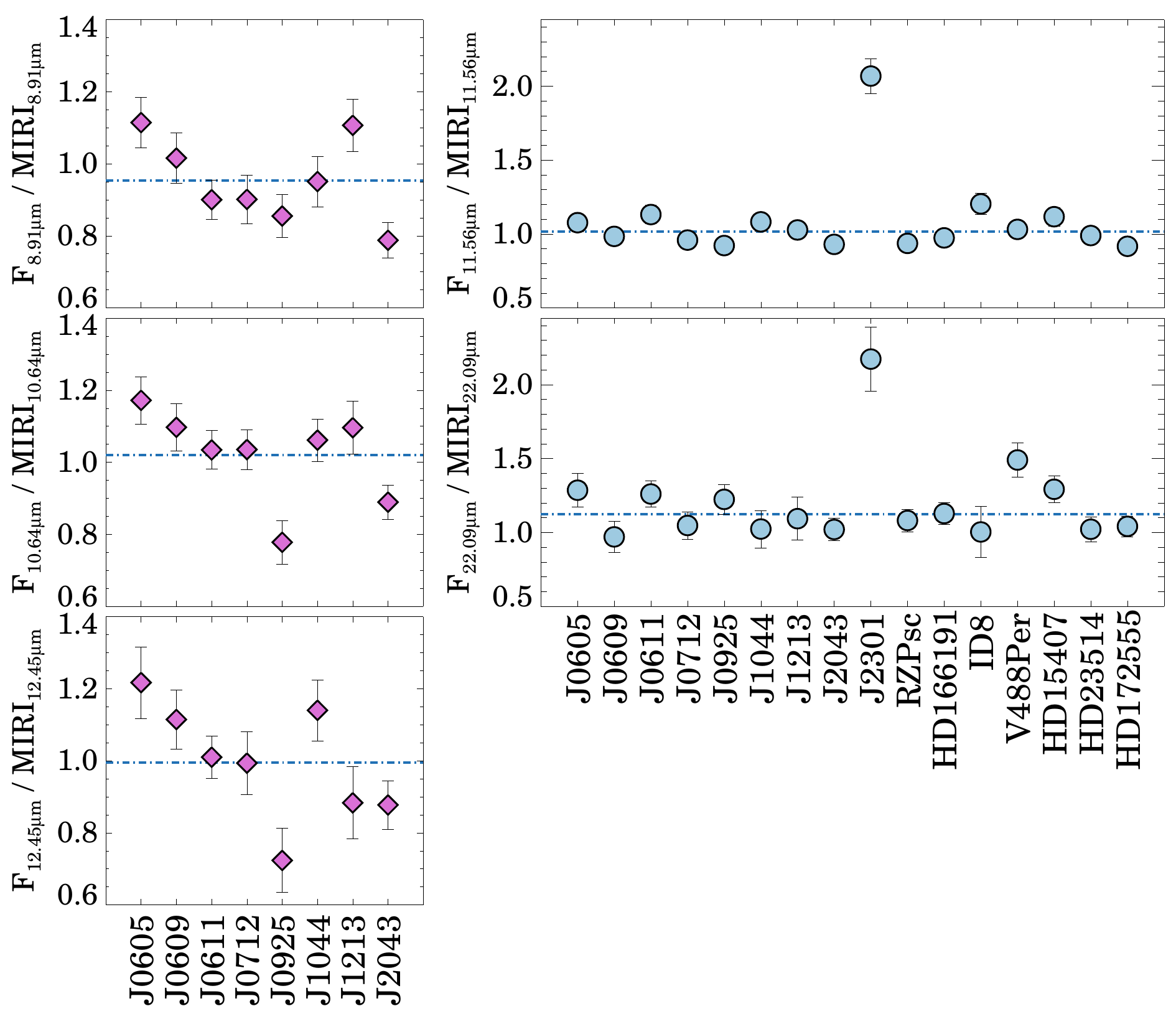}
    \caption{This figure shows the flux ratio between the previous 10/20 \micron\ photometry and the synthetic values derived from the JWST/MIRI/MRS spectra. The left panel displays the results for the VLT/VISIR narrow-band photometry, while the right panel depicts the results using WISE W3/W4 data}
    \label{fig:midir_var2}
\end{figure*}

We further investigate the stability of the 10 \micron\ feature by comparing the synthetic photometry derived from the JWST/MIRI/MRS spectrum to other existing 10--20 \micron\ photometry from IRAS, AKARI, Spitzer, MSX, and ground-based mid-infrared instruments as outlined in \citet{gorlova07,schneider13,melis21,rieke21_v488per,moor21} and \citet{su25_hd23514}. For the 16 EDDs observed with JWST, we assume a 2\% uncertainty in the flux calibration, which is added quadratically when evaluating the significance of change  (see Figure \ref{fig:midir_var} for examples). As anticipated, we observe both brightening and fading behaviors as listed in Table \ref{tab:measuredindices}. However, fading behavior is detected more frequently across the entire sample. Figure \ref{fig:midir_var2} displays the flux ratios between earlier and later JWST measurements for VLT/VISIR narrow-band and WISE W3/W4 photometry. There are more systems with ratios greater than 1 (indicating fading) than those with ratios less than 1 (indicating brightening). The robust average of the measured flux ratios (indicated by the horizontal line in Figure \ref{fig:midir_var2}) is around 1, except for WISE/W4, which has a value of 1.12. Although the difference in absolute flux calibration might be a factor, it should affect all WISE bands, not just W4; therefore, we consider the general fading in WISE/W4 to be generic. This fading trend among the sample is expected if the EDD phenomenon is linked to the aftermath of a significant past event.

The consistent trend observed in the VLT/VISIR narrow-band photometry for eight out of the 16 systems suggests that variability does not alter the overall shape of the 10 \micron\ feature; instead, the changes are primarily attributed to the underlying continuum rather than the solid-state features from the small grains. While photometry alone is not definitive proof of this behavior, a similar conclusion was drawn from comparisons between JWST spectra and those from previous Spitzer or ground-based mid-infrared spectroscopic studies. Nearly identical dust features in the giant impact (silica-rich) disks around HD\,172555 and HD\,23514 have been reported by \citet{samland25_hd172555,su25_hd23514}, with the former exhibiting 5\% brightening at 9 \micron\ and the latter remaining stable at 9 \micron\ but showing a 10\% fading at 20 \micron. Here we report that the JWST spectrum of HD\,15407, the third silica-rich EDD discovered by Spitzer, shows $\sim$15--20\% fading compared to the Spitzer one taken 15 years earlier (Figure \ref{fig:midir_var}). The 10 \micron\ properties derived from the Spitzer HD\,15407 spectrum are: FWHM = 3.15$\pm$0.12, $W_{10} =$3.03$\pm$0.40, $P_{10} =$2.04$\pm$0.06, $O_{10} =$1.31$\pm$0.04, $S_{10} =$7.62$\pm$0.69, $S_{10,s} =$6.08$\pm$0.17 -- all very similar to the values from the JWST spectrum (Table \ref{tab:measuredindices}), demonstrating the effectiveness of using dust indices. Additionally, both RZ\,Psc and J2301 have VLT/VISIR N-band grism spectra \citep{kennedy17_rzpsc,melis21} taken in 2016, and the feature morphology remains very similar to that of the JWST spectra taken in 2024. For J2301, although the dust composition as measured in the small grains remains stable, the $W_{10}$ value drops by $\sim$55\% over a span of 8 years \citep{melis26}. V488\,Per, a silica-rich system identified in this study, has two epochs of JWST MIRI/MRS spectra taken six months apart that appear to be identical within calibration uncertainty. A silica-poor system, ID\,8, has four high-quality mid-infrared spectra including one from Spitzer/IRS, proving a baseline of 17 years. All four spectra exhibit similar dust mineralogy (i.e., without changes in dust composition)  but reveal varying degrees of wavelength-dependent variation in the dust continuum (Su et al.\ in prep.).

In summary, we found no clear correlation between the degree of disk variation and the impact-produced dust mineralogy or system age and multiplicity. Nonetheless, the EDD systems with a high level of optically thin dust (i.e., $W_{10}>$12 \micron) that are predominantly rich in forsterite dust tend to exhibit greater variability by a factor of 5, compared to the others. There appears to be no significant change in the 10 \micron\ mineralogy, suggesting that the composition of small grains formed during past violent events remains unchanged over decadal timescales. The infrared variability in EDDs exhibits a wide range of behaviors, depending on the observed wavelengths, which may different aspects of the impact-produced debris and reflect varying sampling times in the aftermath.

\facilities{JWST (MIRI), Spitzer (IRS), WISE}

\begin{acknowledgments}
Work on this paper was supported by grants 80NSSC24K1571 under the NASA ADAP program, and JWST PID\,3189 program to the Space Science Institute, and partially supported by 80NSSC18K0555 from NASA Goddard Space Flight Center to the University of Arizona. 
AM is supported by the Hungarian National Research, Development and Innovation Ofﬁce \'Elvonal grant KKP-143986. AK is supported by the NKFIH NKKP grant ADVANCED 149943. Project no.149943 has been implemented with the support provided by the Ministry of Culture and Innovation of Hungary from the National Research, Development and Innovation Fund, financed under the NKKP ADVANCED funding scheme. AAS is supported by the Heising-Simons Foundation through a 51 Pegasi b Fellowship. RM and IP acknowledge partial support from the National Aeronautics and Space Administration under agreement No.\ 80NSSC21K0593 program "Alien Earths". APJ acknowledges support from the Towson University Fisher College of Science and Mathematics through the Fisher Endowed Chair of physical and biological sciences.

This research has made use of the NASA/IPAC Infrared
Science Archive, which is funded by the National Aeronautics
and Space Administration and operated by the California
Institute of Technology. 

KYL thanks Scott Kenyon for suggestions on editorial clarity and Chiyoe Koike for providing the laboratory measurements of silica. 

The data presented in this article were obtained from the
Mikulski Archive for Space Telescopes (MAST) at the Space
Telescope Science Institute. These observations are associated
with JWST programs 1206, 1282, and 3189. The standard pipeline-processed products can be accessed via DOI:10.17909/14qb-nm35.

\end{acknowledgments}


\end{document}